\pdfoutput=1

\documentclass[11pt]{article}

\usepackage[final]{acl}

\usepackage{times}
\usepackage{latexsym}

\usepackage[T1]{fontenc}

\usepackage[utf8]{inputenc}

\usepackage{microtype}

\usepackage{inconsolata}

\usepackage{graphicx}

\usepackage{booktabs}
\usepackage{multirow}
\usepackage{amsfonts,amsmath}
\usepackage[capitalise,nameinlink]{cleveref}
\usepackage{tikz}
\usetikzlibrary{arrows,angles,quotes,external}
\usepackage[most]{tcolorbox}
\tcbuselibrary{raster}
\usepackage[frozencache,cachedir=minted-cache]{minted}
\usepackage{fontawesome}
\usepackage{makecell}
\usepackage{xcolor}
\usepackage{colortbl}
\usepackage{pgfplots}
\usepackage{xspace}
\usepackage{graphics}
\usepackage{multicol}
\usepackage{listings}
\usepackage{caption}
\usepackage[framemethod=TikZ]{mdframed}
\usetikzlibrary{matrix}
\usepackage{float}

\newtcbox{\inlinebox}[1][]{enhanced,
 box align=base,
 nobeforeafter,
 colback=floralwhite,
 size=small,
 left=0pt,
 right=0pt,
 boxsep=2pt,
 #1}

\usetikzlibrary{shadows}

\definecolor{gray1}{gray}{0.05}
\definecolor{greenish}{RGB}{250, 255, 250}

\newmdenv[
    tikzsetting= {fill=greenish},
    skipabove=0.33em,
    skipbelow=0.33em,
    linewidth=1pt,
    innerleftmargin=4pt,
    innerrightmargin=4pt,
    innertopmargin=2pt,
    innerbottommargin=2pt,
    linecolor=gray1,
    roundcorner=2pt, 
    shadow=true,
    shadowsize=4pt,
    shadowcolor=gray1
]{answerbox}

\newenvironment{result}
{\begin{answerbox}}
{\end{answerbox}}

\definecolor{darkspringgreen}{rgb}{0.09, 0.45, 0.27}
\definecolor{bostonuniversityred}{rgb}{0.8, 0.0, 0.0}

\newcommand{\hbs}[1]{\hspace*{#1\baselineskip}}

\newcommand{\httm}[3][.2]{\ifmmode\mathchoice%
  {\colorbox{#2}{\hbs{#1}\textcolor{white}{$\displaystyle{#3}$}\hbs{#1}}}%
  {\colorbox{#2}{\hbs{#1}\textcolor{white}{$\textstyle{#3}$}\hbs{#1}}}%
  {\colorbox{#2}{\hbs{#1}\textcolor{white}{$\scriptstyle{#3}$}\hbs{#1}}}%
  {\colorbox{#2}{\hbs{#1}\textcolor{white}{$\scriptscriptstyle{#3}$}\hbs{#1}}}
  \else{\colorbox{#2}{\hbs{#1}\textcolor{white}{#3\vphantom{\"Aq}$\mathop{\vphantom{\int}}$}\hbs{#1}}}\fi}

\definecolor{emerald}{rgb}{0.31, 0.78, 0.47}
\definecolor{babyblue}{rgb}{0.54, 0.81, 0.94}
\definecolor{bananayellow}{rgb}{1.0, 0.88, 0.21}
\definecolor{floralwhite}{rgb}{1.0, 0.98, 0.94}
\definecolor{amber}{rgb}{1.0, 0.75, 0.0}
\definecolor{bostonuniversityred}{rgb}{0.8, 0.0, 0.0}
\definecolor{darkgray}{rgb}{0.18, 0.31, 0.31}
\definecolor{ghostwhite}{rgb}{0.97, 0.97, 1.0}
\definecolor{deepchestnut}{rgb}{0.38, 0.25, 0.32}

\definecolor{mpigreen} {RGB} {0, 129, 122}
\definecolor{lacamlilac} {RGB} {107,93,153}
\definecolor{lacamlilac2} {RGB} {93, 109, 152}
\definecolor{lacamlightlilac} {RGB} {174, 166, 201}
\definecolor{lacamdarklilac} {RGB} {51, 10, 102}
\definecolor{lacamgold} {RGB} {255, 87, 0}

\definecolor{lacamgreen} {RGB} {72, 175, 115}
\definecolor{lacamgold5} {RGB} {255, 87, 0}
\definecolor{violet} {RGB} {119, 111, 178}
\definecolor{darkred} {HTML} {67000C}

\definecolor{petroil1} {RGB} {36, 178, 189}
\definecolor{petroil2} {RGB} {36, 165, 175}
\definecolor{petroil4} {RGB} {30, 132, 149}
\definecolor{petroil6} {RGB} {23, 101, 115}

\definecolor{gold2} {RGB} {255, 130, 0}
\definecolor{gold4} {RGB} {250, 100, 0}
\definecolor{gold6} {RGB} {245, 90, 0}

\definecolor{tomato0} {HTML} {e37676}
\definecolor{tomato1} {HTML} {F44336}
\definecolor{tomato2} {HTML} {E53935}
\definecolor{tomato21} {HTML} {eb4c49}
\definecolor{tomato3} {HTML} {D32F2F}
\definecolor{tomato4} {HTML} {C62828}
\definecolor{tomato41} {HTML} {b52121}
\definecolor{tomato5} {HTML} {B71C1C}

\definecolor{peas1} {HTML} {009688}
\definecolor{peas2} {HTML} {00897B}
\definecolor{peas3} {HTML} {00796B}
\definecolor{peas4} {HTML} {00695C}
\definecolor{peas5} {HTML} {004D40}

\definecolor{bgrey0} {HTML} {78909C}
\definecolor{bgrey1} {HTML} {607D8B}
\definecolor{bgrey2} {HTML} {546E7A}
\definecolor{bgrey3} {HTML} {455A64}
\definecolor{bgrey4} {HTML} {37474F}
\definecolor{bgrey5} {HTML} {263238}

\definecolor{olive0} {HTML} {C5E1A5}
\definecolor{olive1} {HTML} {AED581}
\definecolor{olive2} {HTML} {9CCC65}
\definecolor{olive3} {HTML} {8BC34A}
\definecolor{olive4} {HTML} {7CB342}
\definecolor{olive5} {HTML} {689F38}

\definecolor{pink0} {HTML} {FCE4EC}
\definecolor{pink1} {HTML} {F8BBD0}
\definecolor{pink2} {HTML} {F48FB1}
\definecolor{pink3} {HTML} {F06292}
\definecolor{pink4} {HTML} {EC407A}
\definecolor{pink5} {HTML} {FF80AB}

\definecolor{brown0} {HTML} {D7CCC8}
\definecolor{brown1} {HTML} {BCAAA4}
\definecolor{brown2} {HTML} {A1887F}
\definecolor{brown3} {HTML} {8D6E63}
\definecolor{brown4} {HTML} {795548}
\definecolor{brown5} {HTML} {6D4C41}
\definecolor{brown6} {HTML} {5D4037}

\definecolor{yellow0} {HTML} {CDDC39}
\definecolor{yellow1} {HTML} {9E9D24}
\definecolor{yellow3} {HTML} {FFBD2A}
\definecolor{yellow4} {HTML} {FFB000}
\definecolor{yellow5} {HTML} {FFD600}
\definecolor{yellow6} {HTML} {D6C67B}

\newcommand{\dataset}{\text{SeqCoBench}\xspace}

\makeatletter
\newcommand{\srcsize}{\@setfontsize{\srcsize}{7pt}{7pt}}
\makeatother

\newcommand{\typeone}{\scalebox{.9}{\httm[.1]{pink}{\textbf{type-1}}}\xspace}

\newcommand{\typetwo}{\scalebox{.9}{\httm[.1]{olive2}{\textbf{type-2}}}\xspace}

\newcommand{\typethree}{\scalebox{.9}{\httm[.1]{yellow3}{\textbf{type-3}}}\xspace}

\newcommand{\typefour}{\scalebox{.9}{\httm[.1]{brown4}{\textbf{type-4}}}\xspace}

\NewDocumentCommand\emojiwrench{}{
    \includegraphics[scale=0.04]{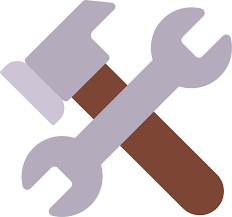}
}

\title{What can Large Language Models Capture\\ about Code Functional Equivalence?}

\author{Nickil Maveli \quad Antonio Vergari\textsuperscript{\emojiwrench} \quad Shay B. Cohen\textsuperscript{\emojiwrench} \\
  School of Informatics, University of Edinburgh \\
  10 Crichton Street, Edinburgh, EH8 9AB \\
  \texttt{\{nickil.maveli,avergari,scohen\}@ed.ac.uk}}

\newcommand{\ignore}[1]{}

\renewcommand{\paragraph}[1]{\noindent{\textbf{#1}}}

\begin{document}
\maketitle

\begin{abstract}
Code-LLMs, large language models pre-trained on code corpora, have shown great progress in learning rich representations of the structure and syntax of code, successfully using it to generate or classify code fragments.
At the same time, 
understanding if they are able to do so because they capture code semantics, and how well, is still an open question. 
In this paper, we tackle this problem by 
introducing \dataset, a benchmark for systematically assessing how Code-LLMs can capture code functional equivalence. 
\dataset{} contains over 20 code transformations that either preserve or alter the semantics of Python 
programs. 
We conduct extensive evaluations in different settings, including zero-shot 
and parameter-efficient finetuning methods on state-of-the-art (Code)-LLMs to see if they can discern semantically equivalent or different pairs of programs in \dataset.
We find that the performance gap between these LLMs and classical match-based retrieval scores is \emph{minimal}, with both approaches showing a concerning lack of depth in understanding code semantics.\footnote{Our code and dataset is available at \url{https://github.com/Nickil21/SeqCoBench}.
\\
\textsuperscript{\emojiwrench}Shared supervision.
}
\end{abstract}

\section{Introduction}
\label{sec:intro}

Comprehending the semantics of code is crucial to generate new code accurately as well as to understand and verify existing code. 
Capturing code semantics would entail the ability to predict \textit{code functional equivalence}, i.e., the property of two functions to produce the same outputs when given the same inputs, yielding the same observable behaviour, even if their implementations differ syntactically. 
In other words, functionally equivalent functions are interchangeable from the perspective of a program's functionality.

\begin{figure*}[!t]
    \centering
    \includegraphics[width=1.\linewidth]{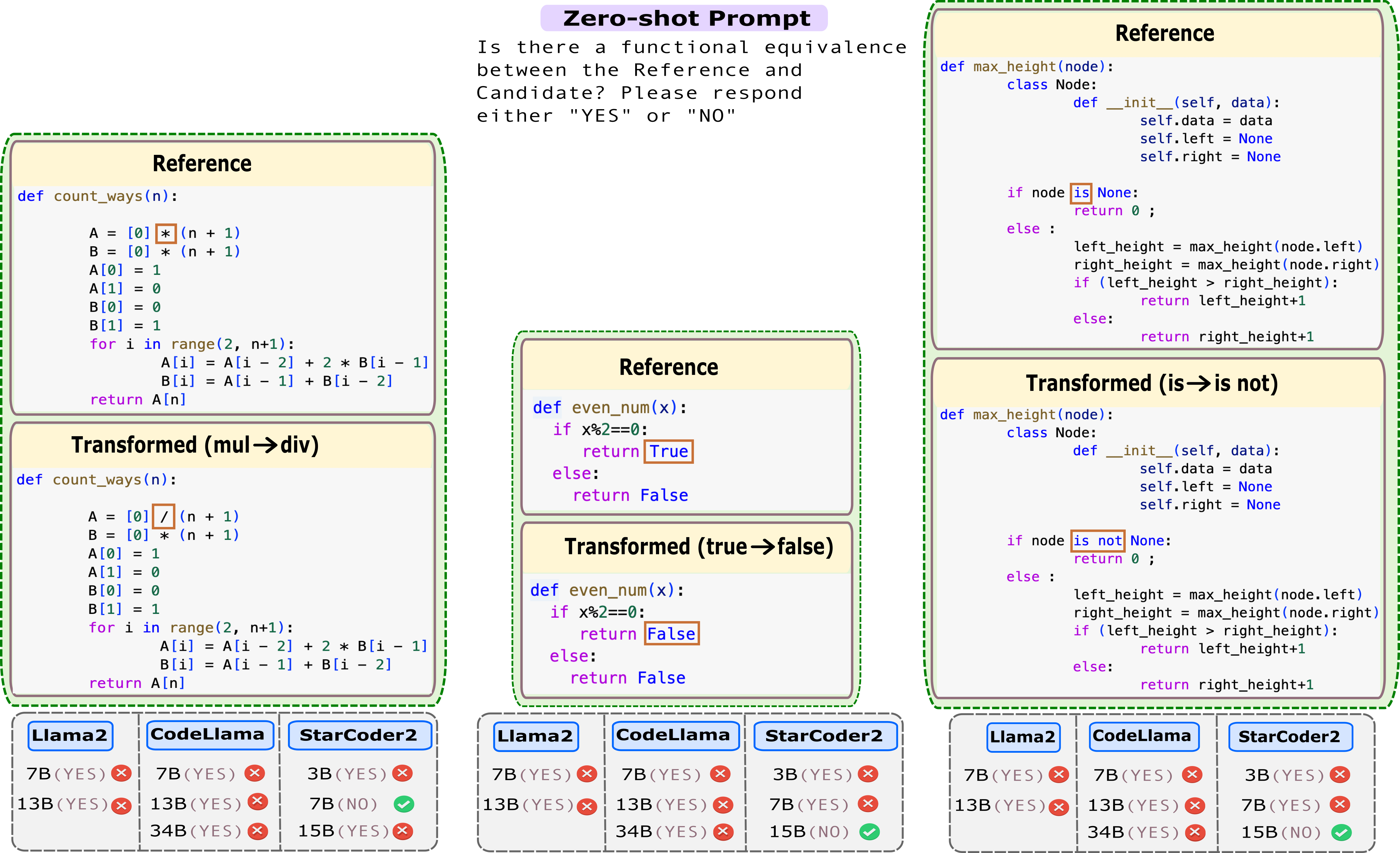}
    \caption{\textbf{State-of-the-art (Code)-LLMs struggle to understand subtle changes in one program syntax that, however, dramatically alter its semantics} as shown by three semantic-altering transformations from \dataset involving arithmetic, boolean and identity operator misuses (\cref{sec:sa-trans}).}
    \label{fig:code-llms-struggle}
\end{figure*}

Identifying such functional equivalences is important for software development and formal verification, as it enhances software quality by detecting redundant code, encouraging reusability, preventing bug spread~\citep{mondal_bug-proneness_2018}, and boosting developer productivity.
For example, when code is refactored~\citep{10479398} or optimized~\citep{shypula2024learning}, it is desirable to automatically confirm that the new and old implementations behave the same. 
In static analysis, it is useful for reducing the risk of unexpected behaviour in a piece of code~\citep{ding-etal-2023-static}, reducing the effort required to verify a system's correctness and to find errors or inconsistencies in the code.

While determining the equivalence between two code segments is an undecidable problem in general~\citep{poonen_undecidable_2014}, in practice, this can be partially achieved by focusing on a narrower input and code domain and by running unit tests on it.
These execution-based code evaluation strategies have become increasingly widespread for evaluating \textit{code generation tasks}, such as program synthesis, code translation and code summarization ~\citep{huang-etal-2022-execution,wang-etal-2023-execution} with Code-LLMs, LLMs pre-trained on large code corpora~\citep{roziere_code_2024,li_starcoder_2023}. 

However, execution-based evaluation comes with drawbacks.
Firstly, it cannot scale to complex codebases that resemble real-world software domains. Currently, the test cases apply mostly to closed-domain problems having limited coverage due to the presence of either built-in functions~\citep{li_competition-level_2022} or handpicked libraries from a specific field~\citep{lai_ds-1000_2023}. 
Secondly, it is infeasible to cover all possible inputs, edge cases, and execution traces, and 
while passing tests is a good proxy for functional correctness, it does not necessarily imply the model truly understands the semantics behind the code.
This leaves us with the open question: \textit{how much are Code-LLMs that are remarkable at code generation able to identify aspects of semantics such as code functional equivalence?}

In this work, we try to answer this question by introducing the \textbf{\textit{\underline{Se}mantic \underline{Eq}uivalence \underline{Code} \underline{Bench}mark}}~(\dataset).
\dataset is a challenging benchmark comprising Python programs generated by applying semantic transformations to a reference program, with the objective of preserving (resp. altering) its semantics while altering (resp. preserving) most of its syntax.
When called to evaluate if the two pieces of code in a pair from \dataset are functionally equivalent or not,  state-of-the-art (Code)-LLMs can get confused, as illustrated in \cref{fig:code-llms-struggle}. Our findings indicate that Code-LLMs have a weak sense of code semantics that breaks when we introduce subtle variations in our \dataset dataset, which helps to systematically measure this effect.

To summarize our contributions:
\textbf{(a)}
We design \dataset to comprise various semantic transformations (\cref{sec:dataset}) according to different types of code clones (\cref{sec:background}) and to understand better which fragment of the syntax-semantics spectrum LLMs capture. 
\textbf{(b)} We use \dataset to extensively evaluate not only state-of-the-art (Code)-LLMs but also classical match-based metrics to capture code functional equivalence (\cref{sec:experiments}).
This includes experiments on zero-shot learning and parameter-efficient fine-tuning (PEFT) settings, noting that Code-LLMs struggle on \dataset and perform on par with classical match-based similarity metrics.
\textbf{(c)}
We investigate which transformations are most challenging to reason with on our benchmark.

\section{Related Work}
\label{sec:related-work}

\paragraph{Code generation benchmarks.}
The common benchmarks, HumanEval~\citep{chen_evaluating_2021} and Mostly Basic Python Problems (MBPP; ~\citealt{austin_program_2021}), help evaluate Python code synthesis based on functional correctness on relatively simple functions. HumanEval comprises 164 human-curated Python programming challenges proposed by OpenAI. Each task contains a docstring, function signature, function body, and a set of unit tests. Whereas MBPP consists of 974 crowd-sourced and hand-crafted Python functions. Each task contains a natural language description, code solution, and three test cases. 

To increase test coverage, EvalPlus~\citep{liu_is_2023} augments HumanEval test cases to automatically generate and diversify additional test inputs. To tackle dataset contamination, LiveCodeBench~\citep{jain_livecodebench_2024} gathers new coding problems not seen during model training, and EvoEval~\citep{xia_top_2024} which uses LLMs to transform existing benchmarks into novel coding tasks. Our approach focuses on the equivalence of the code itself rather than evaluating the correctness of a generated solution against a predefined problem.

Some benchmarks test LLMs' capability to automate real-world software development processes. SWE-Bench~\citep{jimenez2024swebench} automatically resolves GitHub issues by generating code patches that pass the existing test cases. RepoEval~\citep{zhang-etal-2023-repocoder} evaluates repository-level code completion tasks at various levels of granularity. In contrast, we evaluate functional equivalence without considering the broader context of a codebase or real-world software engineering tasks. While we focus on determining whether two code snippets are functionally equivalent, CRUXEval~\citep{gu_cruxeval_2024} specifically evaluates a model's reasoning skills in predicting inputs/outputs for code understanding and execution abilities.

\paragraph{Code evaluation metrics.}
We can broadly categorise code evaluation metrics (CEM) into two main types: reference-based metrics, which compare the generated code to a known reference, and reference-free metrics, which assess the quality of the generated code without relying on a reference by executing them on test cases. In this study, we focus on reference-based metrics. 

Reference-based metrics typically include \emph{match-based} metrics, which rely on lexical exact token matching, and \emph{LLM-based} metrics, which employ models pre-trained on code. Match-based metrics include n-gram matching metrics like BLEU~\citep{papineni-etal-2002-bleu}, ROUGE~\citep{lin-2004-rouge}, etc., and also incorporate syntactic and semantic properties like CodeBLEU~\citep{ren_codebleu_2020}, CrystalBLEU~\citep{eghbali_crystalbleu_2023}, etc. LLM-based metrics include embedding-based metrics like BERTScore~\citep{zhang_bertscore_2020}, CodeBERTScore~\citep{zhou_codebertscore_2023}, CodeScore~\citep{dong_codescore_2023}, etc. We provide the details of the different evaluation metrics in Appendix~\ref{appendix:eval-metric}. There is a need to build robust CEM as existing metrics show a weak correlation with functional correctness on HumanEval as shown in Appendix~\ref{appendix:func-correctness}.
We cover notable work not already mentioned in Appendix~\ref{appendix:extended-related-work}.

\section{Functional Equivalence of Programs}
\label{sec:background}

Determining the semantic equivalence of two code snippets is a challenging task that \textit{lies on a spectrum}, as different codes can compute the same function, but in different ways.
Consider a space of programs $\mathcal{P}$, each explicitly accepting an input $x \in X$ and outputting an output $y \in Y$. 
The semantics we attach to such programs is based on their input-output mapping, i.e., the function $f: X \rightarrow Y$ they implement. 
However, while two different programs $p, p^{\prime}\in\mathcal{P}$ can be \textit{functionally equivalent} if they implement the same input-output mapping, they can concretely implement such a mapping in a very different way.

For example, two pieces of code can have different syntactic variations -- from minor changes as inserting whitespaces to renaming variables -- as well as implement two different algorithms that however encode the same function.
This can be done simply by using different library APIs or dependencies, different abstraction levels, or algorithms with different time and space complexity, e.g., sorting a list of integers can be equally done with mergesort or insertion sort and both will pass the same unit tests that check for input-output consistency.

We operationalise the question of whether LLMs can capture different functional equivalence relationships in this spectrum through the notion of \textit{code clones} \citep{saini2018code}.
Detecting code clones is a proxy task for checking functional equivalence that is highly relevant in software engineering, e.g., to retrieve similar code snippets for code search~\citep{sun-etal-2023-backdooring} or detect duplicates within a codebase~\citep{yang_c_2023}. 
Classifying code clones into types can help us systematize possible functional equivalence classes. 
Following previous work~\cite{roy2007survey, bellon_comparison_2007}, we categorise code clones into four types based on their complexity and degree of similarity.

\noindent\typeone clones comprise two almost syntactically identical pieces of code that differ only for minor variations in the layout, e.g., by the presence of whitespace and comments.

\noindent\typetwo clones resemble two syntactically identical code fragments except for variations in variable and function names, identifiers, literal values, types, etc. 
They also comprise \typeone differences.

\noindent\typethree clones include two syntactically identical code fragments except for additions, deletions, modifications of several statements, and the differences already specified for \typetwo clones.

\noindent\typefour clones, also known as semantic clones. 
They exhibit identical functional behaviour despite having different syntax, control flow, data flow, or programming languages.

This typology of clones helps us devise diverse groups of transformations that can preserve or alter the semantics of a program while also modifying its syntax.
Such transformations enable us to create a challenging benchmark to systematically evaluate whether LLMs can capture functional equivalence and inspect at which level they can disentangle syntax from semantics, as discussed next.

\section{The \dataset~Dataset}
\label{sec:dataset}
Traditionally, code benchmarks for evaluating LLMs have focused on assessing their ability to generate single-function programs based on natural language descriptions.
This is done for example in popular benchmarks such as HumanEval~\citep{chen_evaluating_2021} and Mostly Basic Python Problems (MBPP; ~\citealt{austin_program_2021}), which comprise human-curated  
Python code snippets with a docstring, function signature, function body, and a set of unit tests to check if generated code satisfies specifications.

Instead, as we want to evaluate
the ability of LLMs to capture functional equivalence between already existing pieces of code, we construct our 
\dataset
by creating pairs of code snippets that are labelled to be functionally equivalent or not.
More formally, given a program $p_i$, we generate the tuple $(p_i, t(p_i), \ell_i)$ where $t$ is a \textit{semantic-preserving} (SP; \cref{sec:sp-trans}) or \textit{semantic-altering} (SA; \cref{sec:sa-trans}) transformation that generates a new code snippet whose semantics changes accordingly and $\ell_i\in\{0,1\}$ is a label indicating if the pair has been generating through a SP (1) or SA (0) transformation.

We build \dataset{} by applying a set of SP or SA transformations to programs appearing in MBPP,
they provide unit tests that help us check if our transformations correctly operate on the program semantics.
We prepare the train/valid/test splits following a 60/16/24 ratio and ensure that there is no overlapping of the original code across different data splits.

\subsection{Semantic-Preserving Transformations}
\label{sec:sp-trans}

Given a program $p$,
 we aim to generate a new code snippet $p^{\prime}$ that is \textbf{\textit{functionally equivalent}} to $p$ (i.e., they encode the same input-output mapping $f$)  \textbf{\textit{while maximizing the token-level differences between the original code}} where appropriate. 
 To this end, we consider the four SP transformations.
 We group them by the corresponding clone type (\cref{sec:background}) while trying to cover all types.
Note that we omit \typeone clones as they typically include comment- and docstring-level perturbations written in natural language and are not technically part of code semantics. They are used for documentation purposes only and do not affect the execution or behaviour of the code itself. 

While adversarial perturbations that alter the function's intended meaning can cause the LLM to ignore the function body completely and instead give more emphasis to the perturbed entity's instructions, reformulating the perturbations using text augmentation strategies such as back-translation~\citep{wang-etal-2023-recode}, synonym substitutions, etc., can make it less challenging for LLMs. We create these transformations using the NatGen package~\cite{chakraborty_natgen_2022}. We leave out transformations on the original code where the necessary requirements to perform the transformation are not met (e.g., lack of for/while loop or boolean operators in the code snippets).

 \paragraph{Rename Variables (RV) \typetwo} 
We primarily use three different adaptations. \textit{Naive:} It renames the most common variable name to VAR\_i. \textit{CB:} It identifies the variable name that appears most frequently in the partial code snippet and then substitutes all occurrences of that variable name throughout the prompt with a new name suggested by CodeBERT. \textit{RN:}
It identifies the variable name that occurs most frequently within the given partial code snippet and then generates a \emph{random} string composed of an equal mix of alphabetic and numeric characters. Finally, it substitutes all instances of the most commonly used variable name with this newly generated random string.

\paragraph{Dead Code Insertion (DCI) \typethree} 
It creates unreachable code blocks at a random location. These could be unused variables or redundant assignments. We place these statements in a block around either a looping (e.g., for, while) or a branching structure (e.g., if-else, switch), if any.

\begin{tcolorbox}[title=Dead Code Insertion Example,sidebyside,righthand width=.77\linewidth,left=1pt,right=1pt,top=1pt,bottom=1pt,sidebyside gap=2mm,width=\linewidth]
  \makeatletter
  \addcontentsline{lol}{subsection}{\kvtcb@title}
  \makeatother

\begin{minted}[fontsize=\scriptsize]{python}
x = 5
y = x + 2
print(y)
\end{minted}

\tcblower

\begin{minted}[fontsize=\scriptsize]{python}
x = 5
z = 10 # Dead code  
y = x + 2
if False: 
    print("This will never execute") # Dead code
print(y)
\end{minted}

\end{tcolorbox}

\paragraph{Operand Swap (OS) \typethree} 
It swaps the first occurrence of boolean operators and, if needed, changes the operator to preserve semantic equivalence. 
For example, if the original code had the condition \texttt{a > b}, the transformation could change it to \texttt{b < a}, swapping the operands "a" and "b" while also changing the operator from ">" to "<" to preserve the same logical meaning.

\paragraph{Loop Transformation (LT) \typefour} 
It converts the first occurrence of for-loop into its equivalent while-loop and vice-versa. 

In the for$\rightarrow$while case, we initialize the counter outside the loop and use a while condition that checks the loop counter against the termination condition. Then, we increment/decrement the counter inside the body, which remains unchanged. 
For the reversed case, we initialize the loop counter in the for-loop statement and include the termination condition taken from the while-loop. The loop counter increment/decrement is merged in the for loop, and the body remains unchanged.

\begin{tcolorbox}[title=Loop Transformation Example,sidebyside,righthand width=.5\linewidth,left=1pt,right=1pt,top=1pt,bottom=1pt,sidebyside gap=5mm,width=\linewidth]
  \makeatletter
  \addcontentsline{lol}{subsection}{\kvtcb@title}
  \makeatother

\begin{minted}[fontsize=\scriptsize]{python}
total = 0
for i in range(n): 
    total += i
\end{minted}

\tcblower

\begin{minted}[fontsize=\scriptsize]{python}
total = 0, i = 0
while i < n: 
    total += i
    i += 1
\end{minted}

\end{tcolorbox}

Loop transformation is challenging as loops often contain critical logic and control flow determining the code's functionality. Modifying or transforming loops risks changing the program's intended behaviour. In contrast, simpler code modifications like dead code insertion, variable renaming, or operand swaps have more localized effects and require less global reasoning about data dependencies or code semantics. LLMs can perform these transformations more reliably by learning from the training data patterns. We show the structure of transformations for a representative program taken from the dataset in Appendix~\ref{appendix:dataset-examples}.

\subsection{Semantic-Altering Transformations}
\label{sec:sa-trans}
In this case, our goal is to generate a program $t(p)$ that is functionally not equivalent to the original code $p$ while \textbf{\textit{maximizing token-level similarity to the original code}}.
In other words, generate pairs of programs that are \textit{not} clones but might fool a superficial comparison.
Accordingly, we consider six families of SA transformations. 
As before, we leave out transformations on the original code where the necessary requirements to perform the transformation are not met (e.g., unavailability of identity or boolean operators).

\paragraph{Arithmetic Operators Misuse (AOM).}
We search for the first occurrence of an arithmetic operator and modify it to its semantic opposite counterpart. 
For example, we replace 
\mintinline{python}{a + b} with \mintinline{python}{a - b}, \mintinline{python}{a * b} with \mintinline{python}{a / b}, and the other way around. 
Similarly, we replace augmented assignment operators \mintinline{python}{a += b} to \mintinline{python}{a -= b} and \mintinline{python}{a *= b} to \mintinline{python}{a /= b}.

\paragraph{Dissimilar Code Selection (DCS).}
We randomly select five distinct code snippets from the base dataset, excluding the original code $p$, and create five additional code pairs using $p$ as a reference.

\paragraph{Identity Operators Misuse (IOM).}
We look for the first occurrence of an identity operator 
and adjust it to its corresponding semantic opposite.
For instance, we replace \mintinline{python}{a is c} with \mintinline{python}{a is not c}, and vice-versa.

\paragraph{Boolean Operators Misuse (BOM).}
It searches for the first occurrence of a boolean literal 
and replaces it with its logical negation. 
For example, we replace the keyword \mintinline{python}{True} with \mintinline{python}{False}, and vice-versa.

\paragraph{Logical Operators Misuse (LOM).}
It scans for the first occurrence of a logical operator (e.g., \mintinline{python}{and}, \mintinline{python}{or}) and swaps it with another operator.
For instance, we interchange \mintinline{python}{a > 1 and a < 5} with \mintinline{python}{a > 1 or a < 5}, and vice-versa.

\paragraph{Comparison Operators Misuse (COM).}
It searches for the first occurrence of a comparison operator and replaces it with its logical opposite operator type. For e.g., we replace \mintinline{python}{a > b} with \mintinline{python}{a < b}, 
\mintinline{python}{a >= b} with 
\mintinline{python}{a <= b},
\mintinline{python}{a == b} with \mintinline{python}{a != b}, and vice-versa.

Table~\ref{table:statistics} summarizes the dataset statistics 
after applying the transformations to the code fragments in MBPP and splitting into train/valid/test. Furthermore, we clarify how the motivation to construct \dataset differs from code clone detection and code obfuscation techniques in the Appendix~\ref{appendix:other-code-tasks}.
\begin{table}[!t]
  \centering
  
  \resizebox{0.95\linewidth}{!}{
    \begin{tabular}{lcccc}
    \toprule
    \textbf{Split} & \textbf{Size} & \textbf{Unique Functions} & \multicolumn{2}{c}{\textbf{Transformations}} \\
    \cmidrule(lr){4-5}
    & & & SP & SA \\
    \midrule
    Train & 7214 & 565 & 3085 & 4129 \\
    Valid & 2943 & 229 & 1291 & 1652 \\
    Test & 4415 & 341 & 1860 & 2555 \\
    \bottomrule
  \end{tabular}
  }
  \caption{An overview of \dataset~statistics. Transformation-wise breakdown of counts is shown in the Appendix~\ref{appendix:transformation-counts}.}
  \label{table:statistics}
\end{table}

While many of the transformations, especially the SP ones, overlap with BiFi~\citep{yasunaga2021break} and are well-studied in software engineering literature, our motivation is completely different and tackles an altogether different scenario. We focus on a controlled environment where we can understand strictly which simple transformations can confuse LLMs by discriminating syntax from semantics. SP transformations provide a controlled way to introduce changes while maintaining functional equivalence. This allows for a more precise analysis of what the LLM considers significant or insignificant regarding code structure and syntax. From a practical perspective, without grasping the code semantics completely, LLM-generated code may contain subtle logical errors or edge cases that are difficult to detect through surface-level code evaluation, such as identifying subtle bugs introduced during code refactoring that shouldn't alter functionality. This increases the importance of careful code review by experienced developers.

\section{Experiments}
\label{sec:experiments}

Through the use of our dataset, we aim to answer the following research questions:
    
    \paragraph{RQ1:} How good are state-of-the-art LLMs at zero-shot classification on \dataset?~\label{rq-1}
    
    \paragraph{RQ2:} Which are the most/least challenging semantic transformations for Code-LLMs?~\label{rq-2}
    
    \paragraph{RQ3:} Can the performance improve with fine-tuning on some transformations?~\label{rq-3}

We demonstrate that Code-LLMs are far from ``solving'' our dataset, leaving it for future work to further use our dataset as a benchmark for analysing the level of code understanding in such LLMs.

\subsection{Models}

To evaluate the benchmark, we choose general and state-of-the-art (Code)-LLMs that have performed well on code-related benchmarks (such as HumanEval) and are open-sourced for commercial use, as our test models. We exclude closed-source LLMs because we want to see the impact of fine-tuning and we are worried about possible data leaks, as HumanEval, MBPP, and relative benchmarks might have already been used to train the largest closed-source LLMs. See the model details in Appendix~\ref{appendix:model-details}.

\subsection{RQ1: Match-based vs. LLM-based Metrics Performance}
\begin{table}[t]
\centering
\resizebox{0.75\linewidth}{!}{
\begin{tabular}{llcc}
\toprule
\textbf{Type} & \textbf{Metric} & \textbf{Size} & \textbf{AP} \\
\midrule
\multirow[c]{8}{*}{\rotatebox{90}{Match-based}} & Rouge1 & -- & 50.91 \\
 & Rouge2 & -- & 50.67 \\
 & RougeL & -- & 48.48 \\
 & Meteor & -- & 52.05 \\
 & ChrF & -- & 55.25 \\
 & BLEU & -- & 48.46 \\
 & CrystalBLEU & -- & 48.35 \\
 & CodeBLEU & -- & 50.65 \\
\midrule
\multirow[c]{16}{*}{\rotatebox{90}{LLM-based}} & Comet & -- & 52.41 \\
 & CodeScore & 126M & 46.48 \\
 & BERTScore & 110M & 54.87 \\
 & CodeBERTScore & 125M & 47.45 \\
 \cmidrule(l){2-4}
 \multicolumn{4}{c}{\textit{Generic}}
 \\
 \cmidrule(l){2-4}
 & \multirow[c]{2}{*}{Llama2} & 7B & 41.33 \\
 & & 13B & 43.32 \\
 \cmidrule(l){2-4}
 \multicolumn{4}{c}{\textit{Code-Specific}}
 \\
\cmidrule(l){2-4}
 & \multirow[c]{3}{*}{CodeLlama} & 7B & 44.30 \\
 & & 13B & \textbf{70.85} \\
 & & 34B & 46.59 \\
 \cmidrule(l){2-4}
 & \multirow[c]{3}{*}{StarCoder2} & 3B & 34.11 \\
 & & 7B & 33.91 \\
 & & 15B & 50.75 \\
\bottomrule
\end{tabular}
}
\caption{LLM-based metrics struggle to differentiate between semantically equivalent and non-equivalent code snippets, sometimes performing worse than surface-level match-based metrics. This indicates a lack of understanding of code semantics and reasoning based on underlying logic.}
\label{table:mbpp-derived-results}
\end{table}

To evaluate the model performance, we measure the area under the precision-recall curve to calculate the average precision (AP) score:
\begin{equation}
    \mathrm{AP}=\sum_n\left(R_n-R_{n-1}\right) P_n,
\end{equation}
\noindent where $P_n$ and $R_n$ correspond to the precision and recall at the n\textsuperscript{th} threshold. 
AP accounts for ranking by rewarding models that rank correct predictions higher and are calculated per class, allowing performance evaluation on individual classes. We note that AP is the area under the precision-recall curve (AUC-PR) curve and is a more accurate metric for slightly imbalanced datasets than the usual AUC-ROC. This imbalance mainly happens when the positive class (SP) is lesser in magnitude than the negative class (SA). This is because AP focuses more on the performance of the positive class and is more sensitive to improvements in the positive class predictions compared to AUC-ROC. At the same time, AUC-ROC can give a false sense of high performance on imbalanced data.

\begin{table*}[ht]
\centering
\renewcommand*{\arraystretch}{1.2}
\resizebox{1.\linewidth}{!}
    {\begin{tabular}{@{\extracolsep{5pt}}lllcccccccccccc@{}}
\toprule
\multicolumn{3}{c}{\textbf{Transformation}} & \multicolumn{4}{c}{\textit{Embedding-based}} & \multicolumn{8}{c}{\textit{Zero-shot-prompt-based}} \\
\cline{1-3} \cline{4-7} \cline{8-15}
\multicolumn{1}{c}{\multirow{2}{*}{\textbf{Type}}} & \multicolumn{1}{c}{\multirow{2}{*}{\textbf{Category}}} & \multicolumn{1}{c}{\multirow{2}{*}{\textbf{Name}}} & \multirow{2}{*}{\textbf{Comet}} & \multirow{2}{*}{\textbf{CS}} & \multirow{2}{*}{\textbf{BS}} & \multirow{2}{*}{\textbf{CBS}} & \multicolumn{2}{c}{\textbf{Llama2}} & \multicolumn{3}{c}{\textbf{CodeLlama}} & \multicolumn{3}{c}{\textbf{StarCoder2}} \\
 \cline{8-9} \cline{10-12} \cline{13-15}
& & & &  &  &  & 7B & 13B & 7B & 13B & 34B & 3B &  7B & 15B \\
\midrule
\multirow[c]{15}{*}{SA} & \multirow[c]{4}{*}{AOM} & $+$ $\rightarrow$ $-$ & 4.46 & \textbf{87.52} & 0.76 & 1.39 & 20.85 & 14.77 & 14.19 & 15.69 & 16.25 & 24.64 & 14.99 & 16.92 \\
 &  & $-$ $\rightarrow$ $+$ & 4.29 & \textbf{88.52} & 0.57 & 3.30 & 23.29 & 13.95 & 14.04 & 20.44 & 19.44 & 24.93 & 17.35 & 17.66 \\
 &  & $\div$ $\rightarrow$ $\times$ & 4.19 & \textbf{81.21} & 0.71 & 2.40 & 13.11 & 12.95 & 13.15 & 17.46 & 16.28 & 11.96 & 12.62 & 17.63 \\
 &  & $\times$ $\rightarrow$ $\div$ & 4.83 & \textbf{94.76} & 1.01 & 1.45 & 17.38 & 17.40 & 15.49 & 15.18 & 16.64 & 22.32 & 14.37 & 18.20 \\
\cmidrule{2-15}
 & \multirow[c]{2}{*}{BOM} & \texttt{False} $\rightarrow$ \texttt{True} & 4.32 & 2.29 & 0.20 & 1.56 & 14.47 & 13.95 & 10.49 & 14.15 & 14.68 & \textbf{21.39} & 13.04 & 14.47 \\
 &  & \texttt{True} $\rightarrow$ \texttt{False} & 4.14 & 1.91 & 0.23 & 1.75 & 14.94 & 12.55 & 9.22 & 13.12 & 13.44 & \textbf{22.17} & 10.04 & 9.73 \\
\cmidrule{2-15}
 & \multirow[c]{4}{*}{COM} & == $\rightarrow$ != & 4.75 & \textbf{85.09} & 0.58 & 1.16 & 23.88 & 14.42 & 15.01 & 13.71 & 15.85 & 23.65 & 12.61 & 13.39 \\
 &  & != $\rightarrow$ == & 3.80 & \textbf{88.77} & 0.20 & 0.43 & 9.53 & 9.30 & 14.66 & 19.70 & 10.11 & 14.66 & 10.87 & 7.19 \\
 &  & $>$ $\rightarrow$ $<$ & 4.55 & \textbf{91.16} & 0.61 & 3.85 & 26.02 & 14.97 & 13.67 & 15.42 & 18.75 & 20.26 & 11.87 & 16.93 \\
 &  & $<$ $\rightarrow$ $>$ & 4.50 & \textbf{30.78} & 0.40 & 1.00 & 11.78 & 16.51 & 13.52 & 14.55 & 12.93 & 13.11 & 14.04 & 17.27 \\
\cmidrule{2-15}
 & DCS & Dissimilar Code Inject & 30.42 & \textbf{38.50} & 13.39 & 14.07 & 23.09 & 31.49 & 22.44 & 17.09 & 21.55 & 32.43 & 15.02 & 16.48 \\
\cmidrule{2-15}
 & \multirow[c]{2}{*}{IOM} & \texttt{is} $\rightarrow$ \texttt{is not} & 2.54 & 3.12 & 0.14 & 0.62 & 12.68 & 9.79 & 10.35 & 12.96 & 13.75 & \textbf{16.40} & 12.36 & 13.57 \\
 &  & \texttt{is not} $\rightarrow$ \texttt{is} & \textbf{50.00} & \textbf{50.00} & \textbf{50.00} & \textbf{50.00} & \textbf{50.00} & \textbf{50.00} & \textbf{50.00} & \textbf{50.00} & \textbf{50.00} & \textbf{50.00} & \textbf{50.00} & \textbf{50.00} \\
\cmidrule{2-15}
 & \multirow[c]{2}{*}{LOM} & \texttt{and} $\rightarrow$ \texttt{or} & 4.30 & \textbf{27.00} & 0.25 & 0.76 & 11.54 & 14.61 & 13.54 & 15.66 & 11.99 & 22.52 & 13.45 & 16.37 \\
 &  & or $\rightarrow$ and & 3.05 & 4.24 & 0.30 & 0.78 & 7.71 & 12.80 & 6.16 & 18.04 & 14.32 & \textbf{24.55} & 13.47 & 16.80 \\
\cmidrule{1-15}
\multirow[c]{6}{*}{SP} & DCI & Dead Code Insert & 11.64 & \textbf{72.95} & 5.79 & 7.24 & 16.02 & 17.05 & 11.99 & 21.98 & 16.72 & 16.39 & 9.28 & 13.99 \\
\cmidrule{2-15}
 & LT & \texttt{for} $\leftrightarrow$ \texttt{while} Loop & 10.37 & \textbf{77.64} & 7.46 & 8.81 & 12.56 & 16.17 & 9.20 & 16.95 & 14.56 & 16.10 & 14.73 & 14.25 \\
\cmidrule{2-15}
 & OS & Operand Swap & 8.20 & \textbf{77.25} & 4.92 & 5.01 & 21.61 & 18.29 & 11.15 & 20.44 & 14.83 & 14.15 & 12.60 & 13.56 \\
\cmidrule{2-15}
 & \multirow[c]{3}{*}{RV} & Rename Variable Cb & 24.46 & \textbf{74.29} & 9.74 & 6.60 & 22.32 & 18.22 & 11.57 & 21.25 & 17.04 & 14.80 & 13.50 & 16.68 \\
 &  & Rename Variable Naive & 18.17 & \textbf{72.38} & 7.88 & 8.57 & 17.33 & 16.51 & 9.34 & 18.87 & 16.27 & 13.36 & 14.45 & 14.34 \\
 &  & Rename Variable Rn & 34.78 & \textbf{70.16} & 12.57 & 11.34 & 19.31 & 20.28 & 11.64 & 19.99 & 17.39 & 13.45 & 12.44 & 16.97 \\
\bottomrule
\end{tabular}
    }
\caption{LLM-based metrics struggle to classify SA transformations due to their susceptibility to subtle input variations. Our findings show that LLM variants specifically trained for coding tasks outperform their more general-purpose counterparts. Here, \textit{CS:} CodeScore, \textit{ BS:} BERTScore, \textit{CBS:} CodeBERTScore.}
\label{table:mbpp-transformation-breakdown-llm}
\end{table*}

We aim to investigate whether LLMs perform significantly better than old-school, syntax-based metrics and incorporate all CEMs proposed in the literature. While BLEU is meant to compare at the ngram-matching level, we also consider their improved variations, which include code-related modifications such as CodeBLEU and CrystalBLEU, which offer an upper limit for the performance of match-based metrics. In addition, we need to know whether relying on LLM-based metrics has a significant advantage (at the cost of latency, memory consumption, etc.) compared to match-based metrics.
Table~\ref{table:mbpp-derived-results} shows the results on the \dataset{} test set.
We observe that CodeLlama (CL) outperforms Llama2 across different model sizes. This can be attributed to the fact that CL models were initialized with Llama2 weights but then further trained on a massive 500B token dataset heavily focused on code and code-related content. This specialized training data allows CL to develop a deeper understanding of programming languages, libraries, and coding conventions than the more general Llama2.

We notice that BERTScore outperforms its code-enhanced metric, CodeBERTScore (CBS). We hypothesize this could be due to multiple reasons. Firstly, while encoding the surrounding context (e.g. natural language descriptions) is beneficial for code generation, in the case of understanding tasks without generation, this additional context encoding in CBS may not provide any advantage and could even introduce noise. Secondly, it leverages pre-trained language models for code like CodeBERT, which heavily relies on the names of variables and functions to understand code semantics. When these names are obfuscated or changed, it struggles to comprehend the underlying logic and meaning of the code~\citep{wang_case_2024}. CL-13B is the best-performing overall, outperforming the larger 34B model. While the 34B model has more parameters and performs better on benchmarks, we speculate specific coding tasks or prompts may better suit the 13B model's capabilities. The smaller model size could lead to better generalization or less overfitting on some particular tasks. Training details are shown in Appendix~\ref{appendix:training-details}. The prompt format is shown in Appendix~\ref{appendix:prompt-template}. 

We report additional experiments that show how few-shot improves performance for some LLMs and how zero-shot CoT prompting affects the model performance in Appendix~\ref{appendix:different-prompting}. We can see that zero-shot CoT prompting has only marginal improvements compared to the standard zero-shot prompting.
We note that zero-shot prompting can be more adaptable across different programming languages without needing to adjust the prompting strategy, as it relies more on the model's general understanding of code functionality. LLMs trained on vast amounts of code can often make accurate judgments about code equivalence without needing to "think through" the problem explicitly, effectively leveraging their pre-trained knowledge.

\begin{result}
    \textbf{Answer to RQ1: }{\emph{
    As there are only marginal improvements, we believe LLM-based metrics are not superior compared to match-based metrics despite their strong contextual understanding abilities.
    Bigger LLMs tend to outperform their smaller counterparts.}}
\end{result}

\subsection{RQ2: Impact of Semantic Transformations}

To assess the performance of various CEMs on each transformation, we consider the true positive label for SP transformations as 1 and 0 for SA transformations. As it becomes a single-class classification per transformation, the precision will always be 1 due to zero false positives. In this case, the recall score measures the fraction of all actual positive instances correctly identified. So, we measure the area under the recall curve (AURC) corresponding to all the chosen thresholds, which varies between 0 and 1 and use the obtained result for our analysis as shown:
\begin{equation}
     \begin{aligned}
        \texttt{R}, \texttt{T} & = \texttt{recall\_curve}(y, \hat{y}, posLabel)\\
        \texttt{AURC} & = \texttt{auc(T,R)}
    \end{aligned}
\end{equation}
where \texttt{recall\_curve} refers to the plot of recall scores against different thresholds. $y$ is the true label, $\hat{y}$ is the score probabilities and $posLabel$ is the label of the positive class. \texttt{auc} calculates the area under the recall curve using the trapezoidal rule. We generally observe that the metrics have difficulty understanding SA transformations compared to the SP transformations.
 We take the row with the smallest to the largest sum to determine the level of difficulty of the transformations. 
 
 Among SA transformations, ``Dissimilar Code Inject'' is the least challenging as unrelated code fragments are less likely to share common variables, functions, or data structures, making it simpler to isolate and compare the code snippets independently. On the other hand, the \texttt{``is $\rightarrow$ is not''} appears to be the most challenging, as it struggles with identity operator misuse. Among SP transformations, ``Operand Swap'' is the most challenging, while ``Rename Variable'' seems the least challenging. Table~\ref{table:mbpp-transformation-breakdown-llm} shows the transformation-wise breakdown of LLM-based metrics. CodeScore (CS), an automatic metric, outperforms zero-shot prompted Code-LLMs, presumably due to using both NL context and the reference. We show the transformation-wise breakdown of match-based metrics in the Appendix (Table~\ref{table:mbpp-transformation-breakdown-match}). 
We often observe a high similarity score among different metrics for an SA variant compared to its SP one, as shown in the Appendix (Figure~\ref{fig:hist-metric-plot}).

\begin{result}
    \textbf{Answer to RQ2: }{\emph{The transformed code containing the least challenging transformations is associated with maximum syntactic differences, whereas the most challenging transformations often occur in similar contexts and are relatively close in the embedding space but have completely opposite behaviour.}}
\end{result}

We note that for AP, the models have to distinguish between SP v/s SA labels, which requires the models to be more selective in their positive predictions, ensuring that it is more often correct when predicting a positive class (i.e., SP). To clarify, CL-13B predicts mostly higher probabilities (> 0.8) for SA transformations than CS; hence, it gets a weaker AURC score. Thus, we can infer that CL-13B does a better job classifying the two snippets as either SP or SA, but once we know the transformation is of a specific type, CS performs better.

\subsection{RQ3: Impact of Finetuning}
\label{sec:rq-3}
While finetuning on a diverse set of transformations can improve performance on seen examples, it does not necessarily guarantee effective generalization to novel, unseen transformations. We propose a leave-one-out evaluation strategy for assessing the performance of PEFT methods on unseen semantic transformations. The approach involves finetuning the model on N-1 transformation for a given category, then evaluating on the held-out $N^{th}$ transformation. We repeat the leave-one-out approach $N$ times to assess performance on each held-out transformation. Table~\ref{table:finetuning-results} shows the PEFT results on the \dataset{} test set using the AURC score. We note that Llama2-7B outperforms CL-7B on most transformations.

\begin{table}[ht!]
\centering
\renewcommand*{\arraystretch}{1.2}
\resizebox{0.92\linewidth}{!}
    {\begin{tabular}{@{\extracolsep{5pt}}lllcc@{}}
\toprule
\multicolumn{2}{c}{\textbf{Transformation}} & \multirow[c]{2}{*}{\textbf{Method}} & \textbf{Llama2} & \textbf{CodeLlama} \\
\cline{1-2}
\textbf{Type} & \textbf{Category} &  & 7B & 7B \\
\midrule
\multirow[c]{12}{*}{SA} & \multirow[c]{2}{*}{AOM} & LoRA & 2.69 & 6.12 \\
& & AdaLoRA & \textbf{40.57} & 9.29 \\
& & Prefix-Tuned & 16.6 & 5.06 \\
\cmidrule(l){2-5}
& \multirow[c]{2}{*}{BOM} & LoRA & 0.37 & 4.04 \\
& & AdaLoRA & 23.32 & 10.91 \\
& & Prefix-Tuned & \textbf{31.16} & 20.68 \\
\cmidrule(l){2-5}
& \multirow[c]{2}{*}{COM} & LoRA & 3.17 & 1.67 \\
& & AdaLoRA & 14.72 & 7.44 \\
& & Prefix-Tuned & \textbf{18.73} & 6.01 \\
\cmidrule(l){2-5}
& \multirow[c]{2}{*}{DCS} & LoRA & \textbf{98.69} & 77.23 \\
& & AdaLoRA & 88.28 & 20.77 \\
& & Prefix-Tuned & 36.66 &  35.17 \\
\cmidrule(l){2-5}
& \multirow[c]{2}{*}{IOM} & LoRA & 0.12 & 3.62 \\
& & AdaLoRA & 19.57 & 6.72 \\
& & Prefix-Tuned & \textbf{32.33} & 11.57 \\
\cmidrule(l){2-5}
& \multirow[c]{2}{*}{LOM} & LoRA & 3.84 & 3.85 \\
& & AdaLoRA & 9.14 & 15.60 \\
& & Prefix-Tuned & \textbf{23.99} & 5.72 \\
\midrule
\multirow[c]{8}{*}{SP} & \multirow[c]{2}{*}{DCI} & LoRA & \textbf{81.03} & 63.44 \\
& & AdaLoRA & 56.75 & 29.50 \\
& & Prefix-Tuned & 16.46 & 37.66 \\
\cmidrule(l){2-5}
& \multirow[c]{2}{*}{LT} & LoRA & \textbf{96.01} & 78.33 \\
& & AdaLoRA & 64.76 & 76.70 \\
& & Prefix-Tuned & 10.21 & 26.09 \\
\cmidrule(l){2-5}
& \multirow[c]{2}{*}{OS} & LoRA & \textbf{91.51} & 82.76 \\
& & AdaLoRA & 59.64 & 75.27 \\
& & Prefix-Tuned & 0.24 & 2.08\\
\cmidrule(l){2-5}
& \multirow[c]{2}{*}{RV} & LoRA & 62.33 & \textbf{80.63} \\
& & AdaLoRA & 26.68 & 70.14 \\
& & Prefix-Tuned & 8.92 & 20.86 \\
\bottomrule
\end{tabular}
}
\caption{When finetuning with SP transformations, the PEFT methods learn to be invariant to these transformations.
In contrast, SA transformations require updating the core weights to learn the new semantics, and the PEFT methods are not expressive enough to capture such fundamental changes.}
\label{table:finetuning-results}
\end{table}

\begin{result}
    \textbf{Answer to RQ3: }{\emph{PEFT improves performance on SP transformations while facing challenges with SA transformations. Among different PEFT methods, LoRA is the most effective one for SP, while PrefixTuning is the most successful for SA.}}
\end{result}

\section{Conclusion}
\label{sec:conclusion}
We propose \dataset, a new challenging benchmark to evaluate how well Code-LLMs capture functional equivalence between code snippets from the code semantics standpoint. 
We compare the performance of LLM- and match-based metrics on the \dataset and find the performance gap to be minimal. We identify semantic transformations for the Code-LLMs from least to most challenging on a spectrum. We conduct extensive evaluations in different settings, including zero-shot (w/ prompting) and using PEFT methods. In the future, we would like to study code semantics in both static and dynamic settings at different granularities by incorporating approximate, operational, and abstract semantics~\citep{ding_semcoder_2024}. Incorporating symbolic reasoning modules or hybrid approaches that combine neural networks with formal logic can be a promising direction.

\section*{Limitations}
\label{sec:limitations}
We investigate open-source LLMs to evaluate for code functional equivalence, so we do not consider ICE-Score~\citep{zhuo-2024-ice} that requires using closed-source LLMs like GPT-3.5 or GPT-4 in this analysis. Closed-source models are opaque, as their inner workings, data sources, and training methodologies are not disclosed, making it hard to draw meaningful comparisons with open-source models. 

In addition, they often come with significant usage costs, and finetuning is not fully supported in experimental access. Currently, the code functions in \dataset{} are exclusively in Python. However, we aim to broaden the scope to encompass a broader range of programming languages and domains. By doing so, we strive to enhance the diversity and applicability of the dataset, making it more comprehensive and versatile for various software engineering tasks and scenarios. 

Moreover, we do not check for compilation of semantic code transformations as we perform a static code evaluation to analyse the code without needing it to be executed or compiled.

Further, we do not account for variations due to prompt and temperature as we do not optimise the prompting format, although we ensure it is kept consistent across different LLMs. 

Finally, we refrain from chaining multiple transformations of the same type (either preserving or altering) to make the analysis straightforward.

\section*{Acknowledgements}
We thank the anonymous reviewers, Ke Wang, and Antonio Miceli Barone for their feedback.
This work was supported by the UKRI Centre for Doctoral Training (CDT) in Natural Language Processing through the UKRI grant (EP/S022481/1). We appreciate using computing resources through the CSD3 cluster at the University of Cambridge and the Baskerville cluster at the University of Birmingham.
AV was supported by the ``UNREAL: Unified Reasoning Layer for Trustworthy ML'' project (EP/Y023838/1) selected by the ERC and funded by UKRI EPSRC.

\bibliography{anthology,references,custom}

\clearpage
\appendix

\section{Details of Evaluation Metrics}
\label{appendix:eval-metric}

\subsection{Based on Lexical Overlap}
These metrics operate on the surface form of the code and account only for an exact lexical token match.

\paragraph{ROUGE~\citep{lin-2004-rouge}}
measures the recall between n-grams in generated and reference code.

\paragraph{BLEU~\citep{papineni-etal-2002-bleu}}
is the geometric mean of the n-gram precision multiplied by a brevity penalty between generated and reference code.

\paragraph{CodeBLEU~\citep{ren_codebleu_2020}}
is a composite metric which is a modification of BLEU and uses the abstract syntax tree and data-flow graph in addition to the surface-level matching.

\paragraph{CrystalBLEU~\citep{eghbali_crystalbleu_2023}}
is again a modification of BLEU which considers the underlying differences between source code and natural language (such as trivially shared n-grams).

\paragraph{Meteor~\citep{denkowski-lavie-2014-meteor}}
is a MT metric which is based on the harmonic mean of unigram precision and recall, with the recall being more weighted.

\paragraph{ChrF~\citep{popovic-2015-chrf}}
is again a MT metric which calculates the precision and recall for character n-gram matches and averages it over 1- to 6-character-n-grams.

\subsection{Based on Pre-trained LLMs}
These metrics rely on LLMs to extract the token embeddings of the hidden layer to calculate the similarity.

\paragraph{COMET~\citep{rei-etal-2020-comet}}
uses a pre-trained multilingual model to encode generated and reference code separately. These embeddings are concatenated to obtain a quality score.

\paragraph{BERTScore~\citep{zhang_bertscore_2020}}
leverages BERT embeddings to compute the pairwise cosine similarity between generated and reference code.

\paragraph{CodeBERTScore~\citep{zhou_codebertscore_2023}}
uses CodeBERT to encode the context (the natural language description or comment) in addition to generated and reference code. However, it does not use the encoded context to compute cosine similarities.

\subsection{Based on Execution}
These metrics compare the execution result of generated code by running tests to check for functional correctness.

\paragraph{Pass@$k$~\citep{chen_evaluating_2021}}
generates $k$ solutions for each problem, which is deemed solved if any of the $k$ samples pass the tests.

\paragraph{CodeScore~\citep{dong_codescore_2023}}
provides a framework, UniCE, to finetune LLMs to learn code execution (such as estimating the PassRatio of test cases) of generated code with unified input.

\paragraph{AvgPassRatio~\citep{hendrycks_measuring_2021}}
computes the average pass rate of test cases.

\section{Details of Correlation Metrics}
\label{appendix:correlation-metric}
\paragraph{Kendall-Tau ($\tau$)}
is a non-parametric statistical measure that quantifies the strength and direction of the rank correlation between two ordinal variables. It is calculated as:

\begin{equation*}
    \tau = \frac{|\text{concordant}| - |\text{discordant}|}{|\text{concordant}| + |\text{discordant}|}
\end{equation*}

where, concordant pairs are in the same relative order in both rankings, whereas, discordant pairs are in the opposite relative order.

\paragraph{Pearson ($r_p$)}
is a statistical measure that quantifies the strength and direction of the linear relationship between two continuous variables. It is calculated as:

\begin{equation*}
    r_p = \frac{\sum_{i=1}^{n} (X_i - \overline{X})(Y_i - \overline{Y})}{\sqrt{\sum_{i=1}^{n} (X_i - \overline{X})^2} \sqrt{\sum_{i=1}^{n} (Y_i - \overline{Y})^2}}
\end{equation*}

where, X and Y corresponds to the values of the reference and candidate variables, respectively.

\paragraph{Spearman ($r_s$)}
is a non-parametric measure of the strength and direction of the monotonic relationship between two ranked variables.

\resizebox{\linewidth}{!}{
\begin{minipage}{1.15\linewidth}
\begin{equation*}
    r_s = \frac{\sum\limits_{i=1}^{n} (R(X_i) - \overline{R(X)})(R(Y_i) - \overline{R(Y)})}{\sqrt{\sum\limits_{i=1}^{n} (R(X_i) - \overline{R(X)})^2 \sum\limits_{i=1}^{n} (R(Y_i) - \overline{R(Y)})^2}}
\end{equation*}
\end{minipage}
}

where, $R(X_i)$ and $R(Y_i)$ are the ranks of the $i^{th}$ observations in reference and candidate variables, respectively.

\section{Training Details}
\label{appendix:training-details}

The whole pipeline takes roughly 24 hours to create the full transformations on a single A100 80GB GPU. The zero-shot inference experiments take roughly 12 GPU hours, and fine-tuning takes about 48 GPU hours.

\subsection{Model Generate}
The configuration specified during the generation step is \texttt{max\_new\_tokens=3} to limit the generated output to 3 new tokens, \texttt{top\_p=0.9} and \texttt{temperature=0.2} to control the randomness of the generated text, \texttt{num\_return\_sequences=3} to generate 3 different output sequences, \texttt{top\_k=5} to consider only the 5 most likely tokens at each step, \texttt{do\_sample=True} to use sampling for text generation.

The provided configuration sets up a model to use 4-bit quantization with NF4 type, bfloat16 compute dtype, and optionally enables double quantization for better accuracy.

In our analysis, we exclude instruction-tuned models of CodeLlama and StarCoder due to the potential for learning similar instructions during the finetuning step, such as the code clone detection task.

\subsection{Confidence Scores from LLMs}
To generate token probabilities, we first generate output sequences from the model using various sampling techniques like top-p, temperature, and top-k. Then, we compute the softmax probabilities of the logits at a specific index to obtain the probabilities of all tokens in the vocabulary. We identify the tokens with probabilities above a certain threshold, decode them, and store their probabilities in a dictionary. Then, we extract the probabilities of the "YES" and "NO" tokens from this dictionary and calculate a score as the ratio of the "YES" probability to the sum of the "YES" and "NO" probabilities. Generating token probabilities from the pre-trained model and analyzing the probabilities of specific tokens provides a way to quantify the model's confidence for a particular output (in this case, "YES" or "NO").

\subsection{Finetuning Experiments}
We use the popular PEFT methods:

\paragraph{Low-rank Adaptation (LoRA;~\citealt{hu2022lora})} introduces two learnable weight matrices, $A$ and $B$, attached to a frozen pre-trained weight matrix, $W$, and considers that these updates have a low rank during adaptation. 

\paragraph{Adaptive Low-rank Adaptation (AdaLoRA;~\citealt{zhang_adalora_2023})} adaptively allocates more parameters to the more important layers and fewer parameters to less important layers, unlike LoRA, which distributes trainable parameters evenly across all layers.

\paragraph{Prefix Tuning~\citep{li-liang-2021-prefix}}
prepends pseudo prefix tokens to the input of a language model.

In the case of LoRA, we use rank as 8, alpha as 32, and dropout as 0.1. For AdaLoRa, we use an initial rank of 12 that will be reduced to 8. The beta1 and beta2 parameters for the Adam optimizer are both set to 0.85. The learning rate will be adjusted between the initial time step of 200 and the final time step of 1000, with a step size 10. We use the alpha value of 32 and a dropout rate of 0.1. For Prefix Tuning, we use 20 virtual tokens.

The training arguments include a per-device batch size and gradient accumulation steps, totalling 2 training epochs. The model undergoes 100 warmup steps and a maximum of 400 steps overall, using a learning rate 3e-4 and enabling fp16 precision. We set logging to occur every 10 steps using the AdamW optimizer. Evaluation and saving occur every 200 steps, with outputs directed to a specified directory and a limit of 3 saved checkpoints. The model does not load the best model at the end of training. We group the sequences by length to speed up training and report results to weights and biases with a run name that includes a timestamp.

\section{Model Details}
\label{appendix:model-details}

\paragraph{Llama2~\citep{touvron_llama_2023}}
is a family of LLMs developed by Meta AI, ranging from 7B to 70B parameters. It is an open-source successor to the original Llama model, offering improved performance through a larger training corpus, longer context length, and the use of grouped-query attention. 

\paragraph{StarCoder2~\citep{lozhkov:starcoder-stack-2}}
is a generative model with 3B, 7B, and 15B parameters trained on over 600 programming languages from the Stack v2, along with natural language sources like Wikipedia, ArXiv, and GitHub issues.

\paragraph{Code Llama~\citep{roziere_code_2024}}
is initialized using pre-trained weights of Llama2 and trained on code-specific datasets. It then undergoes long-context finetuning and can handle repository-level inputs of 100K tokens.

\section{Transformation Counts}
\label{appendix:transformation-counts}

The average line of code for the transformed version is 9.2247. The problems in the benchmark are designed to be solvable by entry-level programmers. Among these questions, 58\% are mathematical (e.g., calculating the volume of a sphere), 43\% involve list processing, 19\% require string manipulation, 9\% focus on integer sequences, and 2\% revolve around using other data structures. Table~\ref{table:transformation-counts} lists the counts of different transformations.

\begin{table}[t]
\centering
\renewcommand*{\arraystretch}{1.2}
\resizebox{1.\linewidth}{!}{
\begin{tabular}{llccc}
\toprule
\multicolumn{2}{c}{\textbf{Transformation}} & \multirow[c]{2}{*}{\textbf{Test}} & \multirow[c]{2}{*}{\textbf{Train}} & \multirow[c]{2}{*}{\textbf{Valid}} \\
\cline{1-2}
\multicolumn{1}{c}{\multirow{1}{*}{\textbf{Type}}} & \multicolumn{1}{c}{\multirow{1}{*}{\textbf{Name}}} &  &  &  \\
\midrule
\multirow[c]{12}{*}{SA} & $+$ $\rightarrow$ $-$ & 147 & 267 & 114 \\
& $-$ $\rightarrow$ $+$ & 133 & 211 & 73 \\
& $\div$ $\rightarrow$ $\times$ & 47 & 82 & 30 \\
& $\times$ $\rightarrow$ $\div$ & 93 & 133 & 53 \\
& \texttt{False} $\rightarrow$ \texttt{True} & 41 & 49 & 23 \\
& \texttt{True} $\rightarrow$ \texttt{False} & 39 & 50 & 21 \\
& == $\rightarrow$ != & 117 & 173 & 62 \\
& != $\rightarrow$ == & 29 & 34 & 26 \\
& $>$ $\rightarrow$ $<$ & 79 & 112 & 38 \\
& $<$ $\rightarrow$ $>$ & 57 & 96 & 39 \\
& Dissimilar Code Inject & 1705 & 2825 & 1140 \\
& \texttt{is} $\rightarrow$ \texttt{is not} & 10 & 11 & 2 \\
& \texttt{is not} $\rightarrow$ \texttt{is} & 1 & 0 & 1 \\
& \texttt{and} $\rightarrow$ \texttt{or} & 44 & 47 & 17 \\
& or $\rightarrow$ and & 13 & 39 & 13 \\
\midrule
\multirow[c]{8}{*}{SP} & Dead Code Insert & 320 & 539 & 224 \\
& \texttt{for} $\leftrightarrow$ \texttt{while} Loop & 313 & 515 & 217 \\
& Operand Swap & 311 & 512 & 217 \\
& Rename Variable Cb & 290 & 489 & 199 \\
& Rename Variable Naive & 313 & 515 & 217 \\
& Rename Variable Rn & 313 & 515 & 217 \\
\bottomrule
\end{tabular}
}
\caption{Breakdown of counts of different transformations across train, validation, and test sets of \dataset.}
\label{table:transformation-counts}
\end{table}

\section{Prompt Template}
\label{appendix:prompt-template}
Figure~\ref{fig:prompt} shows the prompt template used in the zero-shot prompting experiments.

\begin{figure}[htbp]
\hrulefill

"""You are a helpful and honest code assistant expert in Python. Is there a functional equivalence between the Code1 and Code2? Please respond either "YES" or "NO".\\
\\
        \#\#\# Code1:\\
        \{code\_1\}\\

        \#\#\# Code2:\\
        \{code\_2\}\\

        \#\#\# Response:\\
"""
\hrulefill
\caption{Prompt for Code-LLMs on \dataset.}
\label{fig:prompt}
\end{figure}

\section{Additional Prompting Results}
\label{appendix:different-prompting}

Table~\ref{table:closed-source-llms} shows the performance of a few hand-picked closed-source LLMs to assess progress in order to have a more holistic evaluation. We include it to ascertain the range of performance of these sophisticated LLMs on our task.

\begin{table}[t]
    \centering
    \begin{tabular}{lcc}
        \toprule
        \textbf{Model} & \textbf{Size} & \textbf{AP} \\
        \midrule
        gpt-4o-mini & $\sim$8B & 83.73 \\
        deepseek-coder-instruct-v1.5 & 7B & 84.10 \\
        qwen2.5-coder-instruct & 32B & 85.21 \\
        \bottomrule
    \end{tabular}
    \caption{Results for zero-shot prompting on the SoTA closed-source LLMs. While they perform better than open-source LLMs (e.g., StarCoder, CodeLlama, etc.), they still struggle to differentiate between functionally equivalent v/s functionally non-equivalent codes. Our claims surrounding RQ1 still hold, as these are trivial tasks for the sophisticated closed-source LLMs.}
    \label{table:closed-source-llms}
\end{table}

Table~\ref{table:few-shot-results} and \ref{table:zero-shot-cot-results} demonstrate few-shot and zero-shot chain-of-thought (CoT) performance on the \dataset test set. It suggests the model cannot fully infer the task requirements or context directly from the zero-shot prompt. Few-shot examples help bridge this gap by reducing ambiguity and assisting the model in aligning its output to the expected format or logic of the task by recognising patterns in the input-output pairs and applying these patterns to new, unseen data. Since the representative examples resemble the different plausible styles of transformations, the CodeLLMs can learn these patterns but might struggle to understand other variations of transformations (e.g., De Morgan's laws).

\begin{table}[t]
\centering
\resizebox{0.55\linewidth}{!}{
\begin{tabular}{lcc}
\toprule
\textbf{Model} & \textbf{Size} & \textbf{AP} \\
\midrule
 \multicolumn{3}{c}{\textit{Generic}}
 \\
 \cmidrule(l){1-3}
 \multirow[c]{2}{*}{Llama2} & 7B & 55.52 \\
 & 13B & 58.53 \\
 \cmidrule(l){1-3}
 \multicolumn{3}{c}{\textit{Code-Specific}}
 \\
\cmidrule(l){1-3}
 \multirow[c]{3}{*}{CodeLlama} & 7B & 80.20 \\
 & 13B & 85.92 \\
 & 34B & 92.54 \\
 \cmidrule(l){1-3}
 \multirow[c]{3}{*}{StarCoder2} & 3B & 71.84 \\
 & 7B & 56.14 \\
 & 15B & 97.81 \\
\bottomrule
\end{tabular}
}
\caption{Two-shot with one SP and SA demonstration example inserted into the original prompt. By including examples, few-shot prompts offer more context about analysing and comparing code snippets. This helps the model focus on relevant aspects like logic flow, variable usage, and output rather than superficial differences in syntax or formatting.}
\label{table:few-shot-results}
\end{table}

\begin{table}[ht]
\centering
\resizebox{0.55\linewidth}{!}{
\begin{tabular}{lcc}
\toprule
\textbf{Model} & \textbf{Size} & \textbf{AP} \\
\midrule
 \multicolumn{3}{c}{\textit{Generic}}
 \\
 \cmidrule(l){1-3}
 \multirow[c]{2}{*}{Llama2} & 7B & 37.51 \\
 & 13B & 40.55 \\
 \cmidrule(l){1-3}
 \multicolumn{3}{c}{\textit{Code-Specific}}
 \\
\cmidrule(l){1-3}
 \multirow[c]{3}{*}{CodeLlama} & 7B & 36.13 \\
 & 13B & 68.29 \\
 & 34B & 62.58 \\
 \cmidrule(l){1-3}
 \multirow[c]{3}{*}{StarCoder2} & 3B & 39.28 \\
 & 7B & 34.85 \\
 & 15B & 73.93 \\
\bottomrule
\end{tabular}
}
\caption{Zero-shot chain-of-thought results by adding ``Let’s think step by step'' to the original prompt. By prompting the model to think step-by-step, Zero-shot CoT prompting leverages the model's inherent reasoning abilities. We can observe that Zero-shot CoT prompting has comparable performance with the standard zero-shot prompting.}
\label{table:zero-shot-cot-results}
\end{table}

\section{Execution-based Functional Correctness}
\label{appendix:func-correctness}
Functional correctness is assessed by running the generated code against a set of test cases and checking if the output matches the expected results. We use the HumanEval (Python only), and HumanEval-X~\citep{zheng_codegeex_2023} benchmarks to measure the correlation with functional correctness. We filter based on the popularity of the programming language and choose to evaluate on Java, C++, Python, and JavaScript languages. We compute the Pearson, Spearman, and Kendall-Tau correlation coefficients as Pearson captures linear relationships, while the other two capture ordinal relationships, which may be non-linear. Section~\ref{appendix:correlation-metric} provides an overview of different correlation metrics. Table~\ref{table:correlation-multilingual-humaneval} shows correlation coefficients of different metrics with the functional correctness on HumanEval for multiple languages. We notice C++ and Javascript have lower correlation scores than Python and Java. Python and Java are primarily object-oriented languages, while C++ supports both object-oriented and functional programming styles. JavaScript, although object-oriented, has a strong functional programming influence. Generating code that effectively utilizes functional programming constructs can be more challenging for models trained primarily on object-oriented codebases.

\begin{table*}[htbp]
\centering
\resizebox{1.\linewidth}{!}
{\begin{tabular}{llccccccccccccc}
\toprule
\textbf{Type} & \textbf{Metric} & \multicolumn{3}{c}{\textbf{Java}} & \multicolumn{3}{c}{\textbf{C++}} & \multicolumn{3}{c}{\textbf{Python}} & \multicolumn{3}{c}{\textbf{JavaScript}} \\
 \cline{3-5} \cline{6-8} \cline{9-11} \cline{12-14}
 & & $\tau$ & $r_s$ & $r_p$ & $\tau$ & $r_s$ & $r_p$ & $\tau$ & $r_s$ & $r_p$ & $\tau$ & $r_s$ & $r_p$ \\
\midrule
\multirow[c]{8}{*}{Match-Based}
& Rouge1 & {\cellcolor[HTML]{FFFEBE}} \color[HTML]{000000} 0.4982 & {\cellcolor[HTML]{FEE08B}} \color[HTML]{000000} 0.3992 & {\cellcolor[HTML]{FEDA86}} \color[HTML]{000000} 0.3903 & {\cellcolor[HTML]{F88C51}} \color[HTML]{F1F1F1} 0.2466 & {\cellcolor[HTML]{FDC171}} \color[HTML]{000000} 0.3390 & {\cellcolor[HTML]{FDBD6D}} \color[HTML]{000000} 0.3286 & {\cellcolor[HTML]{FED884}} \color[HTML]{000000} 0.3857 & {\cellcolor[HTML]{FDC171}} \color[HTML]{000000} 0.3389 & {\cellcolor[HTML]{FDBD6D}} \color[HTML]{000000} 0.3284 & {\cellcolor[HTML]{F99355}} \color[HTML]{000000} 0.2594 & {\cellcolor[HTML]{FA9857}} \color[HTML]{000000} 0.2664 & {\cellcolor[HTML]{F7844E}} \color[HTML]{F1F1F1} 0.2360 \\
& Rouge2 & {\cellcolor[HTML]{FFF1A8}} \color[HTML]{000000} 0.4569 & {\cellcolor[HTML]{FDC372}} \color[HTML]{000000} 0.3417 & {\cellcolor[HTML]{FDBB6C}} \color[HTML]{000000} 0.3273 & {\cellcolor[HTML]{F67F4B}} \color[HTML]{F1F1F1} 0.2273 & {\cellcolor[HTML]{FCAA5F}} \color[HTML]{000000} 0.2963 & {\cellcolor[HTML]{FCAA5F}} \color[HTML]{000000} 0.2961 & {\cellcolor[HTML]{FDC574}} \color[HTML]{000000} 0.3457 & {\cellcolor[HTML]{FB9D59}} \color[HTML]{000000} 0.2747 & {\cellcolor[HTML]{FA9857}} \color[HTML]{000000} 0.2686 & {\cellcolor[HTML]{F16640}} \color[HTML]{F1F1F1} 0.1888 & {\cellcolor[HTML]{F8864F}} \color[HTML]{F1F1F1} 0.2390 & {\cellcolor[HTML]{ED5F3C}} \color[HTML]{F1F1F1} 0.1786 \\
& RougeL & {\cellcolor[HTML]{FFF8B4}} \color[HTML]{000000} 0.4778 & {\cellcolor[HTML]{FEDE89}} \color[HTML]{000000} 0.3953 & {\cellcolor[HTML]{FEDA86}} \color[HTML]{000000} 0.3880 & {\cellcolor[HTML]{F88950}} \color[HTML]{F1F1F1} 0.2454 & {\cellcolor[HTML]{FDBB6C}} \color[HTML]{000000} 0.3265 & {\cellcolor[HTML]{FDB96A}} \color[HTML]{000000} 0.3235 & {\cellcolor[HTML]{FECC7B}} \color[HTML]{000000} 0.3607 & {\cellcolor[HTML]{FDBD6D}} \color[HTML]{000000} 0.3320 & {\cellcolor[HTML]{FDBB6C}} \color[HTML]{000000} 0.3248 & {\cellcolor[HTML]{F46D43}} \color[HTML]{F1F1F1} 0.2012 & {\cellcolor[HTML]{F99153}} \color[HTML]{000000} 0.2541 & {\cellcolor[HTML]{F57245}} \color[HTML]{F1F1F1} 0.2107 \\
& Meteor & {\cellcolor[HTML]{E9F6A1}} \color[HTML]{000000} 0.5576 & {\cellcolor[HTML]{FEEB9D}} \color[HTML]{000000} 0.4337 & {\cellcolor[HTML]{FEE797}} \color[HTML]{000000} 0.4222 & {\cellcolor[HTML]{FA9B58}} \color[HTML]{000000} 0.2714 & {\cellcolor[HTML]{FDC171}} \color[HTML]{000000} 0.3392 & {\cellcolor[HTML]{FDB96A}} \color[HTML]{000000} 0.3217 & {\cellcolor[HTML]{FEEDA1}} \color[HTML]{000000} 0.4428 & {\cellcolor[HTML]{FEE695}} \color[HTML]{000000} 0.4189 & {\cellcolor[HTML]{FEE491}} \color[HTML]{000000} 0.4124 & {\cellcolor[HTML]{FDAF62}} \color[HTML]{000000} 0.3019 & {\cellcolor[HTML]{FECE7C}} \color[HTML]{000000} 0.3660 & {\cellcolor[HTML]{FDC372}} \color[HTML]{000000} 0.3425 \\
& ChrF & {\cellcolor[HTML]{E9F6A1}} \color[HTML]{000000} 0.5576 & {\cellcolor[HTML]{FEE797}} \color[HTML]{000000} 0.4223 & {\cellcolor[HTML]{FEE18D}} \color[HTML]{000000} 0.4040 & {\cellcolor[HTML]{FDBB6C}} \color[HTML]{000000} 0.3257 & {\cellcolor[HTML]{FECE7C}} \color[HTML]{000000} 0.3647 & {\cellcolor[HTML]{FECA79}} \color[HTML]{000000} 0.3576 & {\cellcolor[HTML]{FEEA9B}} \color[HTML]{000000} 0.4306 & {\cellcolor[HTML]{FED683}} \color[HTML]{000000} 0.3808 & {\cellcolor[HTML]{FED27F}} \color[HTML]{000000} 0.3727 & {\cellcolor[HTML]{FDB365}} \color[HTML]{000000} 0.3088 & {\cellcolor[HTML]{FEC877}} \color[HTML]{000000} 0.3551 & {\cellcolor[HTML]{FDBD6D}} \color[HTML]{000000} 0.3298 \\
& BLEU & {\cellcolor[HTML]{FAFDB8}} \color[HTML]{000000} 0.5148 & {\cellcolor[HTML]{FEE491}} \color[HTML]{000000} 0.4137 & {\cellcolor[HTML]{FEE08B}} \color[HTML]{000000} 0.4005 & {\cellcolor[HTML]{F99153}} \color[HTML]{000000} 0.2565 & {\cellcolor[HTML]{FDAD60}} \color[HTML]{000000} 0.2986 & {\cellcolor[HTML]{FCA85E}} \color[HTML]{000000} 0.2914 & {\cellcolor[HTML]{FEE18D}} \color[HTML]{000000} 0.4049 & {\cellcolor[HTML]{FECA79}} \color[HTML]{000000} 0.3560 & {\cellcolor[HTML]{FDC776}} \color[HTML]{000000} 0.3499 & {\cellcolor[HTML]{FDAD60}} \color[HTML]{000000} 0.2970 & {\cellcolor[HTML]{FDAD60}} \color[HTML]{000000} 0.2970 & {\cellcolor[HTML]{FA9B58}} \color[HTML]{000000} 0.2703 \\
& CrystalBLEU & {\cellcolor[HTML]{FAFDB8}} \color[HTML]{000000} 0.5145 & {\cellcolor[HTML]{FEE593}} \color[HTML]{000000} 0.4145 & {\cellcolor[HTML]{FEE08B}} \color[HTML]{000000} 0.4016 & {\cellcolor[HTML]{F98E52}} \color[HTML]{F1F1F1} 0.2519 & {\cellcolor[HTML]{FCAA5F}} \color[HTML]{000000} 0.2953 & {\cellcolor[HTML]{FCA55D}} \color[HTML]{000000} 0.2877 & {\cellcolor[HTML]{FEE18D}} \color[HTML]{000000} 0.4027 & {\cellcolor[HTML]{FECA79}} \color[HTML]{000000} 0.3562 & {\cellcolor[HTML]{FDC776}} \color[HTML]{000000} 0.3502 & {\cellcolor[HTML]{FCAA5F}} \color[HTML]{000000} 0.2949 & {\cellcolor[HTML]{FCAA5F}} \color[HTML]{000000} 0.2945 & {\cellcolor[HTML]{FA9857}} \color[HTML]{000000} 0.2682 \\
& CodeBLEU & {\cellcolor[HTML]{FFFEBE}} \color[HTML]{000000} 0.4990 & {\cellcolor[HTML]{FDC574}} \color[HTML]{000000} 0.3458 & {\cellcolor[HTML]{FDC372}} \color[HTML]{000000} 0.3417 & {\cellcolor[HTML]{F67A49}} \color[HTML]{F1F1F1} 0.2200 & {\cellcolor[HTML]{E75337}} \color[HTML]{F1F1F1} 0.1600 & {\cellcolor[HTML]{E95538}} \color[HTML]{F1F1F1} 0.1630 & {\cellcolor[HTML]{FED683}} \color[HTML]{000000} 0.3806 & {\cellcolor[HTML]{FDB96A}} \color[HTML]{000000} 0.3219 & {\cellcolor[HTML]{FDB768}} \color[HTML]{000000} 0.3195 & {\cellcolor[HTML]{FDB163}} \color[HTML]{000000} 0.3080 & {\cellcolor[HTML]{FA9656}} \color[HTML]{000000} 0.2643 & {\cellcolor[HTML]{F67F4B}} \color[HTML]{F1F1F1} 0.2271 \\
\cline{1-14}
\multirow[c]{3}{*}{LLM-Based}
& CodeScore & {\cellcolor[HTML]{FDB163}} \color[HTML]{000000} 0.3061 & {\cellcolor[HTML]{FA9857}} \color[HTML]{000000} 0.2686 & {\cellcolor[HTML]{FCA55D}} \color[HTML]{000000} 0.2865 & {\cellcolor[HTML]{DE402E}} \color[HTML]{F1F1F1} 0.1253 & {\cellcolor[HTML]{D93429}} \color[HTML]{F1F1F1} 0.1074 & {\cellcolor[HTML]{D42D27}} \color[HTML]{F1F1F1} 0.0962 & {\cellcolor[HTML]{FCAA5F}} \color[HTML]{000000} 0.2951 & {\cellcolor[HTML]{FCAA5F}} \color[HTML]{000000} 0.2936 & {\cellcolor[HTML]{FDAF62}} \color[HTML]{000000} 0.3012 & {\cellcolor[HTML]{ED5F3C}} \color[HTML]{F1F1F1} 0.1793 & {\cellcolor[HTML]{E95538}} \color[HTML]{F1F1F1} 0.1605 & {\cellcolor[HTML]{E54E35}} \color[HTML]{F1F1F1} 0.1498 \\
& BERTScore & {\cellcolor[HTML]{FFFEBE}} \color[HTML]{000000} 0.4977 & {\cellcolor[HTML]{FED07E}} \color[HTML]{000000} 0.3707 & {\cellcolor[HTML]{FED481}} \color[HTML]{000000} 0.3782 & {\cellcolor[HTML]{F8864F}} \color[HTML]{F1F1F1} 0.2391 & {\cellcolor[HTML]{FDAD60}} \color[HTML]{000000} 0.2969 & {\cellcolor[HTML]{FCAA5F}} \color[HTML]{000000} 0.2954 & {\cellcolor[HTML]{FEC877}} \color[HTML]{000000} 0.3550 & {\cellcolor[HTML]{FDB365}} \color[HTML]{000000} 0.3106 & {\cellcolor[HTML]{FDAD60}} \color[HTML]{000000} 0.2971 & {\cellcolor[HTML]{F7844E}} \color[HTML]{F1F1F1} 0.2357 & {\cellcolor[HTML]{FB9D59}} \color[HTML]{000000} 0.2756 & {\cellcolor[HTML]{F88C51}} \color[HTML]{F1F1F1} 0.2465 \\
& CodeBERTScore & {\cellcolor[HTML]{E0F295}} \color[HTML]{000000} 0.5798 & {\cellcolor[HTML]{FEE28F}} \color[HTML]{000000} 0.4080 & {\cellcolor[HTML]{FEE491}} \color[HTML]{000000} 0.4130 & {\cellcolor[HTML]{FDC171}} \color[HTML]{000000} 0.3389 & {\cellcolor[HTML]{FDC574}} \color[HTML]{000000} 0.3473 & {\cellcolor[HTML]{FDC776}} \color[HTML]{000000} 0.3478 & {\cellcolor[HTML]{FEEFA3}} \color[HTML]{000000} 0.4488 & {\cellcolor[HTML]{FED683}} \color[HTML]{000000} 0.3817 & {\cellcolor[HTML]{FEDE89}} \color[HTML]{000000} 0.3953 & {\cellcolor[HTML]{FEC877}} \color[HTML]{000000} 0.3521 & {\cellcolor[HTML]{FDB96A}} \color[HTML]{000000} 0.3227 & {\cellcolor[HTML]{FED27F}} \color[HTML]{000000} 0.3739 \\
\bottomrule
\end{tabular}
    }
\includegraphics[width=0.8\linewidth]{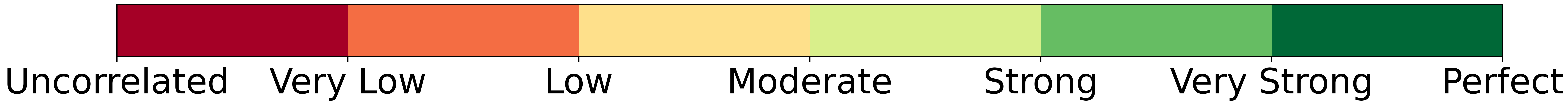}
\vspace*{-0.5em}
\caption{Current LLM-based metrics, like CodeBERTScore and CodeScore, show minimal correlation with the execution-based functional correctness on HumanEval across multiple languages. No metric exceeds an average correlation coefficient of $r = 0.31$, highlighting a significant opportunity for developing better metrics.}
\label{table:correlation-multilingual-humaneval}
\end{table*}

\section{\dataset{} Transformation Examples}
\label{appendix:dataset-examples}

\begin{figure*}[ht]
\setminted{fontsize=\tiny,framesep=1mm}
    
\begin{tcbraster}[raster columns=3, raster rows=2, raster column skip=2mm, raster valign=top]
\begin{tcolorbox}[enhanced,frame hidden,boxrule=1pt,coltitle=black,colbacktitle=babyblue,title=\centering\textcolor{white}{\footnotesize{Original}},colback=floralwhite,borderline={0.75pt}{0pt}{lightgray,dashed},left=.1em, right=.1em, top=.1em, bottom=.1em]
        \begin{minted}[autogobble,numberblanklines=false,frame=lines,breaklines,tabsize=2,breaksymbolleft=]{python}
    def binary_search(item_list,item):
        first = 0
        last = len(item_list)-1
        found = False
        while( first<=last and not found):
            mid = (first + last)//2
            if item_list[mid] == item :
                found = True
            else:
                if item < item_list[mid]:
                    last = mid - 1
                else:
                    first = mid + 1	
        return found
        \end{minted}
    \end{tcolorbox}
    \begin{tcolorbox}[enhanced,frame hidden,boxrule=1pt,coltitle=black,colbacktitle=emerald,title=\centering\textcolor{white}{\footnotesize{Dead code insertion}},colback=floralwhite,borderline={0.75pt}{0pt}{lightgray,dashed},left=.1em, right=.1em, top=.1em, bottom=.1em]
        \begin{minted}[autogobble,numberblanklines=false,frame=lines,breaklines,tabsize=2,breaksymbolleft=]{python}
    def binary_search(item_list, item):
        first = 0
        for _i_3 in range(0):
            first = 0
        last = len(item_list) - 1
        found = False
        while first <= last and not found:
            mid = (first + last) // 2
            if item_list[mid] == item:
                found = True
            else:
                if item < item_list[mid]:
                    last = mid - 1
                else:
                    first = mid + 1
        return found
        \end{minted}
    \end{tcolorbox}
    \begin{tcolorbox}[enhanced,frame hidden,boxrule=1pt,coltitle=black,colbacktitle=emerald,title=\centering\textcolor{white}{\footnotesize{For loop \textrightarrow While loop}},colback=floralwhite,borderline={0.75pt}{0pt}{lightgray,dashed},left=.1em, right=.1em, top=.1em, bottom=.1em]
        \begin{minted}[escapeinside=||,autogobble,numberblanklines=false,frame=lines,breaklines,tabsize=2,breaksymbolleft=]{python}
    def binary_search(item_list, item):
        first = 0
        last = len(item_list) - 1
        found = False
        while first <= last and not found:
            mid = (first + last) // 2
            if item_list[mid] == item:
                found = True
            else:
                if item < item_list[mid]:
                    last = mid - 1
                else:
                    first = mid + 1
        return found
        \end{minted}
    \end{tcolorbox}
    \begin{tcolorbox}[enhanced,frame hidden,boxrule=1pt,coltitle=black,colbacktitle=emerald,title=\centering\textcolor{white}{\footnotesize{Operand swap}},colback=floralwhite,borderline={0.75pt}{0pt}{lightgray,dashed},left=.1em, right=.1em, top=.1em, bottom=.1em]
        \begin{minted}[escapeinside=||,autogobble,numberblanklines=false,frame=lines,breaklines,tabsize=2,breaksymbolleft=]{python}
    def binary_search(item_list, item):
        first = 0
        last = len(item_list) - 1
        found = False
        while |\colorbox{yellow}{last >= first}| and not found:
            mid = (first + last) // 2
            if item_list[mid] == item:
                found = True
            else:
                if item < item_list[mid]:
                    last = mid - 1
                else:
                    first = mid + 1
        return found
        \end{minted}
    \end{tcolorbox}
    \begin{tcolorbox}[enhanced,frame hidden,boxrule=1pt,coltitle=black,colbacktitle=emerald,title=\centering\textcolor{white}{\footnotesize{Rename variables (CB)}},colback=floralwhite,borderline={0.75pt}{0pt}{lightgray,dashed},left=.1em, right=.1em, top=.1em, bottom=.1em]
        \begin{minted}[escapeinside=||,autogobble,numberblanklines=false,frame=lines,breaklines,tabsize=2,breaksymbolleft=]{python}
    def binary_search(item_list, item):
        first = 0
        last = len(item_list) - 1
        found = False
        while first <= last and not found:
            |\colorbox{yellow}{first2}| = (first + last) // 2
            if item_list[|\colorbox{yellow}{first2}|] == item:
                found = True
            else:
                if item < item_list[|\colorbox{yellow}{first2}|]:
                    last = |\colorbox{yellow}{first2}| - 1
                else:
                    first = |\colorbox{yellow}{first2}| + 1
        return found
        \end{minted}
    \end{tcolorbox}
    \begin{tcolorbox}[enhanced,frame hidden,boxrule=1pt,coltitle=black,colbacktitle=emerald,title=\centering\textcolor{white}{\footnotesize{Rename variables (Naive)}},colback=floralwhite,borderline={0.75pt}{0pt}{lightgray,dashed},left=.1em, right=.1em, top=.1em, bottom=.1em]
        \begin{minted}[escapeinside=||,autogobble,numberblanklines=false,frame=lines,breaklines,tabsize=2,breaksymbolleft=]{python}
    def binary_search(item_list, item):
        first = 0
        last = len(item_list) - 1
        found = False
        while first <= last and not found:
            |\colorbox{yellow}{VAR\_0}| = (first + last) // 2
            if item_list[|\colorbox{yellow}{VAR\_0}|] == item:
                found = True
            else:
                if item < item_list[|\colorbox{yellow}{VAR\_0}|]:
                    last = |\colorbox{yellow}{VAR\_0}| - 1
                else:
                    first = |\colorbox{yellow}{VAR\_0}| + 1
        return found
        \end{minted}
    \end{tcolorbox}
    \begin{tcolorbox}[enhanced,frame hidden,boxrule=1pt,coltitle=black,colbacktitle=emerald,title=\centering\textcolor{white}{\footnotesize{Rename variables (RN)}},colback=floralwhite,borderline={0.75pt}{0pt}{lightgray,dashed},left=.1em, right=.1em, top=.1em, bottom=.1em]
        \begin{minted}[escapeinside=||,autogobble,numberblanklines=false,frame=lines,breaklines,tabsize=2,breaksymbolleft=]{python}
    def binary_search(item_list, item):
        first = 0
        last = len(item_list) - 1
        found = False
        while first <= last and not found:
            |\colorbox{yellow}{ztc}| = (first + last) // 2
            if item_list[|\colorbox{yellow}{ztc}|] == item:
                found = True
            else:
                if item < item_list[ztc]:
                    last = |\colorbox{yellow}{ztc}| - 1
                else:
                    first = |\colorbox{yellow}{ztc}| + 1
        return found
        \end{minted}
    \end{tcolorbox}
\end{tcbraster}
\caption{Examples of the output of semantic-preserving transformations.}
\vspace*{-1em}
\label{fig:dataset-example}
\end{figure*}

\clearpage

\begin{figure*}[!ht]
    \centering
    \includegraphics[width=1.\linewidth]{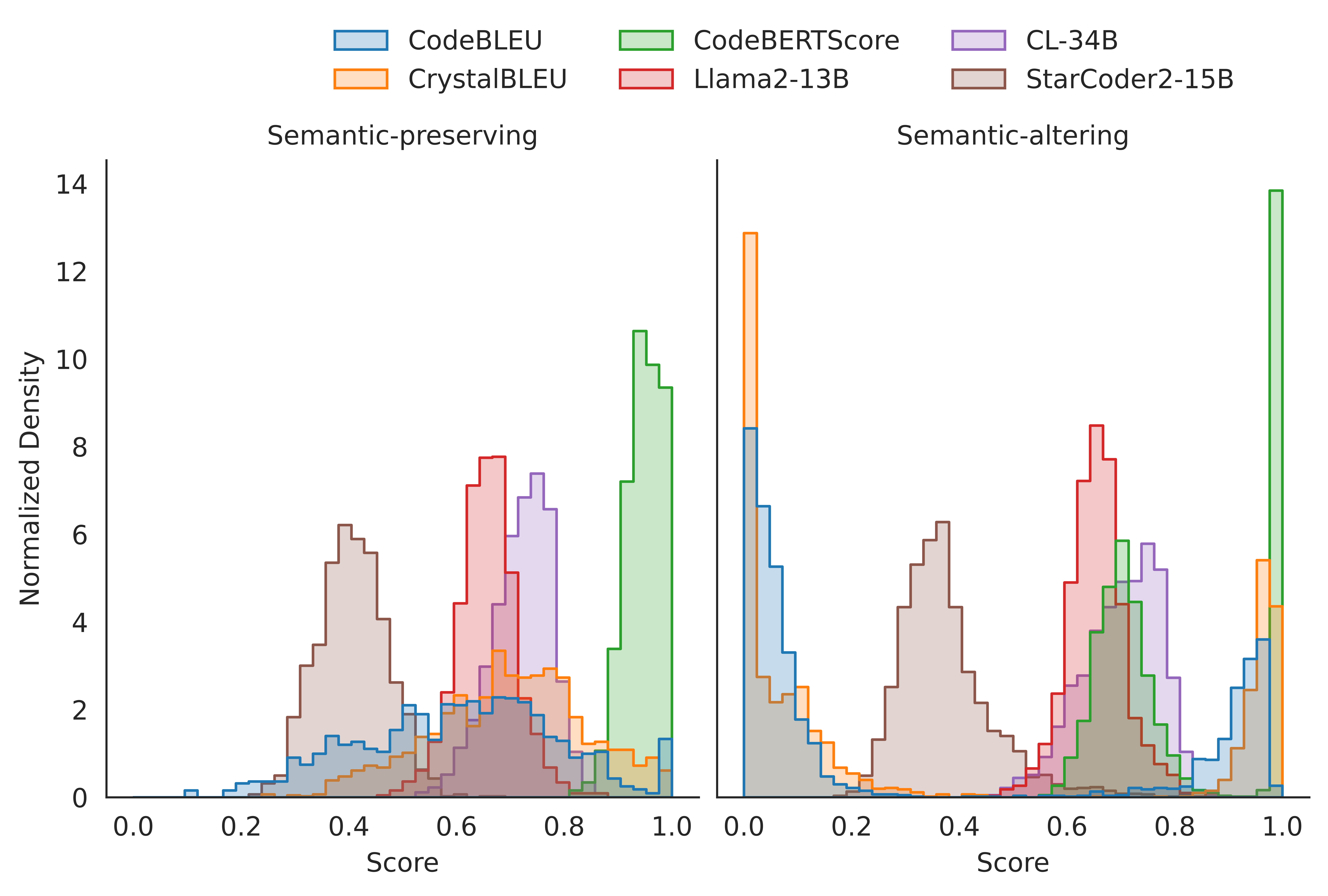}
    \caption{We observe peaks in scores above 0.9 for SA transformations, leading to incorrect semantic similarity calculations. When we consider all the metrics, we observe that the probability distribution of scores for semantic-altering labels is often higher than for SP labels.
    }
    \label{fig:hist-metric-plot}
\end{figure*}

\begin{table*}[!ht]
\centering
\renewcommand*{\arraystretch}{1.3}
\resizebox{1.\linewidth}{!}
    {\begin{tabular}{@{\extracolsep{5pt}}lllcccccccccccc@{}}
\toprule
\multicolumn{3}{c}{\textbf{Transformation}} & \multicolumn{6}{c}{\textit{N-gram-matching}} & \multicolumn{2}{c}{\textit{N-gram-matching w/ code-related features}} \\
\cline{1-3} \cline{4-9} \cline{10-11}
\multicolumn{1}{c}{\textbf{Type}} &  \multicolumn{1}{c}{\textbf{Category}} &  \multicolumn{1}{c}{\textbf{Name}} & \multirow{1}{*}{\textbf{Rouge1}} &  \multirow{1}{*}{\textbf{Rouge2}} &  \multirow{1}{*}{\textbf{RougeL}} &  \multirow{1}{*}{\textbf{Meteor}} &  \multirow{1}{*}{\textbf{ChrF}} &  \multirow{1}{*}{\textbf{BLEU}} &  \multirow{1}{*}{\textbf{CrystalBLEU}} &  \multirow{1}{*}{\textbf{CodeBLEU}} \\
\midrule
\multirow[c]{15}{*}{SA} & \multirow[c]{4}{*}{AOM} & $+$ $\rightarrow$ $-$ & 2.70 & 5.70 & 2.70 & 2.63 & 6.35 & 7.46 & 7.03 & \textbf{26.44} \\
 &  & $-$ $\rightarrow$ $+$ & 2.33 & 4.96 & 2.33 & 3.57 & 5.43 & 13.08 & 11.56 & \textbf{18.81} \\
 &  & $\div$ $\rightarrow$ $\times$ & 2.41 & 5.23 & 2.41 & 2.49 & 3.88 & 6.79 & 7.72 & \textbf{23.25} \\
 &  & $\times$ $\rightarrow$ $\div$ & 5.13 & 8.53 & 5.13 & 3.94 & 6.04 & 14.51 & 12.82 & \textbf{46.31} \\
\cmidrule{2-11}
 & \multirow[c]{2}{*}{BOM} & \texttt{False} $\rightarrow$ \texttt{True} & 1.86 & 3.90 & 1.86 & 2.73 & 3.96 & 7.84 & 10.56 & \textbf{16.84} \\
 &  & \texttt{True} $\rightarrow$ \texttt{False} & 1.84 & 3.86 & 1.84 & 2.71 & 4.37 & 7.77 & 10.49 & \textbf{24.57} \\
\cmidrule{2-11}
 & \multirow[c]{4}{*}{COM} & == $\rightarrow$ != & 2.69 & 5.68 & 2.69 & 2.81 & 4.00 & 8.02 & 10.46 & \textbf{25.46} \\
 &  & != $\rightarrow$ == & 1.07 & 2.22 & 1.07 & 1.08 & 2.65 & 2.91 & 3.01 & \textbf{9.23} \\
 &  & $>$ $\rightarrow$ $<$ & 3.03 & 5.11 & 3.03 & 3.58 & 7.54 & 12.95 & 11.21 & \textbf{25.90} \\
 &  & $<$ $\rightarrow$ $>$ & 3.02 & 5.01 & 3.02 & 2.34 & 7.19 & 6.99 & 6.00 & \textbf{21.38} \\
\cmidrule{2-11}
 & DCS & Dissimilar Code Inject & \textbf{29.34} & 14.84 & 25.02 & 27.92 & 13.11 & 6.33 & 6.32 & 5.45 \\
\cmidrule{2-11}
 & \multirow[c]{2}{*}{IOM} & \texttt{is} $\rightarrow$ \texttt{is not} & 0.86 & 3.16 & 0.86 & 2.16 & 2.96 & 4.87 & 5.62 & \textbf{17.48} \\
 &  & \texttt{is not} $\rightarrow$ \texttt{is} & \textbf{50.00} & \textbf{50.00} & \textbf{50.00} & \textbf{50.00} & \textbf{50.00} & \textbf{50.00} & \textbf{50.00} & \textbf{50.00} \\
\cmidrule{2-11}
 & \multirow[c]{2}{*}{LOM} & \texttt{and} $\rightarrow$ \texttt{or} & 1.29 & 2.67 & 1.29 & 3.86 & 3.50 & 11.20 & 12.45 & \textbf{28.65} \\
 &  & \texttt{or} $\rightarrow$ \texttt{and} & 1.40 & 2.88 & 1.40 & 2.71 & 2.31 & 7.57 & 8.14 & \textbf{10.78} \\
\cmidrule{1-11}
\multirow[c]{6}{*}{SP} & DCI & Dead Code Insert & 22.26 & 27.94 & 20.59 & 21.67 & 31.55 & 33.34 & 33.26 & \textbf{49.88} \\
\cmidrule{2-11}
 & LT & \texttt{for} $\leftrightarrow$ \texttt{while} Loop & 19.13 & 35.11 & 28.82 & 26.28 & 26.00 & 43.63 & 45.20 & \textbf{46.70} \\
\cmidrule{2-11}
 & OS & Operand Swap & 10.75 & 19.42 & 13.29 & 16.57 & 30.10 & 31.19 & 30.89 & \textbf{54.26} \\
\cmidrule{2-11}
 & \multirow[c]{3}{*}{RV} & Rename Variable Cb & 16.62 & 30.72 & 16.10 & 23.85 & 31.57 & 42.46 & 41.95 & \textbf{44.16} \\
 &  & Rename Variable Naive & 16.57 & 30.62 & 16.01 & 23.67 & 30.70 & 42.16 & 41.15 & \textbf{43.56} \\
 &  & Rename Variable Rn & 16.60 & 30.67 & 16.04 & 23.72 & 33.28 & 42.26 & 41.65 & \textbf{43.75} \\
\bottomrule
\end{tabular}
    }
\caption{Transformation-wise breakdown for Match-based metrics in the zero-shot setting using Area under the Recall curve (AURC) metric. 
}
\label{table:mbpp-transformation-breakdown-match}
\end{table*}

\clearpage
\section{Other Work}
\label{appendix:extended-related-work}
\paragraph{Robustness against adversarial attacks.}
There is a strong connection between adversarial attacks and semantic transformations for code. Both involve minor changes to a code input that preserve the original meaning or functionality but cause a machine learning model to make incorrect predictions. We derive inspiration from previous works in software engineering to leverage semantic-preserving code transformations. For instance,~\citet{rabin_generalizability_2021} highlights the importance of considering semantics-preserving transformations when building reliable neural program analysers, and~\citet{compton_embedding_2020} rely on data augmentation methods based on variable renaming to design efficient models.

A few works focused on addressing the robustness problem for the code on attacks and defences.~\citet{miceli-barone-etal-2023-larger} demonstrates how LLMs understand code semantics by approximating $\alpha$-equivalence of Python code and suggest improved defences against identifier substitution attacks.
~\citet{yang_natural_2022} argue that adversarial examples should preserve natural semantics in addition to operational semantics. To identify vulnerabilities of Code-LLMs to adversarial attacks,~\citet{jha_codeattack_2023} introduces CodeAttack that generates adversarial samples for a code fragment, and ~\citet{wang_recode_2023} designs a robustness evaluation benchmark and evaluates Code-LLMs on their ability to generate equivalent code across perturbed prompts while we aim to comprehend and analyze existing code.

\subsection*{Novelty of our approach compared to existing works}
We clarify the motivation and real-world applicability of the proposed task setting.

\paragraph{Motivation.}
By using transformations that preserve semantics, we can better isolate the LLMs ability to understand the code's underlying meaning rather than just its surface-level structure. So, essentially, the task becomes more about detecting equivalence despite superficial differences rather than recognizing that two entirely different implementations achieve the same result. In addition, it allows for a more nuanced sensitivity analysis of the model's behaviour, helping to identify which types of code changes are most likely to affect the model's judgment of equivalence. Our focus is on small lexical changes that preserve semantics or the other way around, not on general control flow restructuring, as even with these small lexical changes, LLMs do not do well. We can most certainly expect them to be able to tackle full restructuring at the moment. We are deepening and further showing issues with LLMs in a similar style to~\citet{miceli-barone-etal-2023-larger} but much more extensively and comprehensively.

\section{Code Clone Detection and Code Obfuscation}
\label{appendix:other-code-tasks}
While there is a degree of overlap \emph{in spirit} between the transformations appearing in code clone detection datasets and our benchmark, our primary goal is different. As stated above, capturing functional equivalence is a harder task and lets us evaluate how LLMs disentangle syntax from semantics. We note that our transformations are designed to be the simplest as possible (e.g., flipping an operator) to detect how much LLMs get confused by transformations that are trivial for us humans. This is a design choice that lets us carry on a finer grain analysis that is not possible with current benchmarks such as BigCloneBench~\citep{svajlenko2014towards} and, as such, should not be rated as \emph{not substantial}. 

Current code clone detection benchmarks, e.g., BigCloneBench, propose a different task than ours: they are designed to identify code fragments that share the same high-level functionality, but that can be very different from semantically equivalent problems. For example, two functions that can have different side-effects, or even having different input and output types are still considered “clones”. Our semantic equivalence definition is more stringent and better captures what LLMs understand about code execution. Moreover,~\citet{9978255} highlight how many clone pairs are falsely tagged as actual clones in BigCloneBench.

We remark that our objective is not to create a dataset to improve LLMs capabilities to detect clones in the real world. Instead, we want a controlled environment where we can exactly understand which simple transformations are able to confuse LLMs by discriminating syntax from semantics. We also argue that simple transformations can be common in real-world scenarios, e.g., out of simple typos and wrong copy-pasting actions that are common in programming. While having real-world large code repositories would be nice, it is a non-trivial task to collect the samples (for e.g., from GitHub) that can be validated using unit tests and at the same time allow us to have a controlled experimental setting. Hence, we rely on existing benchmarks which have samples tested by human programmers for functional correctness to build our benchmark.

We further note that our approach differs from code obfuscation approaches as the main difference is that the former intentionally reduces readability (up to the point that the code can be hard for humans to follow). Our semantic transformations, on the other hand, want to preserve human readability and intelligibility. Our transformations are simple by design and \emph{hard} enough to show certain failure modes of LLMs that are commonly used and deployed. As we state in Section~\ref{sec:dataset} we design them to make the simplest change in syntax or semantics that allows us to measure that code LLMs struggle to disentangle these two aspects. This is a feature, not a bug. Having more complex transformations (e.g., iterative vs recursive implementations, loop unrolling, etc.) would simply imply that LLMs failing on simple transformations are also failing on more complex ones (but without the simplest ones, we would not know what is the smallest perturbation that confuses the LLMs).

\section{Examples}
When evaluating LLMs for their ability to classify code snippets as functionally equivalent or not, it is essential to consider a variety of examples that highlight both the strengths and limitations of these models. Below, we present a selection of examples that demonstrate scenarios where LLMs may find it easy or hard to determine code equivalence for the different transformations in \dataset.

\captionsetup[lstlisting]{labelformat=empty}

\lstset{breaklines=true, postbreak=\mbox{\textcolor{red}{$\hookrightarrow$}\space}}

\begin{figure*}[!ht]
\begin{tcolorbox}[enhanced, title={Dead Code Insertion}, attach boxed title to top center={yshift=-3mm}, subtitle style={
    bottomrule=0.4pt,
    toprule=0pt,
    leftrule=0pt,
    rightrule=0pt,
    colback=babyblue,
    colframe=black,
    borderline south={0.4pt}{0pt}{black,dotted}
  }]
\tcbsubtitle[halign=center]{Easy}
\begin{multicols}{2}
\begin{lstlisting}[language=Python, frame=single, basicstyle=\ttfamily\srcsize, caption=\textbf{Reference}]
def count_ways(n):
    A = [0] * (n + 1) 
    B = [0] * (n + 1) 
    A[0] = 1
    A[1] = 0
    B[0] = 0
    B[1] = 1
    for i in range(2, n+1): 
        A[i] = A[i - 2] + 2 * B[i - 1] 
        B[i] = A[i - 1] + B[i - 2] 
    return A[n]
\end{lstlisting}
\begin{lstlisting}[language=Python, frame=single, basicstyle=\ttfamily\srcsize, caption=\textbf{Transformed}]
def count_ways(n):
    A = [0] * (n + 1)
    B = [0] * (n + 1)
    A[0] = 1
    while False:
        B[0] = 0
    A[1] = 0
    B[0] = 0
    B[1] = 1
    for i in range(2, n + 1):
        A[i] = A[i - 2] + 2 * B[i - 1]
        B[i] = A[i - 1] + B[i - 2]
    return A[n]
\end{lstlisting}
\columnbreak
\vspace*{20pt}
\begin{tikzpicture}[font=\tiny]
\matrix (m) [matrix of nodes, 
             nodes={minimum width=3cm, minimum height=0.7cm},
             row sep=-\pgflinewidth,
             column sep=-\pgflinewidth,
             nodes in empty cells,
             column 1/.style={nodes={draw, anchor=center}},
             column 2/.style={nodes={draw, anchor=center}},
             row 1/.style={nodes={fill=gray!20, font=\bfseries\footnotesize}}
             ] {
    Model          & Prediction \\
    Llama2-7B      & YES {\color{darkspringgreen}\large\faCheck} \\
    Llama2-13B     & YES {\color{darkspringgreen}\large\faCheck}  \\
    CodeLlama-7B   & YES {\color{darkspringgreen}\large\faCheck} \\ 
    CodeLlama-13B  & YES {\color{darkspringgreen}\large\faCheck} \\ 
    CodeLlama-34B  & YES {\color{darkspringgreen}\large\faCheck} \\ 
    StarCoder2-3B  & YES {\color{darkspringgreen}\large\faCheck} \\
    StarCoder2-7B  & NO {\color{bostonuniversityred}\large\faTimes} \\
    StarCoder2-15B & YES {\color{darkspringgreen}\large\faCheck} \\
};
\draw (m-1-1.north west) -- (m-8-1.south west);
\draw (m-1-2.north east) -- (m-8-2.south east);
\draw (m-1-1.north west) -- (m-1-2.north east);
\draw (m-8-1.south west) -- (m-8-2.south east);
\end{tikzpicture}
\end{multicols}

\tcbsubtitle[halign=center]{Hard}
\begin{multicols}{2}
\begin{lstlisting}[language=Python, frame=single, basicstyle=\ttfamily\srcsize, caption=\textbf{Reference}]
def check_min_heap(arr, i):
    if 2 * i + 2 > len(arr):
        return True
    left_child = (arr[i] <= arr[2 * i + 1]) and check_min_heap(arr, 2 * i + 1)
    right_child = (2 * i + 2 == len(arr)) or (arr[i] <= arr[2 * i + 2]  and check_min_heap(arr, 2 * i + 2))
    return left_child and right_child
\end{lstlisting}
\begin{lstlisting}[language=Python, frame=single, basicstyle=\ttfamily\srcsize, caption=\textbf{Transformed}]
def check_min_heap(arr, i):
    if 2 * i + 2 > len(arr):
        return True
        _i_4 = 0
        while _i_4 > _i_4:
            right_child = (2 * i + 2 == len(arr)) or (
                arr[i] <= arr[2 * i + 2] and check_min_heap(arr, 2 * i + 2)
            )
    left_child = (arr[i] <= arr[2 * i + 1]) and check_min_heap(arr, 2 * i + 1)
    right_child = (2 * i + 2 == len(arr)) or (
        arr[i] <= arr[2 * i + 2] and check_min_heap(arr, 2 * i + 2)
    )
    return left_child and right_child
\end{lstlisting}
\columnbreak
\vspace*{20pt}
\begin{tikzpicture}[font=\tiny]
\matrix (m) [matrix of nodes, 
             nodes={minimum width=3cm, minimum height=0.7cm},
             row sep=-\pgflinewidth,
             column sep=-\pgflinewidth,
             nodes in empty cells,
             column 1/.style={nodes={draw, anchor=center}},
             column 2/.style={nodes={draw, anchor=center}},
             row 1/.style={nodes={fill=gray!20, font=\bfseries\footnotesize}}
             ] {
    Model          & Prediction \\
    Llama2-7B      & YES {\color{darkspringgreen}\large\faCheck} \\
    Llama2-13B     & YES  {\color{darkspringgreen}\large\faCheck} \\
    CodeLlama-7B   & YES {\color{darkspringgreen}\large\faCheck} \\ 
    CodeLlama-13B  & NO  {\color{bostonuniversityred}\large\faTimes} \\ 
    CodeLlama-34B  & YES {\color{darkspringgreen}\large\faCheck} \\ 
    StarCoder2-3B  & NO {\color{bostonuniversityred}\large\faTimes} \\
    StarCoder2-7B  & NO {\color{bostonuniversityred}\large\faTimes} \\
    StarCoder2-15B & NO {\color{bostonuniversityred}\large\faTimes} \\
};
\draw (m-1-1.north west) -- (m-8-1.south west);
\draw (m-1-2.north east) -- (m-8-2.south east);
\draw (m-1-1.north west) -- (m-1-2.north east);
\draw (m-8-1.south west) -- (m-8-2.south east);
\end{tikzpicture}
\end{multicols}
\end{tcolorbox}
\caption{In the easy example, the function uses a loop to populate these arrays according to specific recurrence relations, ultimately returning the value of A[n]. In the transformed version of the code, an unnecessary while False statement is included, which does not affect the functionality but disrupts code clarity. Model predictions indicate that various LLMs successfully recognize the function's correctness, while one variant of StarCoder2 does not. In the hard example, the transformed version attempts to introduce a loop but retains the recursive checks for left and right children, albeit with some redundancy in the definitions of \texttt{left\_child} and \texttt{right\_child}. Additionally, the model predictions indicate varying success rates of different LLMs in recognizing the min-heap structure.}
\end{figure*}

\captionsetup[lstlisting]{labelformat=empty}

\lstset{breaklines=true, postbreak=\mbox{\textcolor{red}{$\hookrightarrow$}\space}}

\begin{figure*}[!ht]
\begin{tcolorbox}[enhanced, title={For-While Loop}, attach boxed title to top center={yshift=-3mm}, subtitle style={
    bottomrule=0.4pt,
    toprule=0pt,
    leftrule=0pt,
    rightrule=0pt,
    colback=babyblue,
    colframe=black,
    borderline south={0.4pt}{0pt}{black,dotted}
  }]
\tcbsubtitle[halign=center]{Easy}
\begin{multicols}{2}
\begin{lstlisting}[language=Python, frame=single, basicstyle=\ttfamily\srcsize, caption=\textbf{Reference}]
 def solve(s):
    flg = 0
    idx = 0
    new_str = list(s)
    for i in s:
        if i.isalpha():
            new_str[idx] = i.swapcase()
            flg = 1
        idx += 1
    s = ""
    for i in new_str:
        s += i
    if flg == 0:
        return s[len(s)::-1]
    return s
\end{lstlisting}
\begin{lstlisting}[language=Python, frame=single, basicstyle=\ttfamily\srcsize, caption=\textbf{Transformed}]
def solve(s):
    flg = 0
    idx = 0
    new_str = list(s)
    _i_i = 0
    while _i_i < len(s):
        i = s[_i_i]
        if i.isalpha():
            new_str[idx] = i.swapcase()
            flg = 1
        idx += 1
        _i_i += 1
    s = ""
    for i in new_str:
        s += i
    if flg == 0:
        return s[len(s) :: -1]
    return s
\end{lstlisting}
\columnbreak
\vspace*{20pt}
\begin{tikzpicture}[font=\tiny]
\matrix (m) [matrix of nodes, 
             nodes={minimum width=3cm, minimum height=0.7cm},
             row sep=-\pgflinewidth,
             column sep=-\pgflinewidth,
             nodes in empty cells,
             column 1/.style={nodes={draw, anchor=center}},
             column 2/.style={nodes={draw, anchor=center}},
             row 1/.style={nodes={fill=gray!20, font=\bfseries\footnotesize}}
             ] {
    Model          & Prediction \\
    Llama2-7B      & YES {\color{darkspringgreen}\large\faCheck} \\
    Llama2-13B     & YES {\color{darkspringgreen}\large\faCheck}  \\
    CodeLlama-7B   & YES {\color{darkspringgreen}\large\faCheck} \\ 
    CodeLlama-13B  & YES {\color{darkspringgreen}\large\faCheck} \\ 
    CodeLlama-34B  & YES {\color{darkspringgreen}\large\faCheck} \\ 
    StarCoder2-3B  & YES {\color{darkspringgreen}\large\faCheck} \\
    StarCoder2-7B  & YES {\color{darkspringgreen}\large\faCheck} \\
    StarCoder2-15B & NO {\color{bostonuniversityred}\large\faTimes} \\ 
};
\draw (m-1-1.north west) -- (m-8-1.south west);
\draw (m-1-2.north east) -- (m-8-2.south east);
\draw (m-1-1.north west) -- (m-1-2.north east);
\draw (m-8-1.south west) -- (m-8-2.south east);
\end{tikzpicture}
\end{multicols}

\tcbsubtitle[halign=center]{Hard}
\begin{multicols}{2}
\begin{lstlisting}[language=Python, frame=single, basicstyle=\ttfamily\srcsize, caption=\textbf{Reference}]
def ascii_value_string(str1):
    for i in range(len(str1)):
        return ord(str1[i])
\end{lstlisting}
\begin{lstlisting}[language=Python, frame=single, basicstyle=\ttfamily\srcsize, caption=\textbf{Transformed}]
def ascii_value_string(str1):
    i = 0
    while i < len(str1):
        return ord(str1[i])
        i += 1
\end{lstlisting}
\columnbreak
\begin{tikzpicture}[font=\tiny]
\matrix (m) [matrix of nodes, 
             nodes={minimum width=3cm, minimum height=0.7cm},
             row sep=-\pgflinewidth,
             column sep=-\pgflinewidth,
             nodes in empty cells,
             column 1/.style={nodes={draw, anchor=center}},
             column 2/.style={nodes={draw, anchor=center}},
             row 1/.style={nodes={fill=gray!20, font=\bfseries\footnotesize}}
             ] {
    Model          & Prediction \\
    Llama2-7B      & YES {\color{darkspringgreen}\large\faCheck} \\
    Llama2-13B     & YES  {\color{darkspringgreen}\large\faCheck} \\
    CodeLlama-7B   & YES {\color{darkspringgreen}\large\faCheck} \\ 
    CodeLlama-13B  & NO  {\color{bostonuniversityred}\large\faTimes} \\ 
    CodeLlama-34B  & YES {\color{darkspringgreen}\large\faCheck} \\ 
    StarCoder2-3B  & YES {\color{darkspringgreen}\large\faCheck} \\ 
    StarCoder2-7B  & NO {\color{bostonuniversityred}\large\faTimes} \\
    StarCoder2-15B & NO {\color{bostonuniversityred}\large\faTimes} \\
};
\draw (m-1-1.north west) -- (m-8-1.south west);
\draw (m-1-2.north east) -- (m-8-2.south east);
\draw (m-1-1.north west) -- (m-1-2.north east);
\draw (m-8-1.south west) -- (m-8-2.south east);
\end{tikzpicture}
\end{multicols}
\end{tcolorbox}
\caption{In the easy example, it defines a function that processes a string by swapping the case of its alphabetic characters while preserving the order of non-alphabetic characters. The transformed version of the function uses a \texttt{while} loop instead of a \texttt{for} loop to achieve the same functionality. Most LLMs confirm its correctness. In the hard example, the original implementation uses a \texttt{for} loop to iterate through the string, while the transformed version employs a \texttt{while} loop. Model predictions indicate that various LLMs, including Llama2-7B, Llama2-13B, CodeLlama-7B, and CodeLlama-34B, successfully recognize the function's equivalence, whereas CodeLlama-13B, StarCoder2-7B, and StarCoder2-15B do not.}
\end{figure*}

\captionsetup[lstlisting]{labelformat=empty}

\lstset{breaklines=true, postbreak=\mbox{\textcolor{red}{$\hookrightarrow$}\space}}

\begin{figure*}[!ht]
\begin{tcolorbox}[enhanced, title={Operand Swap}, attach boxed title to top center={yshift=-3mm}, subtitle style={
    bottomrule=0.4pt,
    toprule=0pt,
    leftrule=0pt,
    rightrule=0pt,
    colback=babyblue,
    colframe=black,
    borderline south={0.4pt}{0pt}{black,dotted}
  }]
\tcbsubtitle[halign=center]{Easy}
\begin{multicols}{2}
\begin{lstlisting}[language=Python, frame=single, basicstyle=\ttfamily\srcsize, caption=\textbf{Reference}]
import math

def poly(xs: list, x: float):
    return sum([coeff * math.pow(x, i) for i, coeff in enumerate(xs)])
    
def find_zero(xs: list):
    begin, end = -1., 1.
    while poly(xs, begin) * poly(xs, end) > 0:
        begin *= 2.0
        end *= 2.0
    while end - begin > 1e-10:
        center = (begin + end) / 2.0
        if poly(xs, center) * poly(xs, begin) > 0:
            begin = center
        else:
            end = center
    return begin
\end{lstlisting}
\begin{lstlisting}[language=Python, frame=single, basicstyle=\ttfamily\srcsize, caption=\textbf{Transformed}]
import math

def poly(xs: list, x: float):
    return sum([coeff * math.pow(x, i) for i, coeff in enumerate(xs)])
    
def find_zero(xs: list):
    begin, end = -1.0, 1.0
    while 0 < poly(xs, begin) * poly(xs, end):
        begin *= 2.0
        end *= 2.0
    while end - begin > 1e-10:
        center = (begin + end) / 2.0
        if poly(xs, center) * poly(xs, begin) > 0:
            begin = center
        else:
            end = center
    return begin
\end{lstlisting}
\columnbreak
\vspace*{60pt}
\begin{tikzpicture}[font=\tiny]
\matrix (m) [matrix of nodes, 
             nodes={minimum width=3cm, minimum height=0.7cm},
             row sep=-\pgflinewidth,
             column sep=-\pgflinewidth,
             nodes in empty cells,
             column 1/.style={nodes={draw, anchor=center}},
             column 2/.style={nodes={draw, anchor=center}},
             row 1/.style={nodes={fill=gray!20, font=\bfseries\footnotesize}}
             ] {
    Model          & Prediction \\
    Llama2-7B      & YES {\color{darkspringgreen}\large\faCheck} \\
    Llama2-13B     & YES {\color{darkspringgreen}\large\faCheck}  \\
    CodeLlama-7B   & YES {\color{darkspringgreen}\large\faCheck} \\ 
    CodeLlama-13B  & YES {\color{darkspringgreen}\large\faCheck} \\ 
    CodeLlama-34B  & YES {\color{darkspringgreen}\large\faCheck} \\ 
    StarCoder2-3B  & YES {\color{darkspringgreen}\large\faCheck} \\
    StarCoder2-7B  & YES {\color{darkspringgreen}\large\faCheck} \\
    StarCoder2-15B & NO {\color{bostonuniversityred}\large\faTimes} \\ 
};
\draw (m-1-1.north west) -- (m-8-1.south west);
\draw (m-1-2.north east) -- (m-8-2.south east);
\draw (m-1-1.north west) -- (m-1-2.north east);
\draw (m-8-1.south west) -- (m-8-2.south east);
\end{tikzpicture}
\end{multicols}

\tcbsubtitle[halign=center]{Hard}
\begin{multicols}{2}
\begin{lstlisting}[language=Python, frame=single, basicstyle=\ttfamily\srcsize, caption=\textbf{Reference}]
def triangle_area(a, b, c):
    if a + b <= c or a + c <= b or b + c <= a:
        return -1 
    s = (a + b + c)/2    
    area = (s * (s - a) * (s - b) * (s - c)) ** 0.5
    area = round(area, 2)
    return area
\end{lstlisting}
\begin{lstlisting}[language=Python, frame=single, basicstyle=\ttfamily\srcsize, caption=\textbf{Transformed}]
def triangle_area(a, b, c):
    if a + b <= c or a + c <= b or a >= b + c:
        return -1
    s = (a + b + c) / 2
   
    area = (s * (s - a) * (s - b) * (s - c)) ** 0.5
    area = round(area, 2)
    return area
\end{lstlisting}
\columnbreak
\begin{tikzpicture}[font=\tiny]
\matrix (m) [matrix of nodes, 
             nodes={minimum width=3cm, minimum height=0.7cm},
             row sep=-\pgflinewidth,
             column sep=-\pgflinewidth,
             nodes in empty cells,
             column 1/.style={nodes={draw, anchor=center}},
             column 2/.style={nodes={draw, anchor=center}},
             row 1/.style={nodes={fill=gray!20, font=\bfseries\footnotesize}}
             ] {
    Model          & Prediction \\
    Llama2-7B      & YES {\color{darkspringgreen}\large\faCheck} \\
    Llama2-13B     & YES  {\color{darkspringgreen}\large\faCheck} \\
    CodeLlama-7B   & YES {\color{darkspringgreen}\large\faCheck} \\ 
    CodeLlama-13B  & YES {\color{darkspringgreen}\large\faCheck} \\
    CodeLlama-34B  & YES {\color{darkspringgreen}\large\faCheck} \\ 
    StarCoder2-3B  & NO {\color{bostonuniversityred}\large\faTimes} \\
    StarCoder2-7B  & NO {\color{bostonuniversityred}\large\faTimes} \\
    StarCoder2-15B & NO {\color{bostonuniversityred}\large\faTimes} \\
};
\draw (m-1-1.north west) -- (m-8-1.south west);
\draw (m-1-2.north east) -- (m-8-2.south east);
\draw (m-1-1.north west) -- (m-1-2.north east);
\draw (m-8-1.south west) -- (m-8-2.south east);
\end{tikzpicture}
\end{multicols}
\end{tcolorbox}
\caption{In the easy example, the transformed version of the code slightly modifies the conditions in the function for clarity. Most models, except for StarCoder2-15B, successfully recognize the code's functionality. In the hard example, an error in the conditional check is noted, where a >= b + c should be corrected to b + c <= a, in the transformed version. Model predictions indicate that several models successfully identify the functional equivalence aspect, while others (StarCoder2 variants) do not.}
\end{figure*}

\captionsetup[lstlisting]{labelformat=empty}

\lstset{breaklines=true, postbreak=\mbox{\textcolor{red}{$\hookrightarrow$}\space}}

\begin{figure*}[!ht]
\begin{tcolorbox}[enhanced, title={Rename-Variable (CB) Example}, attach boxed title to top center={yshift=-3mm}, subtitle style={
    bottomrule=0.4pt,
    toprule=0pt,
    leftrule=0pt,
    rightrule=0pt,
    colback=babyblue,
    colframe=black,
    borderline south={0.4pt}{0pt}{black,dotted}
  }]
\tcbsubtitle[halign=center]{Easy}
\begin{multicols}{2}
\begin{lstlisting}[language=Python, frame=single, basicstyle=\ttfamily\srcsize, caption=\textbf{Reference}]
import math
 
def poly(xs: list, x: float):
    return sum([coeff * math.pow(x, i) for i, coeff in enumerate(xs)])
    
def find_zero(xs: list):
    begin, end = -1., 1.
    while poly(xs, begin) * poly(xs, end) > 0:
        begin *= 2.0
        end *= 2.0
    while end - begin > 1e-10:
        center = (begin + end) / 2.0
        if poly(xs, center) * poly(xs, begin) > 0:
            begin = center
        else:
            end = center
    return begin
\end{lstlisting}
\begin{lstlisting}[language=Python, frame=single, basicstyle=\ttfamily\srcsize, caption=\textbf{Transformed}]
import math

def poly(xs: list, x: float):
    return sum([coeff * math.pow(x, i) for i, coeff in enumerate(xs)])
    
def find_zero(xs: list):
    center2, end = -1.0, 1.0
    while poly(xs, center2) * poly(xs, end) > 0:
        center2 *= 2.0
        end *= 2.0
    while end - center2 > 1e-10:
        center = (center2 + end) / 2.0
        if poly(xs, center) * poly(xs, center2) > 0:
            center2 = center
        else:
            end = center
    return center2
\end{lstlisting}
\columnbreak
\vspace*{40pt}
\begin{tikzpicture}[font=\tiny]
\matrix (m) [matrix of nodes, 
             nodes={minimum width=3cm, minimum height=0.7cm},
             row sep=-\pgflinewidth,
             column sep=-\pgflinewidth,
             nodes in empty cells,
             column 1/.style={nodes={draw, anchor=center}},
             column 2/.style={nodes={draw, anchor=center}},
             row 1/.style={nodes={fill=gray!20, font=\bfseries\footnotesize}}
             ] {
    Model          & Prediction \\
    Llama2-7B      & YES {\color{darkspringgreen}\large\faCheck} \\
    Llama2-13B     & YES {\color{darkspringgreen}\large\faCheck}  \\
    CodeLlama-7B   & YES {\color{darkspringgreen}\large\faCheck} \\ 
    CodeLlama-13B  & YES {\color{darkspringgreen}\large\faCheck} \\ 
    CodeLlama-34B  & YES {\color{darkspringgreen}\large\faCheck} \\ 
    StarCoder2-3B  & YES {\color{darkspringgreen}\large\faCheck} \\
    StarCoder2-7B  & YES {\color{darkspringgreen}\large\faCheck} \\
    StarCoder2-15B & NO {\color{bostonuniversityred}\large\faTimes} \\ 
};
\draw (m-1-1.north west) -- (m-8-1.south west);
\draw (m-1-2.north east) -- (m-8-2.south east);
\draw (m-1-1.north west) -- (m-1-2.north east);
\draw (m-8-1.south west) -- (m-8-2.south east);
\end{tikzpicture}
\end{multicols}

\tcbsubtitle[halign=center]{Hard}
\begin{multicols}{2}
\begin{lstlisting}[language=Python, frame=single, basicstyle=\ttfamily\srcsize, caption=\textbf{Reference}]
def triangle_area(a, b, c):
    if a + b <= c or a + c <= b or b + c <= a:
        return -1 
    s = (a + b + c)/2    
    area = (s * (s - a) * (s - b) * (s - c)) ** 0.5
    area = round(area, 2)
    return area
\end{lstlisting}
\begin{lstlisting}[language=Python, frame=single, basicstyle=\ttfamily\srcsize, caption=\textbf{Transformed}]
def triangle_area(a, b2, c):
    if a + b2 <= c or a + c <= b2 or b2 + c <= a:
        return -1
    s = (a + b2 + c) / 2
    area = (s * (s - a) * (s - b2) * (s - c)) ** 0.5
    area = round(area, 2)
    return area
\end{lstlisting}
\columnbreak
\begin{tikzpicture}[font=\tiny]
\matrix (m) [matrix of nodes, 
             nodes={minimum width=3cm, minimum height=0.7cm},
             row sep=-\pgflinewidth,
             column sep=-\pgflinewidth,
             nodes in empty cells,
             column 1/.style={nodes={draw, anchor=center}},
             column 2/.style={nodes={draw, anchor=center}},
             row 1/.style={nodes={fill=gray!20, font=\bfseries\footnotesize}}
             ] {
    Model          & Prediction \\
    Llama2-7B      & YES {\color{darkspringgreen}\large\faCheck} \\
    Llama2-13B     & YES  {\color{darkspringgreen}\large\faCheck} \\
    CodeLlama-7B   & YES {\color{darkspringgreen}\large\faCheck} \\ 
    CodeLlama-13B  & YES {\color{darkspringgreen}\large\faCheck} \\
    CodeLlama-34B  & NO {\color{darkspringgreen}\large\faCheck} \\ 
    StarCoder2-3B  & NO {\color{bostonuniversityred}\large\faTimes} \\
    StarCoder2-7B  & NO {\color{bostonuniversityred}\large\faTimes} \\
    StarCoder2-15B & NO {\color{bostonuniversityred}\large\faTimes} \\
};
\draw (m-1-1.north west) -- (m-8-1.south west);
\draw (m-1-2.north east) -- (m-8-2.south east);
\draw (m-1-1.north west) -- (m-1-2.north east);
\draw (m-8-1.south west) -- (m-8-2.south east);
\end{tikzpicture}
\end{multicols}
\end{tcolorbox}
\caption{In the easy example, the transformed code modifies variable names for clarity without changing the underlying logic. Model predictions indicate that several LLMs (Llama2 and CodeLlama) successfully handle the task, while one (StarCoder2-15B) does not. In the hard example, the transformed version of the function uses a different variable name for one of the sides (b2 instead of b) but maintains the same logic. Model predictions indicate that various LLMs (Llama2 and CodeLlama) successfully solve the task, while others (StarCoder2) do not.}
\end{figure*}

\captionsetup[lstlisting]{labelformat=empty}

\lstset{breaklines=true, postbreak=\mbox{\textcolor{red}{$\hookrightarrow$}\space}}

\begin{figure*}[!ht]
\begin{tcolorbox}[enhanced, title={Rename-Variable (Naive) Example}, attach boxed title to top center={yshift=-3mm}, subtitle style={
    bottomrule=0.4pt,
    toprule=0pt,
    leftrule=0pt,
    rightrule=0pt,
    colback=babyblue,
    colframe=black,
    borderline south={0.4pt}{0pt}{black,dotted}
  }]
\tcbsubtitle[halign=center]{Easy}
\begin{multicols}{2}
\begin{lstlisting}[language=Python, frame=single, basicstyle=\ttfamily\srcsize, caption=\textbf{Reference}]
def text_match_two_three(text):
    import re
    patterns = 'ab{2,3}'
    if re.search(patterns,  text):
        return 'Found a match!'
    else:
        return('Not matched!')
\end{lstlisting}
\begin{lstlisting}[language=Python, frame=single, basicstyle=\ttfamily\srcsize, caption=\textbf{Transformed}]
def text_match_two_three(VAR_0):
    import re
    patterns = "ab{2,3}"
    if re.search(patterns, VAR_0):
        return "Found a match!"
    else:
        return "Not matched!"
\end{lstlisting}
\columnbreak
\begin{tikzpicture}[font=\tiny]
\matrix (m) [matrix of nodes, 
             nodes={minimum width=3cm, minimum height=0.7cm},
             row sep=-\pgflinewidth,
             column sep=-\pgflinewidth,
             nodes in empty cells,
             column 1/.style={nodes={draw, anchor=center}},
             column 2/.style={nodes={draw, anchor=center}},
             row 1/.style={nodes={fill=gray!20, font=\bfseries\footnotesize}}
             ] {
    Model          & Prediction \\
    Llama2-7B      & YES {\color{darkspringgreen}\large\faCheck} \\
    Llama2-13B     & YES {\color{darkspringgreen}\large\faCheck} \\ 
    CodeLlama-7B   & YES {\color{darkspringgreen}\large\faCheck} \\ 
    CodeLlama-13B  & YES {\color{darkspringgreen}\large\faCheck} \\
    CodeLlama-34B  & YES {\color{darkspringgreen}\large\faCheck} \\ 
    StarCoder2-3B  & YES {\color{darkspringgreen}\large\faCheck} \\
    StarCoder2-7B  & YES {\color{darkspringgreen}\large\faCheck} \\
    StarCoder2-15B & YES {\color{darkspringgreen}\large\faCheck} \\
};
\draw (m-1-1.north west) -- (m-8-1.south west);
\draw (m-1-2.north east) -- (m-8-2.south east);
\draw (m-1-1.north west) -- (m-1-2.north east);
\draw (m-8-1.south west) -- (m-8-2.south east);
\end{tikzpicture}
\end{multicols}

\tcbsubtitle[halign=center]{Hard}
\begin{multicols}{2}
\begin{lstlisting}[language=Python, frame=single, basicstyle=\ttfamily\srcsize, caption=\textbf{Reference}]
def all_Bits_Set_In_The_Given_Range(n,l,r):
    num = (((1 << r) - 1) ^ ((1 << (l - 1)) - 1)) 
    new_num = n & num
    if (new_num == 0): 
        return True
    return False
\end{lstlisting}
\begin{lstlisting}[language=Python, frame=single, basicstyle=\ttfamily\srcsize, caption=\textbf{Transformed}]
def all_Bits_Set_In_The_Given_Range(n, l, VAR_0):
    num = ((1 << VAR_0) - 1) ^ ((1 << (l - 1)) - 1)
    new_num = n & num
    if new_num == 0:
        return True
    return False
\end{lstlisting}
\columnbreak
\begin{tikzpicture}[font=\tiny]
\matrix (m) [matrix of nodes, 
             nodes={minimum width=3cm, minimum height=0.7cm},
             row sep=-\pgflinewidth,
             column sep=-\pgflinewidth,
             nodes in empty cells,
             column 1/.style={nodes={draw, anchor=center}},
             column 2/.style={nodes={draw, anchor=center}},
             row 1/.style={nodes={fill=gray!20, font=\bfseries\footnotesize}}
             ] {
    Model          & Prediction \\
    Llama2-7B      & YES {\color{darkspringgreen}\large\faCheck} \\
    Llama2-13B     & NO {\color{bostonuniversityred}\large\faTimes} \\ 
    CodeLlama-7B   & YES {\color{darkspringgreen}\large\faCheck} \\ 
    CodeLlama-13B  & NO {\color{bostonuniversityred}\large\faTimes} \\ 
    CodeLlama-34B  & YES {\color{darkspringgreen}\large\faCheck} \\ 
    StarCoder2-3B  & YES {\color{darkspringgreen}\large\faCheck} \\
    StarCoder2-7B  & NO {\color{bostonuniversityred}\large\faTimes} \\ 
    StarCoder2-15B & NO {\color{bostonuniversityred}\large\faTimes} \\ 
};
\draw (m-1-1.north west) -- (m-8-1.south west);
\draw (m-1-2.north east) -- (m-8-2.south east);
\draw (m-1-1.north west) -- (m-1-2.north east);
\draw (m-8-1.south west) -- (m-8-2.south east);
\end{tikzpicture}
\end{multicols}
\end{tcolorbox}
\caption{In the easy example, the transformed version of the function replaces the parameter name with \texttt{VAR\_0}, but the functionality remains unchanged. All the LLMs successfully predict the function's behavior, indicating that they recognize the code's intent and structure. In the hard example, the function is transformed to use a variable \texttt{VAR\_0} instead of \texttt{r}, and various model predictions indicate whether they agree with the function's expected output, with some models confirming the correctness and others not.}
\end{figure*}

\captionsetup[lstlisting]{labelformat=empty}

\lstset{breaklines=true, postbreak=\mbox{\textcolor{red}{$\hookrightarrow$}\space}}

\begin{figure*}[!ht]
\begin{tcolorbox}[enhanced, title={$-$ $\rightarrow$ $+$}, attach boxed title to top center={yshift=-3mm}, subtitle style={
    bottomrule=0.4pt,
    toprule=0pt,
    leftrule=0pt,
    rightrule=0pt,
    colback=babyblue,
    colframe=black,
    borderline south={0.4pt}{0pt}{black,dotted}
  }]
\tcbsubtitle[halign=center]{Easy}
\begin{multicols}{2}
\begin{lstlisting}[language=Python, frame=single, basicstyle=\ttfamily\srcsize, caption=\textbf{Reference}]
def binomial_Coeff(n,k):
    if k > n : 
       return 0
    if k==0 or k ==n : 
        return 1 
    return binomial_Coeff(n-1,k-1) + binomial_Coeff(n-1,k) 
\end{lstlisting}
\begin{lstlisting}[language=Python, frame=single, basicstyle=\ttfamily\srcsize, caption=\textbf{Transformed}]
def binomial_Coeff(n,k):
    if k > n : 
       return 0
    if k==0 or k ==n : 
        return 1 
    return binomial_Coeff(n+1,k-1) + binomial_Coeff(n-1,k) 
\end{lstlisting}
\columnbreak
\begin{tikzpicture}[font=\tiny]
\matrix (m) [matrix of nodes, 
             nodes={minimum width=3cm, minimum height=0.7cm},
             row sep=-\pgflinewidth,
             column sep=-\pgflinewidth,
             nodes in empty cells,
             column 1/.style={nodes={draw, anchor=center}},
             column 2/.style={nodes={draw, anchor=center}},
             row 1/.style={nodes={fill=gray!20, font=\bfseries\footnotesize}}
             ] {
    Model          & Prediction \\
    Llama2-7B      & YES {\color{bostonuniversityred}\large\faTimes} \\
    Llama2-13B     & YES {\color{bostonuniversityred}\large\faTimes} \\
    CodeLlama-7B   & YES {\color{bostonuniversityred}\large\faTimes} \\
    CodeLlama-13B  & NO {\color{darkspringgreen}\large\faCheck} \\ 
    CodeLlama-34B  & YES {\color{bostonuniversityred}\large\faTimes} \\
    StarCoder2-3B  & YES {\color{bostonuniversityred}\large\faTimes} \\
    StarCoder2-7B  & NO {\color{darkspringgreen}\large\faCheck} \\ 
    StarCoder2-15B & NO {\color{darkspringgreen}\large\faCheck} \\ 
};
\draw (m-1-1.north west) -- (m-8-1.south west);
\draw (m-1-2.north east) -- (m-8-2.south east);
\draw (m-1-1.north west) -- (m-1-2.north east);
\draw (m-8-1.south west) -- (m-8-2.south east);
\end{tikzpicture}
\end{multicols}

\tcbsubtitle[halign=center]{Hard}
\begin{multicols}{2}
\begin{lstlisting}[language=Python, frame=single, basicstyle=\ttfamily\srcsize, caption=\textbf{Reference}]
def count_ways(n):
    A = [0] * (n + 1) 
    B = [0] * (n + 1) 
    A[0] = 1
    A[1] = 0
    B[0] = 0
    B[1] = 1
    for i in range(2, n+1): 
        A[i] = A[i - 2] + 2 * B[i - 1] 
        B[i] = A[i - 1] + B[i - 2] 
    return A[n]
\end{lstlisting}
\begin{lstlisting}[language=Python, frame=single, basicstyle=\ttfamily\srcsize, caption=\textbf{Transformed}]
def count_ways(n):
    A = [0] * (n + 1) 
    B = [0] * (n + 1) 
    A[0] = 1
    A[1] = 0
    B[0] = 0
    B[1] = 1
    for i in range(2, n+1): 
        A[i] = A[i + 2] + 2 * B[i - 1] 
        B[i] = A[i - 1] + B[i - 2] 
    return A[n] 
\end{lstlisting}
\columnbreak
\vspace*{20pt}
\begin{tikzpicture}[font=\tiny]
\matrix (m) [matrix of nodes, 
             nodes={minimum width=3cm, minimum height=0.7cm},
             row sep=-\pgflinewidth,
             column sep=-\pgflinewidth,
             nodes in empty cells,
             column 1/.style={nodes={draw, anchor=center}},
             column 2/.style={nodes={draw, anchor=center}},
             row 1/.style={nodes={fill=gray!20, font=\bfseries\footnotesize}}
             ] {
    Model          & Prediction \\
    Llama2-7B      & YES {\color{bostonuniversityred}\large\faTimes} \\
    Llama2-13B     & YES {\color{bostonuniversityred}\large\faTimes} \\
    CodeLlama-7B   & YES {\color{bostonuniversityred}\large\faTimes} \\
    CodeLlama-13B  & YES {\color{bostonuniversityred}\large\faTimes} \\
    CodeLlama-34B  & YES {\color{bostonuniversityred}\large\faTimes} \\
    StarCoder2-3B  & YES {\color{bostonuniversityred}\large\faTimes} \\
    StarCoder2-7B  & NO {\color{darkspringgreen}\large\faCheck} \\ 
    StarCoder2-15B & YES {\color{bostonuniversityred}\large\faTimes} \\
};
\draw (m-1-1.north west) -- (m-8-1.south west);
\draw (m-1-2.north east) -- (m-8-2.south east);
\draw (m-1-1.north west) -- (m-1-2.north east);
\draw (m-8-1.south west) -- (m-8-2.south east);
\end{tikzpicture}
\end{multicols}
\end{tcolorbox}
\caption{In the easy example, the reference version correctly implements the recursive logic, returning 0 if k is greater than n, 1 if k is 0 or equal to n, and otherwise summing two recursive calls. The transformed version modifies the first recursive call to use n + 1 instead of n - 1, which alters the logic but retains the original base cases. In terms of model predictions, various LLMs show differing capabilities in handling the transformed function, with some successfully predicting the output while others do not. In the hard example, the function uses a loop to fill these sequences based on specific recurrence relations. However, a transformation in the code erroneously modifies the index in the calculation for A[i], changing it from A[i - 2] to A[i + 2], which likely leads to incorrect results. Various LLMs produced erroneous classification, while one variant of StarCoder2 predicted correctly.}
\end{figure*}

\captionsetup[lstlisting]{labelformat=empty}

\lstset{breaklines=true, postbreak=\mbox{\textcolor{red}{$\hookrightarrow$}\space}}

\begin{figure*}[!ht]
\begin{tcolorbox}[enhanced, title={$+$ $\rightarrow$ $-$}, attach boxed title to top center={yshift=-3mm}, subtitle style={
    bottomrule=0.4pt,
    toprule=0pt,
    leftrule=0pt,
    rightrule=0pt,
    colback=babyblue,
    colframe=black,
    borderline south={0.4pt}{0pt}{black,dotted}
  }]
\tcbsubtitle[halign=center]{Easy}
\begin{multicols}{2}
\begin{lstlisting}[language=Python, frame=single, basicstyle=\ttfamily\srcsize, caption=\textbf{Reference}]
def maximum_Sum(list1):
    maxi = -100000
    for x in list1: 
        sum = 0 
        for y in x: 
            sum+= y      
        maxi = max(sum,maxi)     
    return maxi 
\end{lstlisting}
\begin{lstlisting}[language=Python, frame=single, basicstyle=\ttfamily\srcsize, caption=\textbf{Transformed}]
def maximum_Sum(list1):
    maxi = -100000
    for x in list1: 
        sum = 0 
        for y in x: 
            sum-= y      
        maxi = max(sum,maxi)     
    return maxi 
\end{lstlisting}
\columnbreak
\begin{tikzpicture}[font=\tiny]
\matrix (m) [matrix of nodes, 
             nodes={minimum width=3cm, minimum height=0.7cm},
             row sep=-\pgflinewidth,
             column sep=-\pgflinewidth,
             nodes in empty cells,
             column 1/.style={nodes={draw, anchor=center}},
             column 2/.style={nodes={draw, anchor=center}},
             row 1/.style={nodes={fill=gray!20, font=\bfseries\footnotesize}}
             ] {
    Model          & Prediction \\
    Llama2-7B      & YES {\color{bostonuniversityred}\large\faTimes} \\
    Llama2-13B     & YES {\color{bostonuniversityred}\large\faTimes} \\
    CodeLlama-7B   & YES {\color{bostonuniversityred}\large\faTimes} \\
    CodeLlama-13B  & NO {\color{darkspringgreen}\large\faCheck} \\ 
    CodeLlama-34B  & NO {\color{darkspringgreen}\large\faCheck} \\ 
    StarCoder2-3B  & YES {\color{bostonuniversityred}\large\faTimes} \\
    StarCoder2-7B  & NO {\color{darkspringgreen}\large\faCheck} \\ 
    StarCoder2-15B & NO {\color{darkspringgreen}\large\faCheck} \\ 
};
\draw (m-1-1.north west) -- (m-8-1.south west);
\draw (m-1-2.north east) -- (m-8-2.south east);
\draw (m-1-1.north west) -- (m-1-2.north east);
\draw (m-8-1.south west) -- (m-8-2.south east);
\end{tikzpicture}
\end{multicols}

\tcbsubtitle[halign=center]{Hard}
\begin{multicols}{2}
\begin{lstlisting}[language=Python, frame=single, basicstyle=\ttfamily\srcsize, caption=\textbf{Reference}]
def count_ways(n):
    A = [0] * (n + 1) 
    B = [0] * (n + 1) 
    A[0] = 1
    A[1] = 0
    B[0] = 0
    B[1] = 1
    for i in range(2, n+1): 
        A[i] = A[i - 2] + 2 * B[i - 1] 
        B[i] = A[i - 1] + B[i - 2] 
    return A[n]
\end{lstlisting}
\begin{lstlisting}[language=Python, frame=single, basicstyle=\ttfamily\srcsize, caption=\textbf{Transformed}]
def count_ways(n):
    A = [0] * (n - 1) 
    B = [0] * (n + 1) 
    A[0] = 1
    A[1] = 0
    B[0] = 0
    B[1] = 1
    for i in range(2, n+1): 
        A[i] = A[i - 2] + 2 * B[i - 1] 
        B[i] = A[i - 1] + B[i - 2] 
    return A[n]
\end{lstlisting}
\columnbreak
\vspace*{20pt}
\begin{tikzpicture}[font=\tiny]
\matrix (m) [matrix of nodes, 
             nodes={minimum width=3cm, minimum height=0.7cm},
             row sep=-\pgflinewidth,
             column sep=-\pgflinewidth,
             nodes in empty cells,
             column 1/.style={nodes={draw, anchor=center}},
             column 2/.style={nodes={draw, anchor=center}},
             row 1/.style={nodes={fill=gray!20, font=\bfseries\footnotesize}}
             ] {
    Model          & Prediction \\
    Llama2-7B      & YES {\color{bostonuniversityred}\large\faTimes} \\
    Llama2-13B     & YES {\color{bostonuniversityred}\large\faTimes} \\
    CodeLlama-7B   & YES {\color{bostonuniversityred}\large\faTimes} \\
    CodeLlama-13B  & YES {\color{bostonuniversityred}\large\faTimes} \\
    CodeLlama-34B  & YES {\color{bostonuniversityred}\large\faTimes} \\
    StarCoder2-3B  & YES {\color{bostonuniversityred}\large\faTimes} \\
    StarCoder2-7B  & NO {\color{darkspringgreen}\large\faCheck} \\ 
    StarCoder2-15B & YES {\color{bostonuniversityred}\large\faTimes} \\
};
\draw (m-1-1.north west) -- (m-8-1.south west);
\draw (m-1-2.north east) -- (m-8-2.south east);
\draw (m-1-1.north west) -- (m-1-2.north east);
\draw (m-8-1.south west) -- (m-8-2.south east);
\end{tikzpicture}
\end{multicols}
\end{tcolorbox}
\caption{In the easy example, the function sums the inner lists' elements and updates a maximum value in the original version. The transformed version, however, subtracts the elements instead of adding them, which fundamentally alters the function's purpose. The model predictions indicate that various LLMs have some difficulty in classifying them as functionally nonequivalent. In the hard example, the function initializes the base cases for A and B, then iteratively fills these lists based on previously computed values to derive the total ways to reach the n\textsuperscript{th} step. The first version of the function allocates an array A of size n + 1, while the transformed version incorrectly allocates A of size n - 1, which could lead to an index error. Model predictions indicate that various LLMs fail to discern the function's validity, with some exceptions.}
\end{figure*}

\captionsetup[lstlisting]{labelformat=empty}

\lstset{breaklines=true, postbreak=\mbox{\textcolor{red}{$\hookrightarrow$}\space}}

\begin{figure*}[!ht]
\begin{tcolorbox}[enhanced, title={$\times$ $\rightarrow$ $\div$}, attach boxed title to top center={yshift=-3mm}, subtitle style={
    bottomrule=0.4pt,
    toprule=0pt,
    leftrule=0pt,
    rightrule=0pt,
    colback=babyblue,
    colframe=black,
    borderline south={0.4pt}{0pt}{black,dotted}
  }]
\tcbsubtitle[halign=center]{Easy}
\begin{multicols}{2}
\begin{lstlisting}[language=Python, frame=single, basicstyle=\ttfamily\srcsize, caption=\textbf{Reference}]
def find_Volume(l,b,h):
    return ((l * b * h) / 2) 
\end{lstlisting}
\begin{lstlisting}[language=Python, frame=single, basicstyle=\ttfamily\srcsize, caption=\textbf{Transformed}]
def find_Volume(l,b,h):
    return ((l / b * h) / 2)
\end{lstlisting}
\columnbreak
\begin{tikzpicture}[font=\tiny]
\matrix (m) [matrix of nodes, 
             nodes={minimum width=3cm, minimum height=0.7cm},
             row sep=-\pgflinewidth,
             column sep=-\pgflinewidth,
             nodes in empty cells,
             column 1/.style={nodes={draw, anchor=center}},
             column 2/.style={nodes={draw, anchor=center}},
             row 1/.style={nodes={fill=gray!20, font=\bfseries\footnotesize}}
             ] {
    Model          & Prediction \\
    Llama2-7B      & YES {\color{bostonuniversityred}\large\faTimes} \\
    Llama2-13B     & YES {\color{bostonuniversityred}\large\faTimes}  \\
    CodeLlama-7B   & YES {\color{bostonuniversityred}\large\faTimes} \\ 
    CodeLlama-13B  & YES {\color{bostonuniversityred}\large\faTimes} \\ 
    CodeLlama-34B  & NO {\color{darkspringgreen}\large\faCheck} \\
    StarCoder2-3B  & YES {\color{bostonuniversityred}\large\faTimes} \\
    StarCoder2-7B  & NO {\color{darkspringgreen}\large\faCheck} \\
    StarCoder2-15B & NO {\color{darkspringgreen}\large\faCheck} \\
};
\draw (m-1-1.north west) -- (m-8-1.south west);
\draw (m-1-2.north east) -- (m-8-2.south east);
\draw (m-1-1.north west) -- (m-1-2.north east);
\draw (m-8-1.south west) -- (m-8-2.south east);
\end{tikzpicture}
\end{multicols}

\tcbsubtitle[halign=center]{Hard}
\begin{multicols}{2}
\begin{lstlisting}[language=Python, frame=single, basicstyle=\ttfamily\srcsize, caption=\textbf{Reference}]
def count_ways(n):
    A = [0] * (n + 1) 
    B = [0] * (n + 1) 
    A[0] = 1
    A[1] = 0
    B[0] = 0
    B[1] = 1
    for i in range(2, n+1): 
        A[i] = A[i - 2] + 2 * B[i - 1] 
        B[i] = A[i - 1] + B[i - 2] 
    return A[n]
\end{lstlisting}
\begin{lstlisting}[language=Python, frame=single, basicstyle=\ttfamily\srcsize, caption=\textbf{Transformed}]
def count_ways(n):
    A = [0] / (n + 1) 
    B = [0] * (n + 1) 
    A[0] = 1
    A[1] = 0
    B[0] = 0
    B[1] = 1
    for i in range(2, n+1): 
        A[i] = A[i - 2] + 2 * B[i - 1] 
        B[i] = A[i - 1] + B[i - 2] 
    return A[n]
\end{lstlisting}
\columnbreak
\begin{tikzpicture}[font=\tiny]
\matrix (m) [matrix of nodes, 
             nodes={minimum width=3cm, minimum height=0.7cm},
             row sep=-\pgflinewidth,
             column sep=-\pgflinewidth,
             nodes in empty cells,
             column 1/.style={nodes={draw, anchor=center}},
             column 2/.style={nodes={draw, anchor=center}},
             row 1/.style={nodes={fill=gray!20, font=\bfseries\footnotesize}}
             ] {
    Model          & Prediction \\
    Llama2-7B      & YES {\color{bostonuniversityred}\large\faTimes} \\
    Llama2-13B     & YES {\color{bostonuniversityred}\large\faTimes} \\ 
    CodeLlama-7B   & YES {\color{bostonuniversityred}\large\faTimes} \\
    CodeLlama-13B  & YES {\color{bostonuniversityred}\large\faTimes} \\
    CodeLlama-34B  & YES {\color{bostonuniversityred}\large\faTimes} \\ 
    StarCoder2-3B  & YES {\color{bostonuniversityred}\large\faTimes} \\
    StarCoder2-7B  & NO  {\color{darkspringgreen}\large\faCheck} \\
    StarCoder2-15B & YES {\color{bostonuniversityred}\large\faTimes} \\
};
\draw (m-1-1.north west) -- (m-8-1.south west);
\draw (m-1-2.north east) -- (m-8-2.south east);
\draw (m-1-1.north west) -- (m-1-2.north east);
\draw (m-8-1.south west) -- (m-8-2.south east);
\end{tikzpicture}
\end{multicols}
\end{tcolorbox}
\caption{In the easy example, the reference code correctly calculates the volume of a rectangular prism using the formula ($l \times b \times h$), while the transformed code incorrectly divides l by b and then divides the result by 2, which does not match the formula. In the hard example, the function sets base cases for A and B, then iteratively fills the lists using previously computed values. The transformed code contains a minor error in the initialization of list A, where the division operator is incorrectly used instead of the multiplication operator. The model predictions indicate that various LLMs, unsuccessfully classify the function's structure and logic, with some models showing varying levels of accuracy.}
\end{figure*}

\captionsetup[lstlisting]{labelformat=empty}

\lstset{breaklines=true, postbreak=\mbox{\textcolor{red}{$\hookrightarrow$}\space}}

\begin{figure*}[!ht]
\begin{tcolorbox}[enhanced, title={$\div$ $\rightarrow$ $\times$}, attach boxed title to top center={yshift=-3mm}, subtitle style={
    bottomrule=0.4pt,
    toprule=0pt,
    leftrule=0pt,
    rightrule=0pt,
    colback=babyblue,
    colframe=black,
    borderline south={0.4pt}{0pt}{black,dotted}
  }]
\tcbsubtitle[halign=center]{Easy}
\begin{multicols}{2}
\begin{lstlisting}[language=Python, frame=single, basicstyle=\ttfamily\srcsize, caption=\textbf{Reference}]
def is_woodall(x):
    if (x %
        return False
    if (x == 1): 
        return True
    x = x + 1 
    p = 0
    while (x %
        x = x/2
        p = p + 1
        if (p == x): 
            return True
    return False
\end{lstlisting}
\begin{lstlisting}[language=Python, frame=single, basicstyle=\ttfamily\srcsize, caption=\textbf{Transformed}]
def is_woodall(x):
    if (x %
        return False
    if (x == 1): 
        return True
    x = x + 1 
    p = 0
    while (x %
        x = x*2
        p = p + 1
        if (p == x): 
            return True
    return False
\end{lstlisting}
\columnbreak
\vspace*{20pt}
\begin{tikzpicture}[font=\tiny]
\matrix (m) [matrix of nodes, 
             nodes={minimum width=3cm, minimum height=0.7cm},
             row sep=-\pgflinewidth,
             column sep=-\pgflinewidth,
             nodes in empty cells,
             column 1/.style={nodes={draw, anchor=center}},
             column 2/.style={nodes={draw, anchor=center}},
             row 1/.style={nodes={fill=gray!20, font=\bfseries\footnotesize}}
             ] {
    Model          & Prediction \\
    Llama2-7B      & YES {\color{bostonuniversityred}\large\faTimes} \\
    Llama2-13B     & YES {\color{bostonuniversityred}\large\faTimes} \\
    CodeLlama-7B   & YES {\color{bostonuniversityred}\large\faTimes} \\
    CodeLlama-13B  & NO {\color{darkspringgreen}\large\faCheck}  \\ 
    CodeLlama-34B  & YES {\color{bostonuniversityred}\large\faTimes} \\ 
    StarCoder2-3B  & YES {\color{bostonuniversityred}\large\faTimes} \\
    StarCoder2-7B  & NO {\color{darkspringgreen}\large\faCheck}  \\
    StarCoder2-15B & NO {\color{darkspringgreen}\large\faCheck}  \\
};
\draw (m-1-1.north west) -- (m-8-1.south west);
\draw (m-1-2.north east) -- (m-8-2.south east);
\draw (m-1-1.north west) -- (m-1-2.north east);
\draw (m-8-1.south west) -- (m-8-2.south east);
\end{tikzpicture}
\end{multicols}

\tcbsubtitle[halign=center]{Hard}
\begin{multicols}{2}
\begin{lstlisting}[language=Python, frame=single, basicstyle=\ttfamily\srcsize, caption=\textbf{Reference}]
def div_even_odd(list1):
    first_even = next((el for el in list1 if el%
    first_odd = next((el for el in list1 if el%
    return (first_even/first_odd)
\end{lstlisting}
\begin{lstlisting}[language=Python, frame=single, basicstyle=\ttfamily\srcsize, caption=\textbf{Transformed}]
def div_even_odd(list1):
    first_even = next((el for el in list1 if el%
    first_odd = next((el for el in list1 if el%
    return (first_even*first_odd)
\end{lstlisting}
\columnbreak
\begin{tikzpicture}[font=\tiny]
\matrix (m) [matrix of nodes, 
             nodes={minimum width=3cm, minimum height=0.7cm},
             row sep=-\pgflinewidth,
             column sep=-\pgflinewidth,
             nodes in empty cells,
             column 1/.style={nodes={draw, anchor=center}},
             column 2/.style={nodes={draw, anchor=center}},
             row 1/.style={nodes={fill=gray!20, font=\bfseries\footnotesize}}
             ] {
    Model          & Prediction \\
    Llama2-7B      & YES {\color{bostonuniversityred}\large\faTimes} \\
    Llama2-13B     & YES {\color{bostonuniversityred}\large\faTimes} \\
    CodeLlama-7B   & YES {\color{bostonuniversityred}\large\faTimes} \\
    CodeLlama-13B  & YES {\color{bostonuniversityred}\large\faTimes} \\
    CodeLlama-34B  & YES {\color{bostonuniversityred}\large\faTimes} \\
    StarCoder2-3B  & YES {\color{bostonuniversityred}\large\faTimes} \\
    StarCoder2-7B  & NO  {\color{darkspringgreen}\large\faCheck} \\ 
    StarCoder2-15B & NO {\color{darkspringgreen}\large\faCheck} \\ 
};
\draw (m-1-1.north west) -- (m-8-1.south west);
\draw (m-1-2.north east) -- (m-8-2.south east);
\draw (m-1-1.north west) -- (m-1-2.north east);
\draw (m-8-1.south west) -- (m-8-2.south east);
\end{tikzpicture}
\end{multicols}
\end{tcolorbox}
\caption{In the easy example, the function checks if this count equals the modified value of $x$ to determine if it is a Woodall number. The transformed version incorrectly multiplies $x$ by 2 in the loop instead of dividing, which may lead to different results. Model predictions from various LLMs indicate varying success rates in identifying the correctness of the function, with some models confirming the original function as valid while others question the transformed version. In the hard example, it returns the result of dividing the first even number by the first odd number in the original version, while the transformed version changes the operation to multiplication instead. The model predictions indicate that various LLMs unsuccessfully classify the original and transformed functions, with some discrepancies noted for certain versions of StarCoder2.}
\end{figure*}

\captionsetup[lstlisting]{labelformat=empty}

\lstset{breaklines=true, postbreak=\mbox{\textcolor{red}{$\hookrightarrow$}\space}}

\begin{figure*}[!ht]
\begin{tcolorbox}[enhanced, title={$>$ $\rightarrow$ $<$}, attach boxed title to top center={yshift=-3mm}, subtitle style={
    bottomrule=0.4pt,
    toprule=0pt,
    leftrule=0pt,
    rightrule=0pt,
    colback=babyblue,
    colframe=black,
    borderline south={0.4pt}{0pt}{black,dotted}
  }]
\tcbsubtitle[halign=center]{Easy}
\begin{multicols}{2}
\begin{lstlisting}[language=Python, frame=single, basicstyle=\ttfamily\srcsize, caption=\textbf{Reference}]
def find_Nth_Digit(p,q,N) :
    while (N > 0) : 
        N -= 1;  
        p *= 10;  
        res = p // q;  
        p %
    return res;  
\end{lstlisting}
\begin{lstlisting}[language=Python, frame=single, basicstyle=\ttfamily\srcsize, caption=\textbf{Transformed}]
def find_Nth_Digit(p,q,N) :
    while (N < 0) : 
        N -= 1;  
        p *= 10;  
        res = p // q;  
        p %
    return res; 
\end{lstlisting}
\columnbreak
\begin{tikzpicture}[font=\tiny]
\matrix (m) [matrix of nodes, 
             nodes={minimum width=3cm, minimum height=0.7cm},
             row sep=-\pgflinewidth,
             column sep=-\pgflinewidth,
             nodes in empty cells,
             column 1/.style={nodes={draw, anchor=center}},
             column 2/.style={nodes={draw, anchor=center}},
             row 1/.style={nodes={fill=gray!20, font=\bfseries\footnotesize}}
             ] {
    Model          & Prediction \\
    Llama2-7B      & YES {\color{bostonuniversityred}\large\faTimes} \\
    Llama2-13B     & YES {\color{bostonuniversityred}\large\faTimes} \\
    CodeLlama-7B   & YES {\color{bostonuniversityred}\large\faTimes} \\
    CodeLlama-13B  & NO {\color{darkspringgreen}\large\faCheck} \\ 
    CodeLlama-34B  & NO {\color{darkspringgreen}\large\faCheck} \\ 
    StarCoder2-3B  & YES {\color{bostonuniversityred}\large\faTimes} \\
    StarCoder2-7B  & NO {\color{darkspringgreen}\large\faCheck} \\
    StarCoder2-15B & NO {\color{darkspringgreen}\large\faCheck} \\
};
\draw (m-1-1.north west) -- (m-8-1.south west);
\draw (m-1-2.north east) -- (m-8-2.south east);
\draw (m-1-1.north west) -- (m-1-2.north east);
\draw (m-8-1.south west) -- (m-8-2.south east);
\end{tikzpicture}
\end{multicols}

\tcbsubtitle[halign=center]{Hard}
\begin{multicols}{2}
\begin{lstlisting}[language=Python, frame=single, basicstyle=\ttfamily\srcsize, caption=\textbf{Reference}]
def rearange_string(S):
    import heapq
    from collections import Counter
    ctr = Counter(S)
    heap = [(-value, key) for key, value in ctr.items()]
    heapq.heapify(heap)
    if (-heap[0][0]) * 2 > len(S) + 1: 
        return ""
    ans = []
    while len(heap) >= 2:
        nct1, char1 = heapq.heappop(heap)
        nct2, char2 = heapq.heappop(heap)
        ans.extend([char1, char2])
        if nct1 + 1: heapq.heappush(heap, (nct1 + 1, char1))
        if nct2 + 1: heapq.heappush(heap, (nct2 + 1, char2))
    return "".join(ans) + (heap[0][1] if heap else "")
\end{lstlisting}
\begin{lstlisting}[language=Python, frame=single, basicstyle=\ttfamily\srcsize, caption=\textbf{Transformed}]
def rearange_string(S):
    import heapq
    from collections import Counter
    ctr = Counter(S)
    heap = [(-value, key) for key, value in ctr.items()]
    heapq.heapify(heap)
    if (-heap[0][0]) * 2 < len(S) + 1: 
        return ""
    ans = []
    while len(heap) >= 2:
        nct1, char1 = heapq.heappop(heap)
        nct2, char2 = heapq.heappop(heap)
        ans.extend([char1, char2])
        if nct1 + 1: heapq.heappush(heap, (nct1 + 1, char1))
        if nct2 + 1: heapq.heappush(heap, (nct2 + 1, char2))
    return "".join(ans) + (heap[0][1] if heap else "")
\end{lstlisting}
\columnbreak
\vspace*{30pt}
\begin{tikzpicture}[font=\tiny]
\matrix (m) [matrix of nodes, 
             nodes={minimum width=3cm, minimum height=0.7cm},
             row sep=-\pgflinewidth,
             column sep=-\pgflinewidth,
             nodes in empty cells,
             column 1/.style={nodes={draw, anchor=center}},
             column 2/.style={nodes={draw, anchor=center}},
             row 1/.style={nodes={fill=gray!20, font=\bfseries\footnotesize}}
             ] {
    Model          & Prediction \\
    Llama2-7B      & YES {\color{bostonuniversityred}\large\faTimes} \\
    Llama2-13B     & YES {\color{bostonuniversityred}\large\faTimes} \\
    CodeLlama-7B   & YES {\color{bostonuniversityred}\large\faTimes} \\
    CodeLlama-13B  & YES {\color{bostonuniversityred}\large\faTimes} \\ 
    CodeLlama-34B  & YES {\color{bostonuniversityred}\large\faTimes} \\
    StarCoder2-3B  & YES {\color{bostonuniversityred}\large\faTimes} \\
    StarCoder2-7B  & YES {\color{bostonuniversityred}\large\faTimes} \\
    StarCoder2-15B & NO {\color{darkspringgreen}\large\faCheck} \\ 
};
\draw (m-1-1.north west) -- (m-8-1.south west);
\draw (m-1-2.north east) -- (m-8-2.south east);
\draw (m-1-1.north west) -- (m-1-2.north east);
\draw (m-8-1.south west) -- (m-8-2.south east);
\end{tikzpicture}
\end{multicols}
\end{tcolorbox}
\caption{In the easy example, the transformed version incorrectly checks for N being less than 0, which would lead to an infinite loop since N would never become negative. Some models successfully predict the output while others do not. In the hard example, the transformation in the code changes the condition for returning an empty string from > to <, affecting the validity check for the arrangement. The model predictions indicate that various LLMs unsuccessfully recognize the function's intent and structure, with the StarCoder2-15B model predicting correctly.}
\end{figure*}

\captionsetup[lstlisting]{labelformat=empty}

\lstset{breaklines=true, postbreak=\mbox{\textcolor{red}{$\hookrightarrow$}\space}}

\begin{figure*}[!ht]
\begin{tcolorbox}[enhanced, title={$<$ $\rightarrow$ $>$}, attach boxed title to top center={yshift=-3mm}, subtitle style={
    bottomrule=0.4pt,
    toprule=0pt,
    leftrule=0pt,
    rightrule=0pt,
    colback=babyblue,
    colframe=black,
    borderline south={0.4pt}{0pt}{black,dotted}
  }]
\tcbsubtitle[halign=center]{Easy}
\begin{multicols}{2}
\begin{lstlisting}[language=Python, frame=single, basicstyle=\ttfamily\srcsize, caption=\textbf{Reference}]
def areEquivalent(num1,num2):
    import math 
    def divSum(n): 
        sum = 1; 
        i = 2; 
        while(i * i <= n): 
            if (n %
                sum = (sum + i +math.floor(n / i)); 
            i += 1; 
        return sum;      
    return divSum(num1) == divSum(num2); 
\end{lstlisting}
\begin{lstlisting}[language=Python, frame=single, basicstyle=\ttfamily\srcsize, caption=\textbf{Transformed}]
def areEquivalent(num1,num2):
    import math 
    def divSum(n): 
        sum = 1; 
        i = 2; 
        while(i * i >= n): 
            if (n %
                sum = (sum + i +math.floor(n / i)); 
            i += 1; 
        return sum; 
    return divSum(num1) == divSum(num2); 
\end{lstlisting}
\columnbreak
\begin{tikzpicture}[font=\tiny]
\matrix (m) [matrix of nodes, 
             nodes={minimum width=3cm, minimum height=0.7cm},
             row sep=-\pgflinewidth,
             column sep=-\pgflinewidth,
             nodes in empty cells,
             column 1/.style={nodes={draw, anchor=center}},
             column 2/.style={nodes={draw, anchor=center}},
             row 1/.style={nodes={fill=gray!20, font=\bfseries\footnotesize}}
             ] {
    Model          & Prediction \\
    Llama2-7B      & YES {\color{bostonuniversityred}\large\faTimes} \\
    Llama2-13B     & YES {\color{bostonuniversityred}\large\faTimes} \\
    CodeLlama-7B   & YES {\color{bostonuniversityred}\large\faTimes} \\
    CodeLlama-13B  & NO {\color{darkspringgreen}\large\faCheck} \\ 
    CodeLlama-34B  & YES {\color{bostonuniversityred}\large\faTimes} \\
    StarCoder2-3B  & YES {\color{bostonuniversityred}\large\faTimes} \\
    StarCoder2-7B  & NO {\color{darkspringgreen}\large\faCheck} \\
    StarCoder2-15B & NO {\color{darkspringgreen}\large\faCheck} \\ 
};
\draw (m-1-1.north west) -- (m-8-1.south west);
\draw (m-1-2.north east) -- (m-8-2.south east);
\draw (m-1-1.north west) -- (m-1-2.north east);
\draw (m-8-1.south west) -- (m-8-2.south east);
\end{tikzpicture}
\end{multicols}

\tcbsubtitle[halign=center]{Hard}
\begin{multicols}{2}
\begin{lstlisting}[language=Python, frame=single, basicstyle=\ttfamily\srcsize, caption=\textbf{Reference}]
def find_length(string, n):
    current_sum = 0
    max_sum = 0
    for i in range(n): 
        current_sum += (1 if string[i] == '0' else -1) 
        if current_sum < 0: 
            current_sum = 0
        max_sum = max(current_sum, max_sum) 
    return max_sum if max_sum else 0
\end{lstlisting}
\begin{lstlisting}[language=Python, frame=single, basicstyle=\ttfamily\srcsize, caption=\textbf{Transformed}]
def find_length(string, n):
    current_sum = 0
    max_sum = 0
    for i in range(n): 
        current_sum += (1 if string[i] == '0' else -1) 
        if current_sum > 0: 
            current_sum = 0
        max_sum = max(current_sum, max_sum) 
    return max_sum if max_sum else 0
\end{lstlisting}
\columnbreak
\begin{tikzpicture}[font=\tiny]
\matrix (m) [matrix of nodes, 
             nodes={minimum width=3cm, minimum height=0.7cm},
             row sep=-\pgflinewidth,
             column sep=-\pgflinewidth,
             nodes in empty cells,
             column 1/.style={nodes={draw, anchor=center}},
             column 2/.style={nodes={draw, anchor=center}},
             row 1/.style={nodes={fill=gray!20, font=\bfseries\footnotesize}}
             ] {
    Model          & Prediction \\
    Llama2-7B      & YES {\color{bostonuniversityred}\large\faTimes} \\
    Llama2-13B     & YES {\color{bostonuniversityred}\large\faTimes} \\
    CodeLlama-7B   & YES {\color{bostonuniversityred}\large\faTimes} \\
    CodeLlama-13B  & YES {\color{bostonuniversityred}\large\faTimes} \\ 
    CodeLlama-34B  & YES {\color{bostonuniversityred}\large\faTimes} \\
    StarCoder2-3B  & YES {\color{bostonuniversityred}\large\faTimes} \\
    StarCoder2-7B  & NO {\color{darkspringgreen}\large\faCheck} \\ 
    StarCoder2-15B & YES {\color{bostonuniversityred}\large\faTimes} \\
};
\draw (m-1-1.north west) -- (m-8-1.south west);
\draw (m-1-2.north east) -- (m-8-2.south east);
\draw (m-1-1.north west) -- (m-1-2.north east);
\draw (m-8-1.south west) -- (m-8-2.south east);
\end{tikzpicture}
\end{multicols}
\end{tcolorbox}
\caption{In the easy example, the provided code defines a function which checks if two numbers are equivalent based on the sum of their divisors. In the original code, the loop condition checks if $i \times i$ is less than or equal to n, while in the transformed version, this condition is incorrectly changed to $i \times i$ being greater than or equal to n. The model predictions indicate varying results from different LLMs regarding the equivalence of the two versions, with some models affirming the equivalence and others disagreeing. In the hard example, the provided code defines a function that calculates the maximum sum of a modified array derived from a binary string, where `0' is treated as +1 and `1' as -1. The transformed version of the function incorrectly resets current\_sum when it is positive instead of negative, which alters its intended functionality. Various LLMs unsuccessfully predict the behavior of both versions of the function, with some exceptions noted for StarCoder2-7B model.}
\end{figure*}

\captionsetup[lstlisting]{labelformat=empty}

\lstset{breaklines=true, postbreak=\mbox{\textcolor{red}{$\hookrightarrow$}\space}}

\begin{figure*}[!ht]
\begin{tcolorbox}[enhanced, title={== $\rightarrow$ !=}, attach boxed title to top center={yshift=-3mm}, subtitle style={
    bottomrule=0.4pt,
    toprule=0pt,
    leftrule=0pt,
    rightrule=0pt,
    colback=babyblue,
    colframe=black,
    borderline south={0.4pt}{0pt}{black,dotted}
  }]
\tcbsubtitle[halign=center]{Easy}
\begin{multicols}{2}
\begin{lstlisting}[language=Python, frame=single, basicstyle=\ttfamily\srcsize, caption=\textbf{Reference}]
def find_Rotations(str):
    tmp = str + str
    n = len(str) 
    for i in range(1,n + 1): 
        substring = tmp[i: i+n] 
        if (str == substring): 
            return i 
    return n
\end{lstlisting}
\begin{lstlisting}[language=Python, frame=single, basicstyle=\ttfamily\srcsize, caption=\textbf{Transformed}]
def find_Rotations(str):
    tmp = str + str
    n = len(str) 
    for i in range(1,n + 1): 
        substring = tmp[i: i+n] 
        if (str != substring): 
            return i 
    return n 
\end{lstlisting}
\columnbreak
\begin{tikzpicture}[font=\tiny]
\matrix (m) [matrix of nodes, 
             nodes={minimum width=3cm, minimum height=0.7cm},
             row sep=-\pgflinewidth,
             column sep=-\pgflinewidth,
             nodes in empty cells,
             column 1/.style={nodes={draw, anchor=center}},
             column 2/.style={nodes={draw, anchor=center}},
             row 1/.style={nodes={fill=gray!20, font=\bfseries\footnotesize}}
             ] {
    Model          & Prediction \\
    Llama2-7B      & YES {\color{bostonuniversityred}\large\faTimes}\\
    Llama2-13B     & YES {\color{bostonuniversityred}\large\faTimes}\\
    CodeLlama-7B   & YES {\color{bostonuniversityred}\large\faTimes}\\
    CodeLlama-13B  & NO {\color{darkspringgreen}\large\faCheck} \\ 
    CodeLlama-34B  & YES {\color{bostonuniversityred}\large\faTimes}\\
    StarCoder2-3B  & YES {\color{bostonuniversityred}\large\faTimes}\\
    StarCoder2-7B  & NO {\color{darkspringgreen}\large\faCheck} \\ 
    StarCoder2-15B & NO {\color{darkspringgreen}\large\faCheck} \\
};
\draw (m-1-1.north west) -- (m-8-1.south west);
\draw (m-1-2.north east) -- (m-8-2.south east);
\draw (m-1-1.north west) -- (m-1-2.north east);
\draw (m-8-1.south west) -- (m-8-2.south east);
\end{tikzpicture}
\end{multicols}

\tcbsubtitle[halign=center]{Hard}
\begin{multicols}{2}
\begin{lstlisting}[language=Python, frame=single, basicstyle=\ttfamily\srcsize, caption=\textbf{Reference}]
def find_length(string, n):
    current_sum = 0
    max_sum = 0
    for i in range(n): 
        current_sum += (1 if string[i] == '0' else -1) 
        if current_sum < 0: 
            current_sum = 0
        max_sum = max(current_sum, max_sum) 
    return max_sum if max_sum else 0
\end{lstlisting}
\begin{lstlisting}[language=Python, frame=single, basicstyle=\ttfamily\srcsize, caption=\textbf{Transformed}]
def find_length(string, n):
    current_sum = 0
    max_sum = 0
    for i in range(n): 
        current_sum += (1 if string[i] != '0' else -1) 
        if current_sum < 0: 
            current_sum = 0
        max_sum = max(current_sum, max_sum) 
    return max_sum if max_sum else 0
\end{lstlisting}
\columnbreak
\begin{tikzpicture}[font=\tiny]
\matrix (m) [matrix of nodes, 
             nodes={minimum width=3cm, minimum height=0.7cm},
             row sep=-\pgflinewidth,
             column sep=-\pgflinewidth,
             nodes in empty cells,
             column 1/.style={nodes={draw, anchor=center}},
             column 2/.style={nodes={draw, anchor=center}},
             row 1/.style={nodes={fill=gray!20, font=\bfseries\footnotesize}}
             ] {
    Model          & Prediction \\
    Llama2-7B      & YES {\color{bostonuniversityred}\large\faTimes} \\ 
    Llama2-13B     & YES {\color{bostonuniversityred}\large\faTimes} \\ 
    CodeLlama-7B   & YES {\color{bostonuniversityred}\large\faTimes} \\ 
    CodeLlama-13B  & YES {\color{bostonuniversityred}\large\faTimes} \\ 
    CodeLlama-34B  & YES {\color{bostonuniversityred}\large\faTimes} \\ 
    StarCoder2-3B  & YES {\color{bostonuniversityred}\large\faTimes} \\ 
    StarCoder2-7B  & YES {\color{bostonuniversityred}\large\faTimes} \\ 
    StarCoder2-15B & YES {\color{bostonuniversityred}\large\faTimes} \\ 
};
\draw (m-1-1.north west) -- (m-8-1.south west);
\draw (m-1-2.north east) -- (m-8-2.south east);
\draw (m-1-1.north west) -- (m-1-2.north east);
\draw (m-8-1.south west) -- (m-8-2.south east);
\end{tikzpicture}
\end{multicols}
\end{tcolorbox}
\caption{In the easy example, the provided code defines a function identifies the starting index of a substring in a concatenated string, where the substring is a rotation of the original string. The original version checks for equality between the original string and each substring derived from the concatenated string, returning the index when a match is found. The transformed version, however, changes the condition to check for inequality, returning the index when the substring does not match the original string. The model predictions indicate that various LLMs have differing capabilities in recognizing the functionality of the original and transformed code, with some models successfully identifying the original logic while others do not. In the hard example, the provided code snippets define a function that calculates the maximum length of a contiguous subarray in a string, treating `0's and `1's differently. In the original version, the function increments a counter for `0's and decrements for `1's, while the transformed version does the opposite, incrementing for `1's and decrementing for `0's. The model predictions indicate that all LLMs fail to recognize and validate the functionality of both versions of the code.}
\end{figure*}

\captionsetup[lstlisting]{labelformat=empty}

\lstset{breaklines=true, postbreak=\mbox{\textcolor{red}{$\hookrightarrow$}\space}}

\begin{figure*}[!ht]
\begin{tcolorbox}[enhanced, title={!= $\rightarrow$ ==}, attach boxed title to top center={yshift=-3mm}, subtitle style={
    bottomrule=0.4pt,
    toprule=0pt,
    leftrule=0pt,
    rightrule=0pt,
    colback=babyblue,
    colframe=black,
    borderline south={0.4pt}{0pt}{black,dotted}
  }]
\tcbsubtitle[halign=center]{Easy}
\begin{multicols}{2}
\begin{lstlisting}[language=Python, frame=single, basicstyle=\ttfamily\srcsize, caption=\textbf{Reference}]
def removals(arr, n, k):
    def find_ind(key, i, n, k, arr):
        ind = -1
        start = i + 1
        end = n - 1;
        while (start < end):
            mid = int(start + (end - start) / 2)
            if (arr[mid] - key <= k):
                ind = mid
                start = mid + 1
            else:
                end = mid
        return ind
    ans = n - 1
    arr.sort()
    for i in range(0, n):
            j = find_ind(arr[i], i, n, k, arr)
            if (j != -1):
                ans = min(ans, n - (j - i + 1))
    return ans
\end{lstlisting}
\begin{lstlisting}[language=Python, frame=single, basicstyle=\ttfamily\srcsize, caption=\textbf{Transformed}]
def removals(arr, n, k):
    def find_ind(key, i, n, k, arr):
        ind = -1
        start = i + 1
        end = n - 1;
        while (start < end):
            mid = int(start + (end - start) / 2)
            if (arr[mid] - key <= k):
                ind = mid
                start = mid + 1
            else:
                end = mid
        return ind
    ans = n - 1
    arr.sort()
    for i in range(0, n):
        j = find_ind(arr[i], i, n, k, arr)
        if (j == -1):
            ans = min(ans, n - (j - i + 1))
    return ans
\end{lstlisting}
\columnbreak
\vspace*{30pt}
\begin{tikzpicture}[font=\tiny]
\matrix (m) [matrix of nodes, 
             nodes={minimum width=3cm, minimum height=0.7cm},
             row sep=-\pgflinewidth,
             column sep=-\pgflinewidth,
             nodes in empty cells,
             column 1/.style={nodes={draw, anchor=center}},
             column 2/.style={nodes={draw, anchor=center}},
             row 1/.style={nodes={fill=gray!20, font=\bfseries\footnotesize}}
             ] {
    Model          & Prediction \\
    Llama2-7B      & YES {\color{bostonuniversityred}\large\faTimes}\\
    Llama2-13B     & YES {\color{bostonuniversityred}\large\faTimes}\\
    CodeLlama-7B   & YES {\color{bostonuniversityred}\large\faTimes}\\
    CodeLlama-13B  & NO {\color{darkspringgreen}\large\faCheck} \\ 
    CodeLlama-34B  & YES {\color{bostonuniversityred}\large\faTimes}\\
    StarCoder2-3B  & YES {\color{bostonuniversityred}\large\faTimes}\\
    StarCoder2-7B  & NO {\color{darkspringgreen}\large\faCheck} \\ 
    StarCoder2-15B & NO {\color{darkspringgreen}\large\faCheck} \\
};
\draw (m-1-1.north west) -- (m-8-1.south west);
\draw (m-1-2.north east) -- (m-8-2.south east);
\draw (m-1-1.north west) -- (m-1-2.north east);
\draw (m-8-1.south west) -- (m-8-2.south east);
\end{tikzpicture}
\end{multicols}

\tcbsubtitle[halign=center]{Hard}
\begin{multicols}{2}
\begin{lstlisting}[language=Python, frame=single, basicstyle=\ttfamily\srcsize, caption=\textbf{Reference}]
def div_even_odd(list1):
    first_even = next((el for el in list1 if el%
    first_odd = next((el for el in list1 if el%
    return (first_even/first_odd)
\end{lstlisting}
\begin{lstlisting}[language=Python, frame=single, basicstyle=\ttfamily\srcsize, caption=\textbf{Transformed}]
def div_even_odd(list1):
    first_even = next((el for el in list1 if el%
    first_odd = next((el for el in list1 if el%
    return (first_even/first_odd)
\end{lstlisting}
\columnbreak
\begin{tikzpicture}[font=\tiny]
\matrix (m) [matrix of nodes, 
             nodes={minimum width=3cm, minimum height=0.7cm},
             row sep=-\pgflinewidth,
             column sep=-\pgflinewidth,
             nodes in empty cells,
             column 1/.style={nodes={draw, anchor=center}},
             column 2/.style={nodes={draw, anchor=center}},
             row 1/.style={nodes={fill=gray!20, font=\bfseries\footnotesize}}
             ] {
    Model          & Prediction \\
    Llama2-7B      & YES {\color{bostonuniversityred}\large\faTimes}\\
    Llama2-13B     & YES {\color{bostonuniversityred}\large\faTimes}\\
    CodeLlama-7B   & YES {\color{bostonuniversityred}\large\faTimes}\\
    CodeLlama-13B  & YES {\color{bostonuniversityred}\large\faTimes}\\ 
    CodeLlama-34B  & YES {\color{bostonuniversityred}\large\faTimes}\\
    StarCoder2-3B  & YES {\color{bostonuniversityred}\large\faTimes}\\
    StarCoder2-7B  & NO {\color{darkspringgreen}\large\faCheck} \\
    StarCoder2-15B & NO {\color{darkspringgreen}\large\faCheck} \\
};
\draw (m-1-1.north west) -- (m-8-1.south west);
\draw (m-1-2.north east) -- (m-8-2.south east);
\draw (m-1-1.north west) -- (m-1-2.north east);
\draw (m-8-1.south west) -- (m-8-2.south east);
\end{tikzpicture}
\end{multicols}
\end{tcolorbox}
\caption{In the easy example, the transformed code contains a logical error in the condition that checks if j is -1, which lead to incorrect results. The model predictions indicate that some models successfully recognize the logic while others do not. In the hard example, the logic is flawed in the transformed version as it incorrectly searches for the first odd number using the same condition as the even number, leading to an unexpected outcome. Most LLMs recognize the original function's equivalence wrongly, while some models flag the transformed version as incorrect.}
\end{figure*}

\captionsetup[lstlisting]{labelformat=empty}

\lstset{breaklines=true, postbreak=\mbox{\textcolor{red}{$\hookrightarrow$}\space}}

\begin{figure*}[!ht]
\begin{tcolorbox}[enhanced, title={\texttt{True} $\rightarrow$ \texttt{False}}, attach boxed title to top center={yshift=-3mm}, subtitle style={
    bottomrule=0.4pt,
    toprule=0pt,
    leftrule=0pt,
    rightrule=0pt,
    colback=babyblue,
    colframe=black,
    borderline south={0.4pt}{0pt}{black,dotted}
  }]
\tcbsubtitle[halign=center]{Easy}
\begin{multicols}{2}
\begin{lstlisting}[language=Python, frame=single, basicstyle=\ttfamily\srcsize, caption=\textbf{Reference}]
def test_distinct(data):
  if len(data) == len(set(data)):
    return True
  else:
    return False;
\end{lstlisting}
\begin{lstlisting}[language=Python, frame=single, basicstyle=\ttfamily\srcsize, caption=\textbf{Transformed}]
def test_distinct(data):
  if len(data) == len(set(data)):
    return False
  else:
    return False;
\end{lstlisting}
\columnbreak
\begin{tikzpicture}[font=\tiny]
\matrix (m) [matrix of nodes, 
             nodes={minimum width=3cm, minimum height=0.7cm},
             row sep=-\pgflinewidth,
             column sep=-\pgflinewidth,
             nodes in empty cells,
             column 1/.style={nodes={draw, anchor=center}},
             column 2/.style={nodes={draw, anchor=center}},
             row 1/.style={nodes={fill=gray!20, font=\bfseries\footnotesize}}
             ] {
    Model          & Prediction \\
    Llama2-7B      & YES {\color{bostonuniversityred}\large\faTimes} \\
    Llama2-13B     & YES {\color{bostonuniversityred}\large\faTimes} \\
    CodeLlama-7B   & YES {\color{bostonuniversityred}\large\faTimes} \\
    CodeLlama-13B  & NO {\color{darkspringgreen}\large\faCheck} \\ 
    CodeLlama-34B  & YES {\color{bostonuniversityred}\large\faTimes} \\
    StarCoder2-3B  & YES {\color{bostonuniversityred}\large\faTimes} \\
    StarCoder2-7B  & NO {\color{darkspringgreen}\large\faCheck} \\ 
    StarCoder2-15B & NO {\color{darkspringgreen}\large\faCheck} \\ 
};
\draw (m-1-1.north west) -- (m-8-1.south west);
\draw (m-1-2.north east) -- (m-8-2.south east);
\draw (m-1-1.north west) -- (m-1-2.north east);
\draw (m-8-1.south west) -- (m-8-2.south east);
\end{tikzpicture}
\end{multicols}

\tcbsubtitle[halign=center]{Hard}
\begin{multicols}{2}
\begin{lstlisting}[language=Python, frame=single, basicstyle=\ttfamily\srcsize, caption=\textbf{Reference}]
def is_valid_URL(str):
	import re
	regex = ("((http|https)://)(www.)?" +
			"[a-zA-Z0-9@:%
			"{2,256}\\.[a-z]" +
			"{2,6}\\b([-a-zA-Z0-9@:%
			"._\\+~#?&//=]*)")
	p = re.compile(regex)
	if (str == None):
		return False
	if(re.search(p, str)):
		return True
	else:
		return False
\end{lstlisting}
\begin{lstlisting}[language=Python, frame=single, basicstyle=\ttfamily\srcsize, caption=\textbf{Transformed}]
def is_valid_URL(str):
	import re
	regex = ("((http|https)://)(www.)?" +
			"[a-zA-Z0-9@:%
			"{2,256}\\.[a-z]" +
			"{2,6}\\b([-a-zA-Z0-9@:%
			"._\\+~#?&//=]*)")
	p = re.compile(regex)
	if (str == None):
		return False
	if(re.search(p, str)):
		return False
	else:
		return False
\end{lstlisting}
\columnbreak
\vspace*{10pt}
\begin{tikzpicture}[font=\tiny]
\matrix (m) [matrix of nodes, 
             nodes={minimum width=3cm, minimum height=0.7cm},
             row sep=-\pgflinewidth,
             column sep=-\pgflinewidth,
             nodes in empty cells,
             column 1/.style={nodes={draw, anchor=center}},
             column 2/.style={nodes={draw, anchor=center}},
             row 1/.style={nodes={fill=gray!20, font=\bfseries\footnotesize}}
             ] {
    Model          & Prediction \\
    Llama2-7B      & YES {\color{bostonuniversityred}\large\faTimes} \\
    Llama2-13B     & YES {\color{bostonuniversityred}\large\faTimes} \\
    CodeLlama-7B   & YES {\color{bostonuniversityred}\large\faTimes} \\
    CodeLlama-13B  & YES {\color{bostonuniversityred}\large\faTimes} \\
    CodeLlama-34B  & YES {\color{bostonuniversityred}\large\faTimes} \\
    StarCoder2-3B  & YES {\color{bostonuniversityred}\large\faTimes} \\
    StarCoder2-7B  & YES {\color{bostonuniversityred}\large\faTimes} \\ 
    StarCoder2-15B & NO {\color{darkspringgreen}\large\faCheck} \\ 
};
\draw (m-1-1.north west) -- (m-8-1.south west);
\draw (m-1-2.north east) -- (m-8-2.south east);
\draw (m-1-1.north west) -- (m-1-2.north east);
\draw (m-8-1.south west) -- (m-8-2.south east);
\end{tikzpicture}
\end{multicols}
\end{tcolorbox}
\caption{In the easy example, the original code correctly returns True if the length of data matches the length of the set created from data, indicating all elements are unique; otherwise, it returns False. However, the transformed code incorrectly returns False regardless of the input, failing to accurately determine the distinctness of the elements. Model predictions from various LLMs show mixed results, with some models correctly identifying the original function's intent while others misinterpret the transformed version's logic. In the hard example, the provided code defines a function that uses regular expressions to validate whether a given string is a valid URL. The function returns False if the input string is None, and it is supposed to return True if the string matches the regex pattern; however, the transformed code incorrectly returns False for all cases. The model predictions indicate that various LLMs, including Llama2 and CodeLlama, consistently incorrectly validate the function as equivalent except for the StarCoder2-15B model, which predicts correctly.}
\end{figure*}

\captionsetup[lstlisting]{labelformat=empty}

\lstset{breaklines=true, postbreak=\mbox{\textcolor{red}{$\hookrightarrow$}\space}}

\begin{figure*}[!ht]
\begin{tcolorbox}[enhanced, title={\texttt{False} $\rightarrow$ \texttt{True}}, attach boxed title to top center={yshift=-3mm}, subtitle style={
    bottomrule=0.4pt,
    toprule=0pt,
    leftrule=0pt,
    rightrule=0pt,
    colback=babyblue,
    colframe=black,
    borderline south={0.4pt}{0pt}{black,dotted}
  }]
\tcbsubtitle[halign=center]{Easy}
\begin{multicols}{2}
\begin{lstlisting}[language=Python, frame=single, basicstyle=\ttfamily\srcsize, caption=\textbf{Reference}]
def check_equilateral(x,y,z):
    if x == y == z:
        return True
    else:
        return False
\end{lstlisting}
\begin{lstlisting}[language=Python, frame=single, basicstyle=\ttfamily\srcsize, caption=\textbf{Transformed}]
def check_equilateral(x,y,z):
    if x == y == z:
        return True
    else:
        return True
\end{lstlisting}
\columnbreak
\begin{tikzpicture}[font=\tiny]
\matrix (m) [matrix of nodes, 
             nodes={minimum width=3cm, minimum height=0.7cm},
             row sep=-\pgflinewidth,
             column sep=-\pgflinewidth,
             nodes in empty cells,
             column 1/.style={nodes={draw, anchor=center}},
             column 2/.style={nodes={draw, anchor=center}},
             row 1/.style={nodes={fill=gray!20, font=\bfseries\footnotesize}}
             ] {
    Model          & Prediction \\
    Llama2-7B      & YES {\color{bostonuniversityred}\large\faTimes} \\
    Llama2-13B     & YES {\color{bostonuniversityred}\large\faTimes} \\
    CodeLlama-7B   & YES {\color{bostonuniversityred}\large\faTimes} \\
    CodeLlama-13B  & NO {\color{darkspringgreen}\large\faCheck} \\ 
    CodeLlama-34B  & YES {\color{bostonuniversityred}\large\faTimes} \\
    StarCoder2-3B  & YES {\color{bostonuniversityred}\large\faTimes} \\
    StarCoder2-7B  & NO {\color{darkspringgreen}\large\faCheck} \\ 
    StarCoder2-15B & NO {\color{darkspringgreen}\large\faCheck} \\ 
};
\draw (m-1-1.north west) -- (m-8-1.south west);
\draw (m-1-2.north east) -- (m-8-2.south east);
\draw (m-1-1.north west) -- (m-1-2.north east);
\draw (m-8-1.south west) -- (m-8-2.south east);
\end{tikzpicture}
\end{multicols}

\tcbsubtitle[halign=center]{Hard}
\begin{multicols}{2}
\begin{lstlisting}[language=Python, frame=single, basicstyle=\ttfamily\srcsize, caption=\textbf{Reference}]
def binary_search(item_list,item):
    first = 0
    last = len(item_list)-1
    found = False
    while( first<=last and not found):
        mid = (first + last)//2
        if item_list[mid] == item :
            found = True
        else:
            if item < item_list[mid]:
                last = mid - 1
            else:
                first = mid + 1	
    return found
\end{lstlisting}
\begin{lstlisting}[language=Python, frame=single, basicstyle=\ttfamily\srcsize, caption=\textbf{Transformed}]
def binary_search(item_list,item):
    first = 0
    last = len(item_list)-1
    found = True
    while( first<=last and not found):
        mid = (first + last)//2
        if item_list[mid] == item :
            found = True
        else:
            if item < item_list[mid]:
                last = mid - 1
            else:
                first = mid + 1	
    return found
\end{lstlisting}
\columnbreak
\vspace*{10pt}
\begin{tikzpicture}[font=\tiny]
\matrix (m) [matrix of nodes, 
             nodes={minimum width=3cm, minimum height=0.7cm},
             row sep=-\pgflinewidth,
             column sep=-\pgflinewidth,
             nodes in empty cells,
             column 1/.style={nodes={draw, anchor=center}},
             column 2/.style={nodes={draw, anchor=center}},
             row 1/.style={nodes={fill=gray!20, font=\bfseries\footnotesize}}
             ] {
    Model          & Prediction \\
    Llama2-7B      & YES {\color{bostonuniversityred}\large\faTimes} \\
    Llama2-13B     & YES {\color{bostonuniversityred}\large\faTimes} \\
    CodeLlama-7B   & YES {\color{bostonuniversityred}\large\faTimes} \\
    CodeLlama-13B  & YES {\color{bostonuniversityred}\large\faTimes} \\
    CodeLlama-34B  & YES {\color{bostonuniversityred}\large\faTimes} \\
    StarCoder2-3B  & YES {\color{bostonuniversityred}\large\faTimes} \\
    StarCoder2-7B  & YES {\color{bostonuniversityred}\large\faTimes} \\ 
    StarCoder2-15B & YES {\color{bostonuniversityred}\large\faTimes} \\
};
\draw (m-1-1.north west) -- (m-8-1.south west);
\draw (m-1-2.north east) -- (m-8-2.south east);
\draw (m-1-1.north west) -- (m-1-2.north east);
\draw (m-8-1.south west) -- (m-8-2.south east);
\end{tikzpicture}
\end{multicols}
\end{tcolorbox}
\caption{In the easy example, the code checks if a triangle is equilateral by comparing the lengths of its three sides. If all sides are equal, it returns True, indicating an equilateral triangle. However, in the transformed code, the else block always returns True regardless of the condition, making the function always return True, even for non-equilateral triangles. The LLM predictions show inconsistent results, with some models correctly identifying the issue in the transformed code while others do not. In the hard example, the provided code snippet implements a binary search algorithm to determine if a specified item exists within a sorted list. The transformed version of the code mistakenly sets the found variable to True at the start, which would cause the loop to exit immediately, effectively rendering the search ineffective. All the LLMs incorrectly identify the transformed code as functionally equivalent to the original code.}
\end{figure*}

\clearpage
\captionsetup[lstlisting]{labelformat=empty}

\lstset{breaklines=true, postbreak=\mbox{\textcolor{red}{$\hookrightarrow$}\space}}

\begin{figure*}[!ht]
\begin{tcolorbox}[enhanced, title={\texttt{and} $\rightarrow$ \texttt{or}}, attach boxed title to top center={yshift=-3mm}, subtitle style={
    bottomrule=0.4pt,
    toprule=0pt,
    leftrule=0pt,
    rightrule=0pt,
    colback=babyblue,
    colframe=black,
    borderline south={0.4pt}{0pt}{black,dotted}
  }]
\tcbsubtitle[halign=center]{Easy}
\begin{multicols}{2}
\begin{lstlisting}[language=Python, frame=single, basicstyle=\ttfamily\srcsize, caption=\textbf{Reference}]
def check_String(str):
    flag_l = False
    flag_n = False
    for i in str: 
        if i.isalpha(): 
            flag_l = True  
        if i.isdigit(): 
            flag_n = True
    return flag_l and flag_n 
\end{lstlisting}
\begin{lstlisting}[language=Python, frame=single, basicstyle=\ttfamily\srcsize, caption=\textbf{Transformed}]
def check_String(str):
    flag_l = False
    flag_n = False
    for i in str: 
        if i.isalpha(): 
            flag_l = True  
        if i.isdigit(): 
            flag_n = True
    return flag_l or flag_n 
\end{lstlisting}
\columnbreak
\begin{tikzpicture}[font=\tiny]
\matrix (m) [matrix of nodes, 
             nodes={minimum width=3cm, minimum height=0.7cm},
             row sep=-\pgflinewidth,
             column sep=-\pgflinewidth,
             nodes in empty cells,
             column 1/.style={nodes={draw, anchor=center}},
             column 2/.style={nodes={draw, anchor=center}},
             row 1/.style={nodes={fill=gray!20, font=\bfseries\footnotesize}}
             ] {
    Model          & Prediction \\
    Llama2-7B      & YES {\color{bostonuniversityred}\large\faTimes} \\
    Llama2-13B     & YES {\color{bostonuniversityred}\large\faTimes} \\
    CodeLlama-7B   & YES {\color{bostonuniversityred}\large\faTimes} \\
    CodeLlama-13B  & NO {\color{darkspringgreen}\large\faCheck} \\ 
    CodeLlama-34B  & YES {\color{bostonuniversityred}\large\faTimes} \\
    StarCoder2-3B  & YES {\color{bostonuniversityred}\large\faTimes} \\
    StarCoder2-7B  & NO {\color{darkspringgreen}\large\faCheck} \\ 
    StarCoder2-15B & NO {\color{darkspringgreen}\large\faCheck} \\ 
};
\draw (m-1-1.north west) -- (m-8-1.south west);
\draw (m-1-2.north east) -- (m-8-2.south east);
\draw (m-1-1.north west) -- (m-1-2.north east);
\draw (m-8-1.south west) -- (m-8-2.south east);
\end{tikzpicture}
\end{multicols}

\tcbsubtitle[halign=center]{Hard}
\begin{multicols}{2}
\begin{lstlisting}[language=Python, frame=single, basicstyle=\ttfamily\srcsize, caption=\textbf{Reference}]
def is_tree_balanced(root):
    class Node: 
        def __init__(self, data): 
            self.data = data 
            self.left = None
            self.right = None
    def get_height(root): 
        if root is None: 
            return 0
        return max(get_height(root.left), get_height(root.right)) + 1
    if root is None: 
        return True
    lh = get_height(root.left) 
    rh = get_height(root.right) 
    if (abs(lh - rh) <= 1) and is_tree_balanced( 
    root.left) is True and is_tree_balanced( root.right) is True: 
        return True
    return False 
\end{lstlisting}
\begin{lstlisting}[language=Python, frame=single, basicstyle=\ttfamily\srcsize, caption=\textbf{Transformed}]
def is_tree_balanced(root):
    class Node: 
        def __init__(self, data): 
            self.data = data 
            self.left = None
            self.right = None
    def get_height(root): 
        if root is None: 
            return 0
        return max(get_height(root.left), get_height(root.right)) + 1
    if root is None: 
        return True
    lh = get_height(root.left) 
    rh = get_height(root.right) 
    if (abs(lh - rh) <= 1) or is_tree_balanced( 
    root.left) is True and is_tree_balanced( root.right) is True: 
        return True
    return False
\end{lstlisting}
\columnbreak
\vspace*{30pt}
\begin{tikzpicture}[font=\tiny]
\matrix (m) [matrix of nodes, 
             nodes={minimum width=3cm, minimum height=0.7cm},
             row sep=-\pgflinewidth,
             column sep=-\pgflinewidth,
             nodes in empty cells,
             column 1/.style={nodes={draw, anchor=center}},
             column 2/.style={nodes={draw, anchor=center}},
             row 1/.style={nodes={fill=gray!20, font=\bfseries\footnotesize}}
             ] {
    Model          & Prediction \\
    Llama2-7B      & YES {\color{bostonuniversityred}\large\faTimes} \\
    Llama2-13B     & YES {\color{bostonuniversityred}\large\faTimes} \\
    CodeLlama-7B   & YES {\color{bostonuniversityred}\large\faTimes} \\
    CodeLlama-13B  & YES {\color{bostonuniversityred}\large\faTimes} \\
    CodeLlama-34B  & YES {\color{bostonuniversityred}\large\faTimes} \\
    StarCoder2-3B  & YES {\color{bostonuniversityred}\large\faTimes} \\
    StarCoder2-7B  & NO {\color{darkspringgreen}\large\faCheck} \\
    StarCoder2-15B & YES {\color{bostonuniversityred}\large\faTimes} \\ 
};
\draw (m-1-1.north west) -- (m-8-1.south west);
\draw (m-1-2.north east) -- (m-8-2.south east);
\draw (m-1-1.north west) -- (m-1-2.north east);
\draw (m-8-1.south west) -- (m-8-2.south east);
\end{tikzpicture}
\end{multicols}
\end{tcolorbox}
\caption{In the easy example, the function returns True only if both conditions are met (using and) in the original version, while the transformed version incorrectly uses or, meaning it will return True if either condition is satisfied. The model predictions indicate varying responses from different LLMs regarding the correctness of the transformed function, with some models confirming its validity and others rejecting it. In the hard example, the transformed version of the code introduces a logical change in the balance condition, using an `or' instead of an `and' operator. Most LLMs misclassify the code's correctness except for one instance.}
\end{figure*}

\captionsetup[lstlisting]{labelformat=empty}

\lstset{breaklines=true, postbreak=\mbox{\textcolor{red}{$\hookrightarrow$}\space}}

\begin{figure*}[!ht]
\begin{tcolorbox}[enhanced, title={\texttt{or} $\rightarrow$ \texttt{and}}, attach boxed title to top center={yshift=-3mm}, subtitle style={
    bottomrule=0.4pt,
    toprule=0pt,
    leftrule=0pt,
    rightrule=0pt,
    colback=babyblue,
    colframe=black,
    borderline south={0.4pt}{0pt}{black,dotted}
  }]
\tcbsubtitle[halign=center]{Easy}
\begin{multicols}{2}
\begin{lstlisting}[language=Python, frame=single, basicstyle=\ttfamily\srcsize, caption=\textbf{Reference}]
def count_char_position(str1):
    count_chars = 0
    for i in range(len(str1)):
        if ((i == ord(str1[i]) - ord('A')) or 
            (i == ord(str1[i]) - ord('a'))): 
            count_chars += 1
    return count_chars 
\end{lstlisting}
\begin{lstlisting}[language=Python, frame=single, basicstyle=\ttfamily\srcsize, caption=\textbf{Transformed}]
def count_char_position(str1):
    count_chars = 0
    for i in range(len(str1)):
        if ((i == ord(str1[i]) - ord('A')) and 
            (i == ord(str1[i]) - ord('a'))): 
            count_chars += 1
    return count_chars 
\end{lstlisting}
\columnbreak
\begin{tikzpicture}[font=\tiny]
\matrix (m) [matrix of nodes, 
             nodes={minimum width=3cm, minimum height=0.7cm},
             row sep=-\pgflinewidth,
             column sep=-\pgflinewidth,
             nodes in empty cells,
             column 1/.style={nodes={draw, anchor=center}},
             column 2/.style={nodes={draw, anchor=center}},
             row 1/.style={nodes={fill=gray!20, font=\bfseries\footnotesize}}
             ] {
    Model          & Prediction \\
    Llama2-7B      & YES {\color{bostonuniversityred}\large\faTimes} \\
    Llama2-13B     & YES {\color{bostonuniversityred}\large\faTimes} \\
    CodeLlama-7B   & YES {\color{bostonuniversityred}\large\faTimes} \\
    CodeLlama-13B  & NO {\color{darkspringgreen}\large\faCheck} \\ 
    CodeLlama-34B  & YES {\color{bostonuniversityred}\large\faTimes} \\
    StarCoder2-3B  & YES {\color{bostonuniversityred}\large\faTimes} \\
    StarCoder2-7B  & NO {\color{darkspringgreen}\large\faCheck} \\ 
    StarCoder2-15B & NO {\color{darkspringgreen}\large\faCheck} \\ 
};
\draw (m-1-1.north west) -- (m-8-1.south west);
\draw (m-1-2.north east) -- (m-8-2.south east);
\draw (m-1-1.north west) -- (m-1-2.north east);
\draw (m-8-1.south west) -- (m-8-2.south east);
\end{tikzpicture}
\end{multicols}

\tcbsubtitle[halign=center]{Hard}
\begin{multicols}{2}
\begin{lstlisting}[language=Python, frame=single, basicstyle=\ttfamily\srcsize, caption=\textbf{Reference}]
def snake_to_camel(word):
    import re
    return ''.join(x.capitalize() or '_' for x in word.split('_'))
\end{lstlisting}
\begin{lstlisting}[language=Python, frame=single, basicstyle=\ttfamily\srcsize, caption=\textbf{Transformed}]
def snake_to_camel(word):
    import re
    return ''.join(x.capitalize() and '_' for x in word.split('_'))
\end{lstlisting}
\columnbreak
\begin{tikzpicture}[font=\tiny]
\matrix (m) [matrix of nodes, 
             nodes={minimum width=3cm, minimum height=0.7cm},
             row sep=-\pgflinewidth,
             column sep=-\pgflinewidth,
             nodes in empty cells,
             column 1/.style={nodes={draw, anchor=center}},
             column 2/.style={nodes={draw, anchor=center}},
             row 1/.style={nodes={fill=gray!20, font=\bfseries\footnotesize}}
             ] {
    Model          & Prediction \\
    Llama2-7B      & YES {\color{bostonuniversityred}\large\faTimes} \\
    Llama2-13B     & YES {\color{bostonuniversityred}\large\faTimes} \\
    CodeLlama-7B   & YES {\color{bostonuniversityred}\large\faTimes} \\
    CodeLlama-13B  & YES {\color{bostonuniversityred}\large\faTimes} \\
    CodeLlama-34B  & YES {\color{bostonuniversityred}\large\faTimes} \\
    StarCoder2-3B  & YES {\color{bostonuniversityred}\large\faTimes} \\
    StarCoder2-7B  & NO {\color{darkspringgreen}\large\faCheck} \\
    StarCoder2-15B & YES {\color{bostonuniversityred}\large\faTimes} \\ 
};
\draw (m-1-1.north west) -- (m-8-1.south west);
\draw (m-1-2.north east) -- (m-8-2.south east);
\draw (m-1-1.north west) -- (m-1-2.north east);
\draw (m-8-1.south west) -- (m-8-2.south east);
\end{tikzpicture}
\end{multicols}
\end{tcolorbox}
\caption{In the easy example, the condition checks if the index $i$ is equal to the ASCII value of the character at that index minus the ASCII value of `A' or `a' in the original version. In the transformed version, the logical operator changes from `or' to `and', which alters the intended functionality. The model predictions indicate varying results across different LLMs, with some confirming the correctness of the original code and others failing to recognize the transformed version. In the hard example, the original implementation uses \texttt{capitalize()} to format each word after splitting the input string by underscores, while the transformed version incorrectly uses `and' instead of `or', which alters the intended behavior. The model predictions indicate that various LLMs struggle to determine the function's purpose, while one variant of StarCoder2 succeeds.}
\end{figure*}

\captionsetup[lstlisting]{labelformat=empty}

\lstset{breaklines=true, postbreak=\mbox{\textcolor{red}{$\hookrightarrow$}\space}}

\begin{figure*}[!ht]
\begin{tcolorbox}[enhanced, title={\texttt{is} $\rightarrow$ \texttt{is not}}, attach boxed title to top center={yshift=-3mm}, subtitle style={
    bottomrule=0.4pt,
    toprule=0pt,
    leftrule=0pt,
    rightrule=0pt,
    colback=babyblue,
    colframe=black,
    borderline south={0.4pt}{0pt}{black,dotted}
  }]
\tcbsubtitle[halign=center]{Easy}
\begin{multicols}{2}
\begin{lstlisting}[language=Python, frame=single, basicstyle=\ttfamily\srcsize, caption=\textbf{Reference}]
def sort_mixed_list(mixed_list):
    int_part = sorted([i for i in mixed_list if type(i) is int])
    str_part = sorted([i for i in mixed_list if type(i) is str])
    return int_part + str_part
\end{lstlisting}
\begin{lstlisting}[language=Python, frame=single, basicstyle=\ttfamily\srcsize, caption=\textbf{Transformed}]
def sort_mixed_list(mixed_list):
    int_part = sorted([i for i in mixed_list if type(i) is not int])
    str_part = sorted([i for i in mixed_list if type(i) is str])
    return int_part + str_part
\end{lstlisting}
\columnbreak
\begin{tikzpicture}[font=\tiny]
\matrix (m) [matrix of nodes, 
             nodes={minimum width=3cm, minimum height=0.7cm},
             row sep=-\pgflinewidth,
             column sep=-\pgflinewidth,
             nodes in empty cells,
             column 1/.style={nodes={draw, anchor=center}},
             column 2/.style={nodes={draw, anchor=center}},
             row 1/.style={nodes={fill=gray!20, font=\bfseries\footnotesize}}
             ] {
    Model          & Prediction \\
    Llama2-7B      & YES {\color{bostonuniversityred}\large\faTimes} \\
    Llama2-13B     & YES {\color{bostonuniversityred}\large\faTimes} \\
    CodeLlama-7B   & YES {\color{bostonuniversityred}\large\faTimes} \\
    CodeLlama-13B  & YES {\color{bostonuniversityred}\large\faTimes} \\
    CodeLlama-34B  & YES {\color{bostonuniversityred}\large\faTimes} \\
    StarCoder2-3B  & YES {\color{bostonuniversityred}\large\faTimes} \\
    StarCoder2-7B  & NO {\color{darkspringgreen}\large\faCheck} \\ 
    StarCoder2-15B & NO {\color{darkspringgreen}\large\faCheck} \\ 
};
\draw (m-1-1.north west) -- (m-8-1.south west);
\draw (m-1-2.north east) -- (m-8-2.south east);
\draw (m-1-1.north west) -- (m-1-2.north east);
\draw (m-8-1.south west) -- (m-8-2.south east);
\end{tikzpicture}
\end{multicols}

\tcbsubtitle[halign=center]{Hard}
\begin{multicols}{2}
\begin{lstlisting}[language=Python, frame=single, basicstyle=\ttfamily\srcsize, caption=\textbf{Reference}]
def is_tree_balanced(root):
    class Node: 
        def __init__(self, data): 
            self.data = data 
            self.left = None
            self.right = None
    def get_height(root): 
        if root is None: 
            return 0
        return max(get_height(root.left), get_height(root.right)) + 1
    if root is None: 
        return True
    lh = get_height(root.left) 
    rh = get_height(root.right) 
    if (abs(lh - rh) <= 1) and is_tree_balanced( 
    root.left) is True and is_tree_balanced( root.right) is True: 
        return True
    return False
\end{lstlisting}
\begin{lstlisting}[language=Python, frame=single, basicstyle=\ttfamily\srcsize, caption=\textbf{Transformed}]
def is_tree_balanced(root):
    class Node: 
        def __init__(self, data): 
            self.data = data 
            self.left = None
            self.right = None
    def get_height(root): 
        if root is not None: 
            return 0
        return max(get_height(root.left), get_height(root.right)) + 1
    if root is None: 
        return True
    lh = get_height(root.left) 
    rh = get_height(root.right) 
    if (abs(lh - rh) <= 1) and is_tree_balanced( 
    root.left) is True and is_tree_balanced( root.right) is True: 
        return True
    return False
\end{lstlisting}
\columnbreak
\vspace*{50pt}
\begin{tikzpicture}[font=\tiny]
\matrix (m) [matrix of nodes, 
             nodes={minimum width=3cm, minimum height=0.7cm},
             row sep=-\pgflinewidth,
             column sep=-\pgflinewidth,
             nodes in empty cells,
             column 1/.style={nodes={draw, anchor=center}},
             column 2/.style={nodes={draw, anchor=center}},
             row 1/.style={nodes={fill=gray!20, font=\bfseries\footnotesize}}
             ] {
    Model          & Prediction \\
    Llama2-7B      & YES {\color{bostonuniversityred}\large\faTimes} \\
    Llama2-13B     & YES {\color{bostonuniversityred}\large\faTimes} \\
    CodeLlama-7B   & YES {\color{bostonuniversityred}\large\faTimes} \\
    CodeLlama-13B  & YES {\color{bostonuniversityred}\large\faTimes} \\
    CodeLlama-34B  & YES {\color{bostonuniversityred}\large\faTimes} \\
    StarCoder2-3B  & YES {\color{bostonuniversityred}\large\faTimes} \\
    StarCoder2-7B  & YES {\color{bostonuniversityred}\large\faTimes} \\
    StarCoder2-15B & YES {\color{bostonuniversityred}\large\faTimes} \\ 
};
\draw (m-1-1.north west) -- (m-8-1.south west);
\draw (m-1-2.north east) -- (m-8-2.south east);
\draw (m-1-1.north west) -- (m-1-2.north east);
\draw (m-8-1.south west) -- (m-8-2.south east);
\end{tikzpicture}
\end{multicols}
\end{tcolorbox}
\caption{In the easy example, the transformed version incorrectly filters for non-integer types instead of integers, which disrupts the intended functionality. The model predictions indicate that several LLMs fail to correctly identify the reference code's functionality, while some versions of StarCoder2 infer correctly. In the hard example, the transformed code contains a minor error in the \texttt{get\_height()} function, where it incorrectly returns 0 when the root is not None. The model predictions indicate that all LLMs fail in assessing for functional equivalence.}
\end{figure*}

\end{document}